\documentclass{aastex}
\usepackage{spr-astr-addons}
\usepackage{url}
\usepackage{graphicx}
\usepackage{epsfig}
\usepackage{epstopdf}
\usepackage{amsmath}

\RequirePackage{color}
\begin{document}

\title{\textbf{Nuclear structure properties and decay rates of molybdenum isotopes }} \slugcomment{Not to appear in Nonlearned J., 45.}

\shorttitle{Short article title} \shortauthors{Autors et al.}
\author{Jameel-Un Nabi\altaffilmark{1}}
\affil{Faculty of Engineering Sciences, GIK Institute of Engineering
Sciences and Technology, Topi 23640, Swabi, Khyber Pakhtunkhwa,
Pakistan} \email{jameel@giki.edu.pk}
\author{Tuncay Bayram\altaffilmark{2}}
\affil{Department of Physics, Faculty of Science, Karadeniz
Technical University, Trabzon, Turkey}\email{t.bayram@ktu.edu.tr}
 \altaffiltext{1}{Corresponding
author email : jameel@giki.edu.pk} \altaffiltext{2}{email :
t.bayram@ktu.edu.tr}

\begin{abstract}
Electron capture and $\beta^{-}$ decay are the
dominant decay processes during late
 phases of evolution of heavy stars. Previous simulation results show that weak rates on isotopes of Molybdenum (Mo) have a meaningful contribution during the development  of phases of stars before they go supernova. The relative abundance coupled with the stellar weak rates on Mo isotopes may change the lepton-to-baryon content of the core material. 
Here we report on the calculation of nuclear structure properties of $^{82-138}$Mo isotopes employing the RMF model. Later we calculate the weak decay rates of these isotopes. We use the pn-QRPA model to compute these rates. 
In the first step, the ground-state
nuclear properties  of Mo isotopes such as binding energy per nucleon,
neutron and proton separation energies, charge radii, total electric quadrupole
moments and deformation parameter of electric quadrupole moments have been calculated using density dependent version of RMF model with DD-PC1 and DD-ME2 functionals. The
calculated electric quadrupole deformation parameters have been used in a deformed pn-QRPA calculation in the second phase of this work to calculate half-lives and weak decay rates for these Mo isotopes in stellar matter. 
We calculate the 
electron capture and $\beta$-decay rates over an extensive range of
temperature (0.01$\times10^{9}$ K to 30$\times10^{9}$ K) and
density  ($10$ to $10^{11}$) g/cm$^{3}$.
Our study can prove useful for simulation of presupernova evolution processes of  stars.
\end{abstract}
\section{Introduction}
The production of energy
in stars \citep{a1}, the associated nucleosynthesis
\citep{a2} and supernova explosion dynamics \citep{a3} are still not fully understood. To date these processes are extensively investigated by astrophysicists in an attempt to understand how our universe works. It is the weak
interaction mediated rates which dictate the terms and conditions for the process of
nucleosynthesis and the dynamics of supernova explosions. The study of charge-changing transitions in stellar matter is one of the important inputs for core collapse simulation
\citep{a4,a4+}. The charge-changing transitions greatly effect the late evolutionary phases of
massive stars. Electron capture (EC),  positron
capture (PC) and $\beta$-decay of nuclei in stellar core influence these transformations.

Fermi and Gamow-Teller (GT) transitions govern
both $\beta$-decay and EC rates.  GT strength functions are required to calculate weak decay rates. The  $BGT_+$
strength transforms a proton into a neutron, whereas the transformation of a neutron to proton is accomplished by the
 $BGT_-$ strength function.


The proton neutron quasi particle random phase approximation (pn-QRPA) is commonly used to perform  calculation of stellar weak rates of heavy nuclei \citep{a12}. These calculations are fully microscopic in nature. They do not make use of the so-called Brink-Axel hypothesis \citep{a14} used by many other genre of stellar weak rate calculations.
It was reported by authors \citep{a13} that Brink-Axel hypothesis is a rather compromised estimate for computing the weak decay rates. Many authors successfully used the pn-QRPA model in the past. \citep{a18}
used the QRPA continuum approach constructed on density functional
theory. To calculate   $\beta^{-}$ decay in axially deformed
even-even nuclei \citep{a19} applied
the method of finite amplitude by modifying QRPA with Skyrme energy
density functionals. For calculation of stellar EC 
\citep{a20} modified the QRPA model by inserting self-consistent mean
field Skyrme Hartree-Fock. The first
attempt to use the pn-QRPA model to compute weak rates  for
nuclei with A ranging from 18 -- 100 was done by  \citep{a12}. 

One can find macroscopic and microscopic nuclear models
for correct prediction of masses and deformations of
nuclei \citep{greiner1996}. In particular, Density Functional Theory
(DFT) is a widely used nuclear model for carrying out various properties of nuclei in wide mass region \citep{ring1997, lalazissis1999, bayram2013b} . Skyrme \citep{vautherin1972} and Gogny type \citep{decharge1980} interactions are successful \citep{lalazissis2009}. In the beginning relativistic mean field (RMF) model was established as to be quantum field theory of nuclear matter~\citep{walecka1974}. Later, it turned out to a covariant form of DFT because of its additional density dependence introduced by Ref.~\citep{boguta1977} for better description of nuclear surface properties of nuclear matter. RMF model has gained much attention and is applied for prediction of nuclear properties of finite nuclei \citep{geng2005, agbemava2014, lu2015}. The 
RMF model can predict deformations of nuclei well~\citep{ring1996,geng2003}. Regarding this point we have employed a hybrid calculations on Mo isotopic chain by using 
RMF+pn-QRPA models. Neutron-rich Mo isotopes play a role in the nucleosynthesis of heavy nuclei. Their masses and decay properties can be taken 
as inputs to model astrophysical  $r$-process investigations. In our previous study~\citep{yilmaz2011} non-linear version of RMF model was applied for calculation of 
masses, radii and deformations for  even-even $^{84-110}$Mo nuclei. Also shape evolution of even-even Mo nuclei were investigated by using potential energy curves as a function of quadrupole moment deformation parameter ($\beta_{2}$) in the RMF model. In the present study more reliable functionals DD-ME2 \citep{lalazissis2005} and DD-PC1 \citep{niksic2008} for RMF model have been employed for calculation of ground-state masses, sizes and quadrupole deformations of Mo nuclei starting from neutron drip-line to proton drip-line. The calculated  $\beta_{2}$ values were later used as input parameter in the pn-QRPA model for calculation of terrestrial and stellar decay rates of molybdenum isotopes. The calculation of electron capture (EC) and $\beta^{-}$ decay rates in stellar environments along with its astrophysical significance  are further discussed.

The paper is designed as follows. Section 2 briefly describes the necessary theoretical framework used in our calculation. We present and discuss our results in Section 3. Conclusions are stated in Section~4.  


\section{Theoretical Formalism}

\subsection{The RMF Model}

Briefly nucleons interact with each other via exchange of various 
mesons and photons in the RMF model \citep{walecka1974}.  
A detailed discussions of RMF theory and its applications can be found in Refs. \citep{gambhir1990, typel1999, ring1996, vretenar2005, bayram2018, tian2009, bayram2013b}. In the RMF model $\sigma$, $\omega$ and $\rho$ mesons are
widely considered. In the simplest version of the model 
interactions of mesons among themselves were not considered but it was 
understood that model did not provide incompressibility of nuclear matter much. Because of this, a self interaction term of the $\sigma$ mesons was added by Ref.~\citep{boguta1977}. Later more reliable versions of RMF model were introduced by means of handling of interactions such as non-linear self interaction of the $\omega$ and $\rho$ mesons and density dependent meson-nucleon couplings which can be found in Refs. \citep{pena16, Sugahara94, Lenske95, Piekarewicz02}. In this paper, the functionals DD-PC1 and DD-ME2 have been launched for calculation of ground-state nuclear properties of Mo isotopes. In this section RMF model is briefly described only by means of medium dependent vertices.     

The RMF model starts with an effective Lagrangian $\mathcal{L}$ density to obtain equations of motion for describing nuclear properties of nuclei. It includes terms for free nucleons, mesons and nucleon-meson interactions for producing 
nuclear properties of nuclei. The Lagrangian density can be given as  

\begin{eqnarray}
\footnotesize
\mathcal{L}=\bar\psi(i\gamma~.~\partial - m)\psi +
\frac{1}{2} (\partial\sigma)^{2}
-\frac{1}{2}m_{\sigma}^{2}\sigma^{2} 
-\frac{1}{4}\boldsymbol{\Omega}_{\mu\nu}\boldsymbol{\Omega}^{\mu\nu}\nonumber\\
+\frac{1}{2}m_{\omega}^{2}\omega^{2} -
\frac{1}{4}\overrightarrow{\boldsymbol{R}}_{\mu\nu}\overrightarrow{\boldsymbol{R}}^{\mu\nu}
+ \frac{1}{2}m_{\rho}^{2}\overrightarrow{\rho}^{2} \label{Lagrangian} \\
- \frac{1}{4}\boldsymbol{F}_{\mu\nu}\boldsymbol{F}^{\mu\nu} - g_{\sigma}\bar{\psi}\sigma\psi - g_{\omega}\bar{\psi}\gamma~.~\omega\psi\nonumber\\
- g_{\sigma}\bar{\psi}\gamma~.~\overrightarrow{\rho}\overrightarrow{\tau}\psi
- A\frac{1-\tau_{3}}{2}\psi,\nonumber 
\end{eqnarray}

\noindent where the masses of mesons (fields) are denoted by
$m_{\sigma}$ ($\sigma$), $m_{\omega}$ ($\omega$) and $m_{\rho}$
($\rho$). The meson-nucleon couplings of $\sigma$, $\omega$ and $\rho$ are represented by $g_\sigma$, $g_\omega$ and $g_\rho$, respectively. 
$\psi$ represents Dirac spinor for nucleons with mass ($m$) and 
bold type symbols donate space vectors. Arrows indicate isospin vectors. In this equation, first term is for free nucleons 
and last four terms are for meson-nucleon interactions while rest 
are for free mesons. Field tensors related with $\omega$, $\rho$ and 
photon vector fields are given by the equation 

\begin{eqnarray}
\boldsymbol{\Omega}^{\mu\nu}=\partial^{\mu}\omega^{\nu}-
\partial^{\nu}\omega^{\mu}, \nonumber \label{Omega}\\
\overrightarrow{\boldsymbol{R}}^{\mu\nu}=\partial^{\mu}\overrightarrow{\rho}^{\nu}-
\partial^{\nu}\overrightarrow{\rho}^{\mu}, \label{Rho}\\
\boldsymbol{F}^{\mu\nu}=\partial^{\mu}A^{\nu}-\partial^{\nu}A^{\mu}
\nonumber. \label{Photon}
\end{eqnarray}

The Lagrangian density stated in Eq.~(\ref{Lagrangian}) remains invariant under parity
transformation. Coupling constants and unknown 
meson masses in Eq.~(\ref{Lagrangian}) can be adjusted by using experimental data 
for reliable production of nuclear properties of nuclei. Application of variational 
principle by using $\mathcal{L}$ in Eq.~(\ref{Lagrangian}) produces equations of motion for the fields which are Dirac and Klein-Gordon like equations. These set of coupled equations can be solved for symmetric, axially deformed and triaxially deformed cases iteratively. In this paper, prescriptions of Ref.~\cite{niksic2014} have been used for RMF model calculation.   
\subsection{The pn-QRPA model}

We used the following Hamiltonian for solving the Schroedinger Equation of our system
\begin {equation} \label{Ham}
H^{qrpa}= H^{sp}+V^{pairing}+V^{pp}_{GT}+V^{ph}_{GT},
\end {equation}
here $H^{sp}$ is the single particle Hamiltonian.
Pairing forces were represented by the second term  within the framework of Bardeen–Cooper–Schrieffer (BCS)
approximation. The model included GT force with separable
particle-hole ($ph$) and particle-particle ($pp$) matrix elements. The last two terms represent the
particle-particle $(pp)$ and  particle-hole $(ph)$ GT forces. The wave functions and energy eigenvalues of single particle were
computed employing Nilsson model \citep{a32}. The $ph$ interaction
constant was characterized by constant $\chi$ whereas the $pp$ interaction was set by the parameter $\kappa$ in our calculation. These parameters were fine tuned in order to reproduce the measured half-lives of the Mo isotopes.
We used $\chi$= $64.6/A$ which
shows 1/A dependence \citep{a33} and $\kappa$= 5.6/A. Other parameters employed in the pn-QRPA model were the pairing
gaps ( $\bigtriangleup_n$ and $\bigtriangleup_p$),
 Q values, the Nilsson potential parameters (NPP) and the nuclear deformation
parameter ($\beta_{2}$). The Nilsson oscillator constant
was represented by $\hbar\omega$=41/$A^{1/3}$, whereas NPP were selected from \citep{a34}. Traditional choice of pairing gaps were adopted in our calculation
\begin {equation}
\bigtriangleup_n = \bigtriangleup_p = 12/\sqrt{A}(MeV).
\end {equation}
$\beta_{2}$ values for the Mo isotopes were computed using the RMF model described earlier.   We took Q-values from  \citep{a47}. For solution of the pn-QRPA Hamiltonian Eq.~(\ref{Ham}) one may see \citep{a39}.

The EC and $\beta$-decay rates of Mo isotopes transforming from
$i{th}$ parent state to the $j{th}$ daughter state in stellar matter
were determined by
\begin{equation} \label{ECBD}
\lambda^{EC(\beta^{-})}_{ij} = \ln2\frac{f_{ij}^{EC(\beta^{-})}(T,\rho,E_{f})}{(ft)_{ij}},
\end{equation}
The  $ft_{ij}$ in Eq.~(\ref{ECBD}) is connected to the reduced
transition probability ($B_{ij}$) of the charge-changing transitions through
\begin {equation}\label{ft}
ft_{ij} ={D}/B_{ij},
\end {equation}
where $D$ = 6143 $s$ \citep{Har09} and 
the $B_{ij}$ is given by
\begin {equation}
B_{ij} = ((g_{A}/g_{V})^2 B(GT)_{ij})+B(F)_{ij}.
\end{equation}
In Eq.~(7) the value of $(g_{A}/g_{V})$ was taken as -1.2694 \citep{Nak10}. The reduced GT
($\Delta J^{\pi}$ =1$^{+}$) transition probabilities were given by
\begin{equation}\label{bgt}
B(GT)_{ij} = \frac{1}{2J_{i}+1} \mid <j
\parallel \sum_{l}\tau_{\pm}^{l}\vec{\sigma}^{l} \parallel i> \mid ^{2},
\end{equation}
whereas the reduced Fermi ($\Delta J^{\pi}$ =0$^{+}$) transition
probabilities were given by
\begin{equation}
B(F)_{ij} = \frac{1}{2J_{i}+1} \mid<j \parallel \sum_{l}\tau_{\pm}^{l}
\parallel i> \mid ^{2},
\end{equation}
where $\sigma$ are the spin  and $\tau$ are the isospin  operators (raising and lowering).

For the case of $\beta$-decay the phase space integrals
represented by $f_{ij}$  in Eq.~(\ref{ECBD}) were given by (using natural units, $c=\hbar=m_{e}=1$)
\begin{equation}\label{psbd}
f_{ij}^{\beta^{-}} = \int_{1}^{w_{m}} w \sqrt{w^{2}-1} (w_{m}-w)^{2} F(+ Z,w)
(1-Z_-) dw,
\end{equation}
where $w$ is the k.e.  of the electron inclusive of its
rest mass and $w_{m}$ is the total $\beta$-decay energy ($ w_{m}
= m_{i}+E_{i}-m_{j}-E_{j}$, where $m_{i}$ and $E_{i}$ are the mass
and energy eigenvalues of the parent, and $m_{j}$ and
$E_{j}$ of the daughter, respectively).

For EC the phase space integrals were given by
\begin{equation}\label{psec}
f_{ij}^{EC} = \int _{w_{l} }^{\infty }w\sqrt{w^{2} -1}  (w_{m}
+w)^{2} F(+ Z,w)Z_{- } dw,
\end{equation}
where  $w_{l}$ denotes the total capture threshold energy
for capture. The Fermi functions appearing in Eq.~(10) and Eq.~(11)  were computed 
according to the procedure used by \citep{a37}. $Z_{-}$ is the
 distribution function for electrons and is given by
\begin{equation}
Z_{-} =\left[\exp \left(\frac{w-1-E_{f} }{kT} \right)+1\right]^{-1}.
\end{equation}
Here $E_{f}$ is
the Fermi energy,  $k$ is
the Boltzmann constant and $T$ is the temperature.

The total EC and $\beta$-decay rates were given by
\begin {equation}
\lambda^{EC(\beta^{-})} = \sum_{ij}P_{i} \lambda^{EC(\beta^{-})}_{ij},
\end {equation}
here $P_{i}$ shows the occupation probability of parent excited
state. Convergence was ensured in our calculation of total weak rates. 

\section {Results and Discussion}

\begin{figure*}[htbp]
\begin{center}
\includegraphics[height=0.5\textwidth]{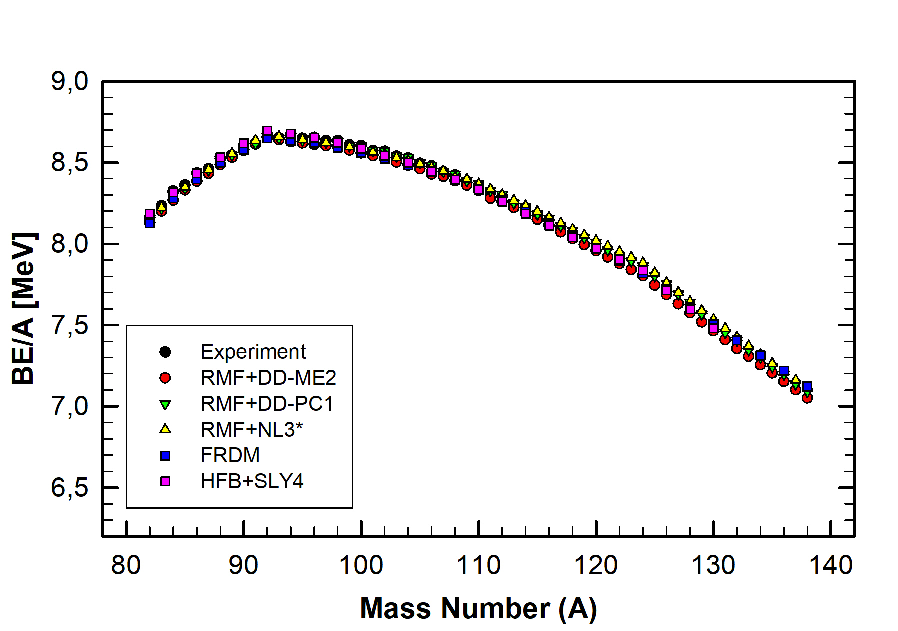}
\caption{The calculated BE/A values for isotopic chain of Mo nuclei using RMF model with DD-ME2 and DD-PC1 interactions. The results of RMF model with NL3* interaction \citep{bayram2013a}, HFB theory with Sly4 parameter set \citep{stoitsov2003}, FRDM \citep{moller1997} and latest experimental data \citep{wang2016} are also shown for comparison.}
\label{Graph_rmf1}
\end{center}
\end{figure*}

\begin{figure*}[htbp]
\begin{center}
\includegraphics[height=0.8\textwidth]{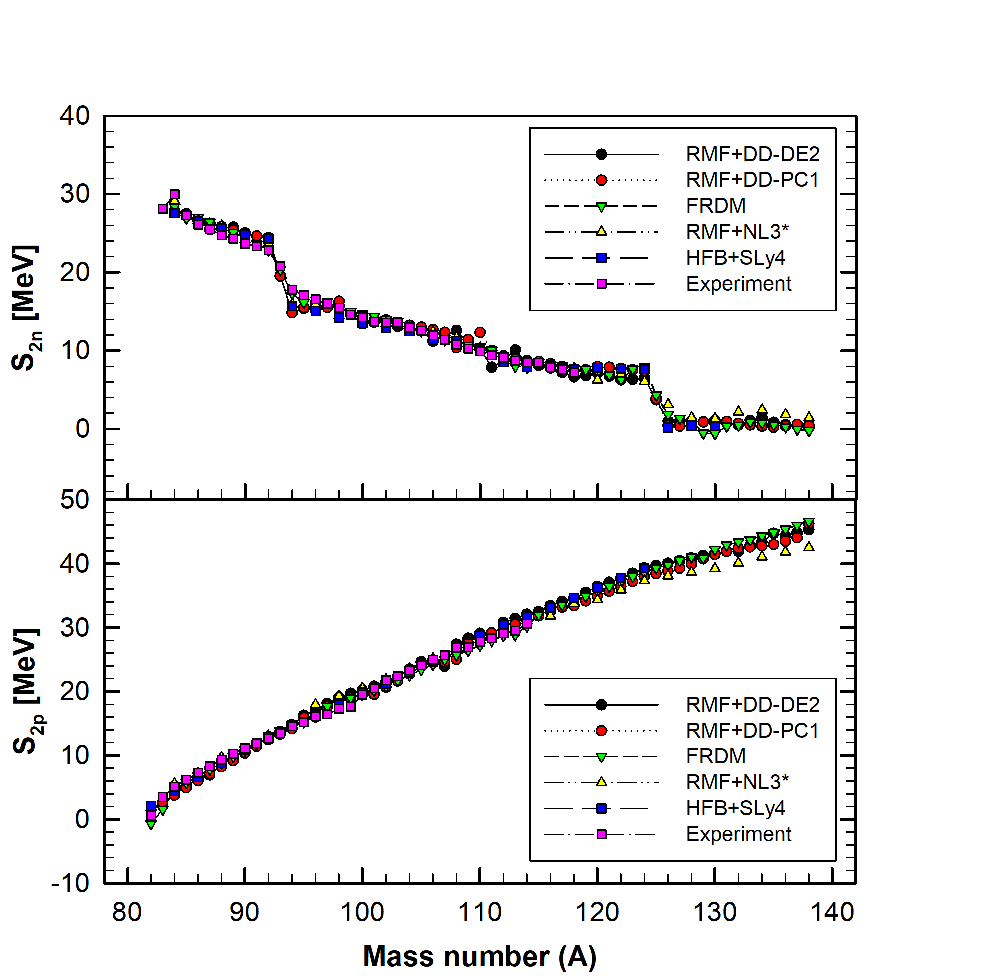}
\caption{The calculated two-neutron (upper panel) and two-proton (lower panel) separation energies for $^{82-138}$Mo using RMF model with DD-ME2 and DD-PC1 interactions. Theoretical predictions from RMF model with NL3* interaction \citep{bayram2013a}, HFB theory with SLy4 parameter set \citep{stoitsov2003}, FRDM \citep{moller1997} and experimental data~\citep{wang2016} are also shown.}
\label{Graph_rmf2}
\end{center}
\end{figure*}

The ground-state BE/A (binding energies per nucleon) of 
$^{82-138}$Mo isotopes have been calculated by using DD-ME2 
and DD-PC1 interactions. Axially symmetric case was considered in these calculations. Pairing correlations were handled by employing BCS formalism with constant
gap approximation. In the axially symmetric case there can be found oblate and prolate deformation of nuclei. Therefore, calculations were done for both prolate and oblate configuration of nuclei and the configuration with lowest binding energy was considered as ground state of nuclei. As can be seen in Fig.~\ref{Graph_rmf1} the BE/A of $^{82-138}$Mo isotopes are in agreement with experimental data as well as predictions of FRDM and HFB models. All calculations and 
experimental data clearly show a bend at mass number $A=92$ because of magic neutron number $N=50$ which is related with shell closure. It can be expected that one and two neutron separation energies may change suddenly at magic neutron number in an isotopic chain of nuclei because shell closure makes nucleon separation energy more bigger than those of neighboring isotopes. This 
point may be taken as a check for the success of nuclear models. In Fig.~\ref{Graph_rmf2}, calculated two-neutron and two-poton separation energies of $^{82-138}$Mo isotopes are shown in comparison with available experimental data~\citep{wang2016}. Also, the predictions of RMF model with NL3* parameter set~\citep{bayram2013a}, HFB theory with SLy4 parameter set~\citep{stoitsov2003} and FRDM \citep{moller1997} are shown for comparison. Two-neutron separation energies ($S_{2n}$) were determined by using binding energy differences of isotopes ($S_{2n} = BE(Z,N)-BE(Z,N-2)$). In a similar manner we  
calculated two-proton separation energies ($S_{2p}$) by using the basic formula $S_{2p} = BE(Z,N)-BE(Z-2,N)$ in the presented results. All theoretical models show shell closures at mass numbers $A=92$ ($N=50$) and $A=124$ ($N=82$) in agreement with experimental data.  

\begin{figure*}[htbp]
\begin{center}
\includegraphics[height=0.5\textwidth]{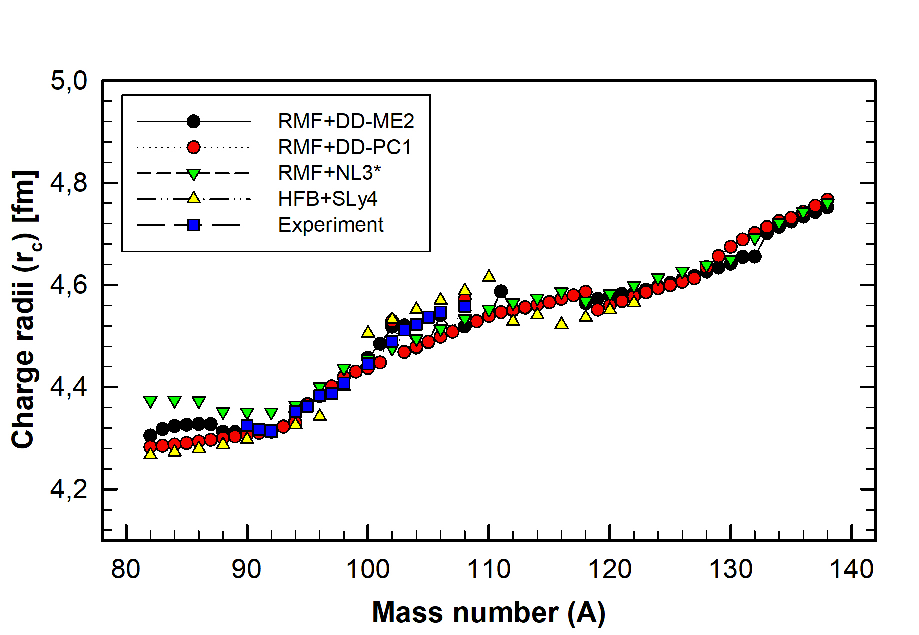}
\caption{The calculated rms charge radii of Mo isotopes using RMF model with
DD-ME2 and DD-PC1 interactions in comparison with the results of HFB method \citep{stoitsov2003},  RMF model with NL3* interaction~\citep{bayram2013a} and available experimental data~\citep{angeli2013}.}
\label{Graph_rmf3}
\end{center}
\end{figure*}

One of the important quantities of nuclei is its nuclear charge radii. Our calculations for charge radii of Mo isotopes are shown in Fig.~\ref{Graph_rmf3} together with available experimental data~\citep{angeli2013}. Also, the predictions of RMF model with non-linear NL3* parametrization~\citep{bayram2013a} and  HFB theory~\citep{stoitsov2003} are shown for comparison. It is clearly seen from Fig.~\ref{Graph_rmf3} that the calculated results with DD-ME2 interaction have better consistency with the experimental data than remaining theoretical models for Mo isotopes. In agreement with our previous discussion on shell closure at neutron number $N=50$, one can  see an elbow at mass number $A=92$ in the curves of experimental data and the calculated results of DD-ME2 interaction.

Furthermore electric quadrupole moments and associated deformation parameters ($\beta_{2}$) were calculated in RMF model in the present study. The $\beta_{2}$ values calculated in the RMF model were used as input parameter in the  pn-QRPA model calculation of terrestrial and stellar weak decay rates of Mo isotopes. Our calculated results with both DD-ME2 and DD-PC1 functionals for ground-state properties of $^{82-138}$Mo isotopes have been listed in Tables~\ref{me2} and \ref{pc1}. Generally DD-ME2 interaction gives closer results to experimental data than the DD-PC1 interaction for Mo isotopes by means of binding energy, neutron and proton separation energies and rms charge radii. By considering this point we later used $\beta_{2}$ values calculated with DD-ME2 interaction as the input model parameter in our pn-QRPA rate calculation.

The GT strength distribution functions of selected Mo isotopes were computed using the pn-QRPA model. The same model was later used to calculate terrestrial half-lives and  $\beta$-decay/EC on these selected nuclei.  We selected a total of 55 isotopes of
molybdenum with mass range $^{82-94}$Mo, $^{96}$Mo, $^{98-138}$Mo,
for the calculation of allowed weak rates along with calculation of
GT strength and half -lives. These isotopes contains
both stable ($^{92}$Mo, $^{94}$Mo, $^{96}$Mo, $^{98}$Mo and
$^{100}$Mo) and unstable species. These unstable isotopes
of Mo include both kinds of neutron rich and neutron deficient
species. All results were quenched by a factor of 0.6 (also used in previous calculations \citep{a44,a46}).

\begin{figure*}[htbp]
\begin{center}
\includegraphics[height=0.35\textwidth]{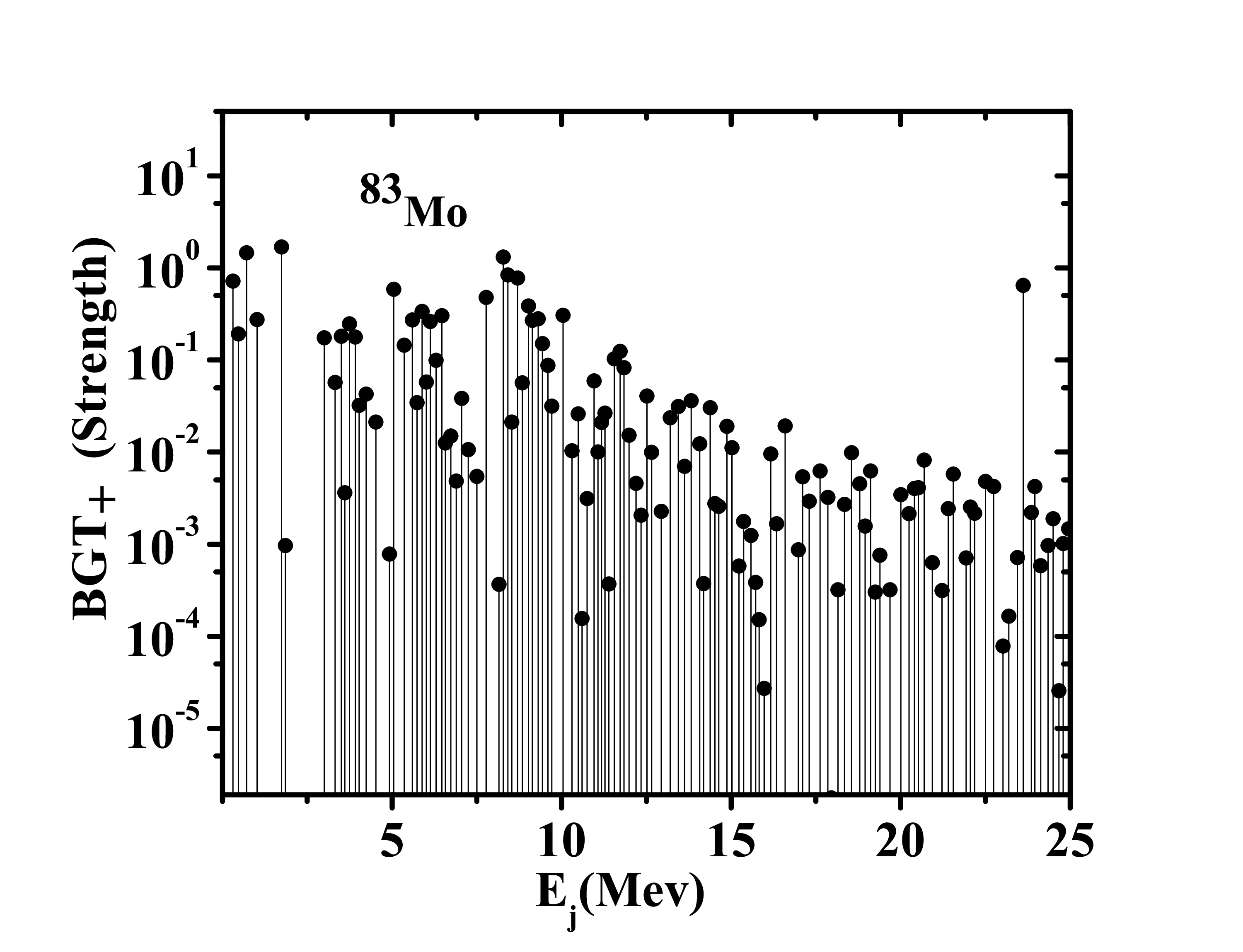}
\includegraphics[height=0.35\textwidth]{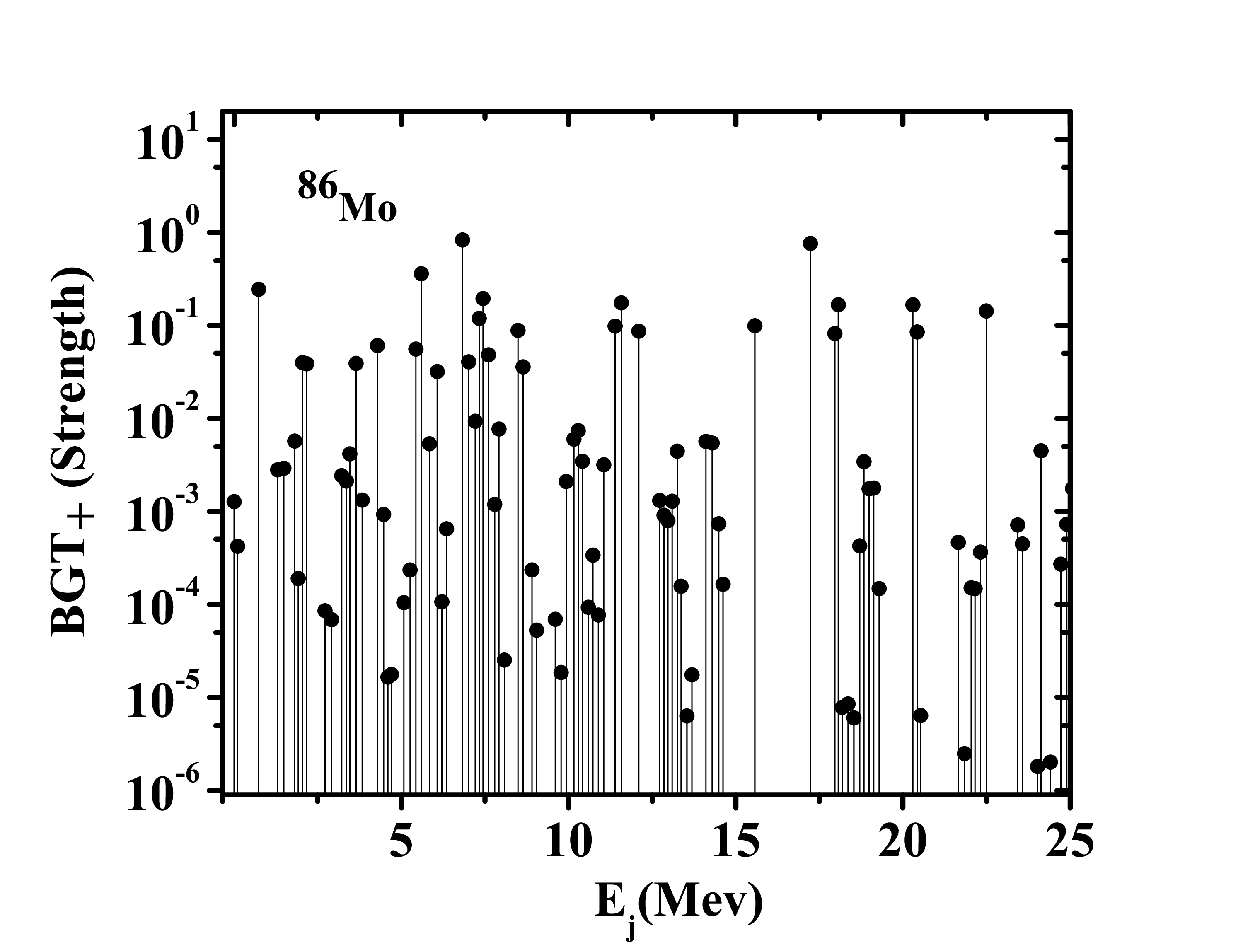}
\includegraphics[height=0.35\textwidth]{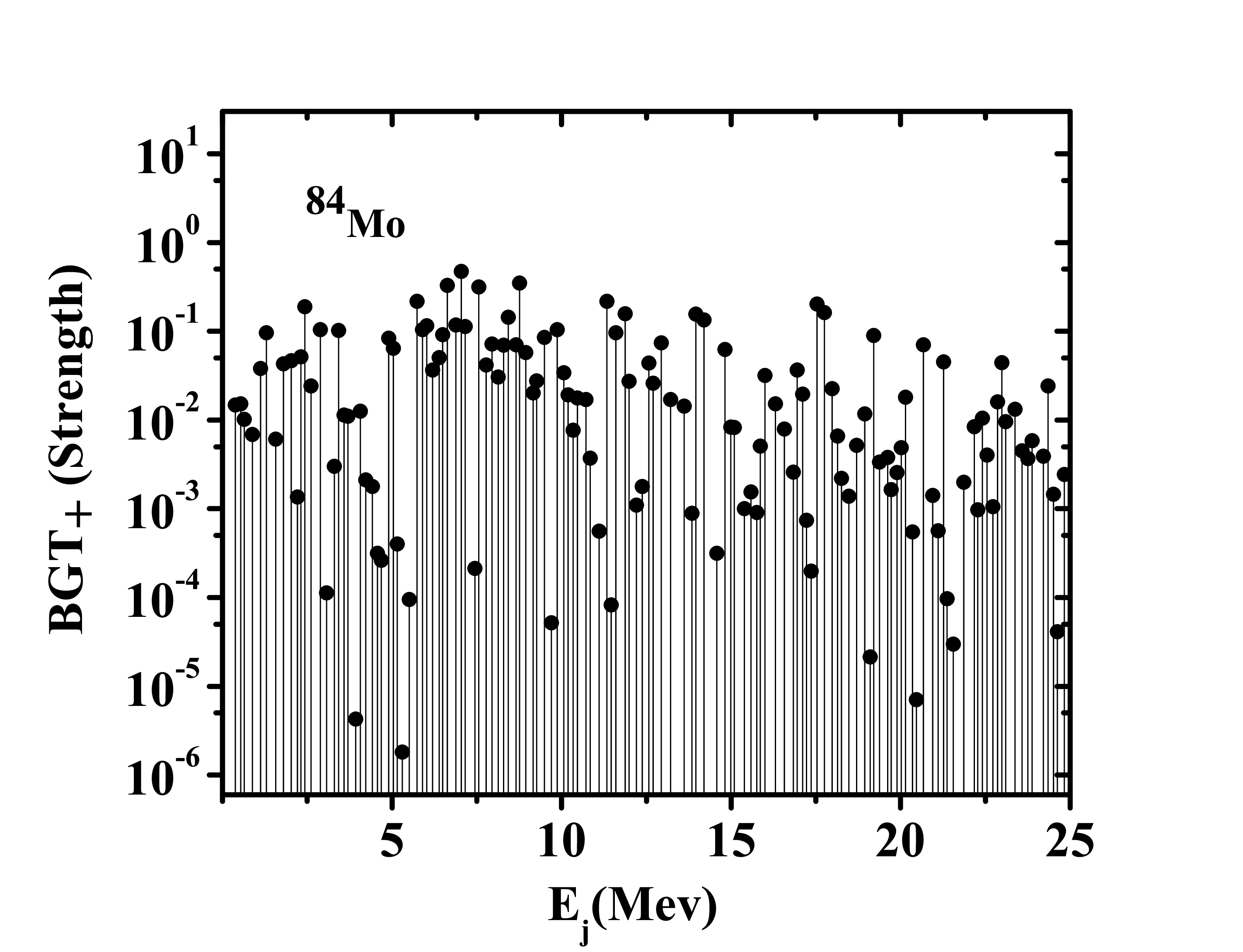}
\includegraphics[height=0.35\textwidth]{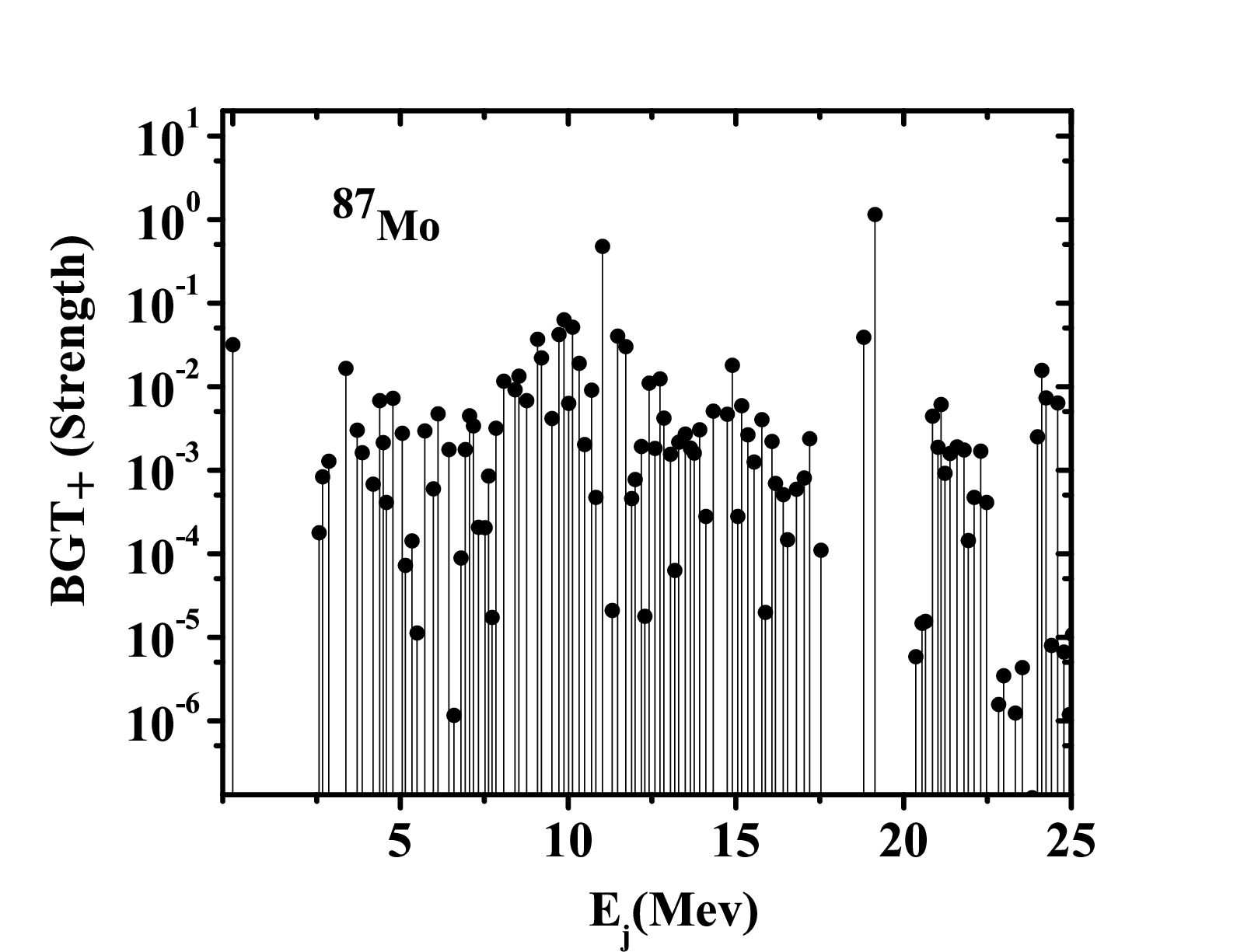}
\includegraphics[height=0.35\textwidth]{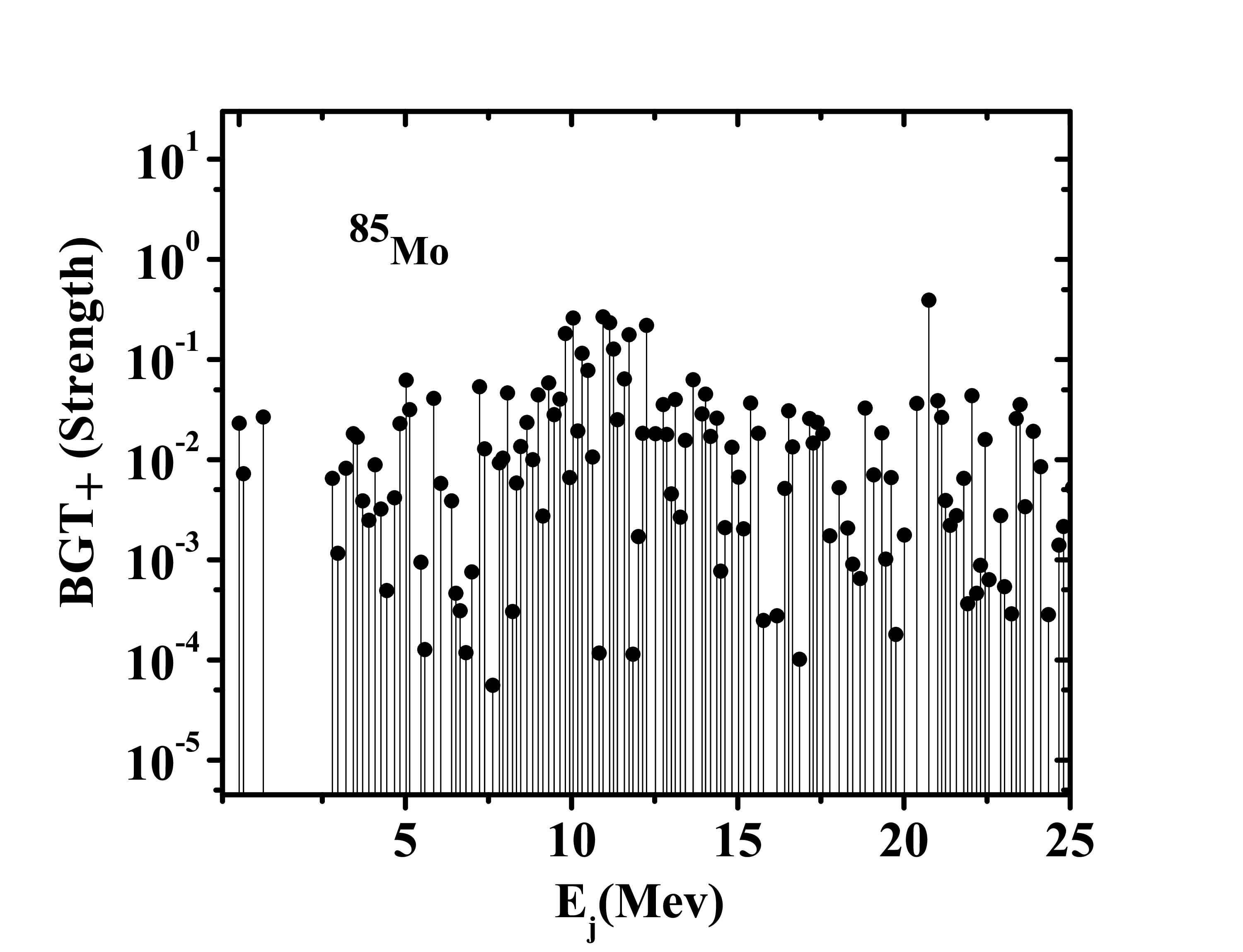}
\includegraphics[height=0.35\textwidth]{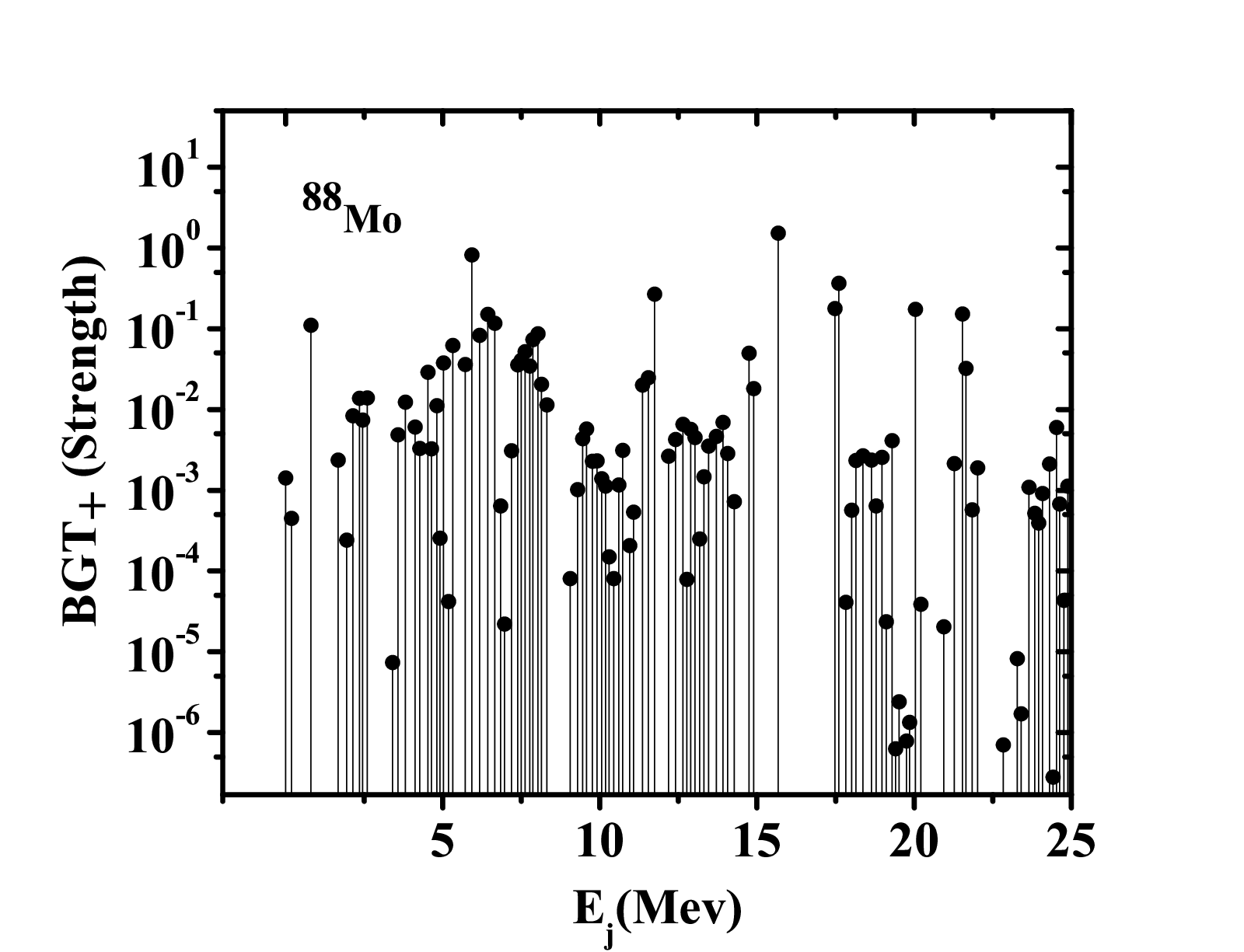}
\caption{The pn-QRPA calculated  GT strength distributions of $^{83-88}$Mo in daughter nuclei in the EC direction.} \label{GTEC}
\end{center}
\end{figure*}

\begin{figure*}[htbp]
\begin{center}
\includegraphics[height=0.35\textwidth]{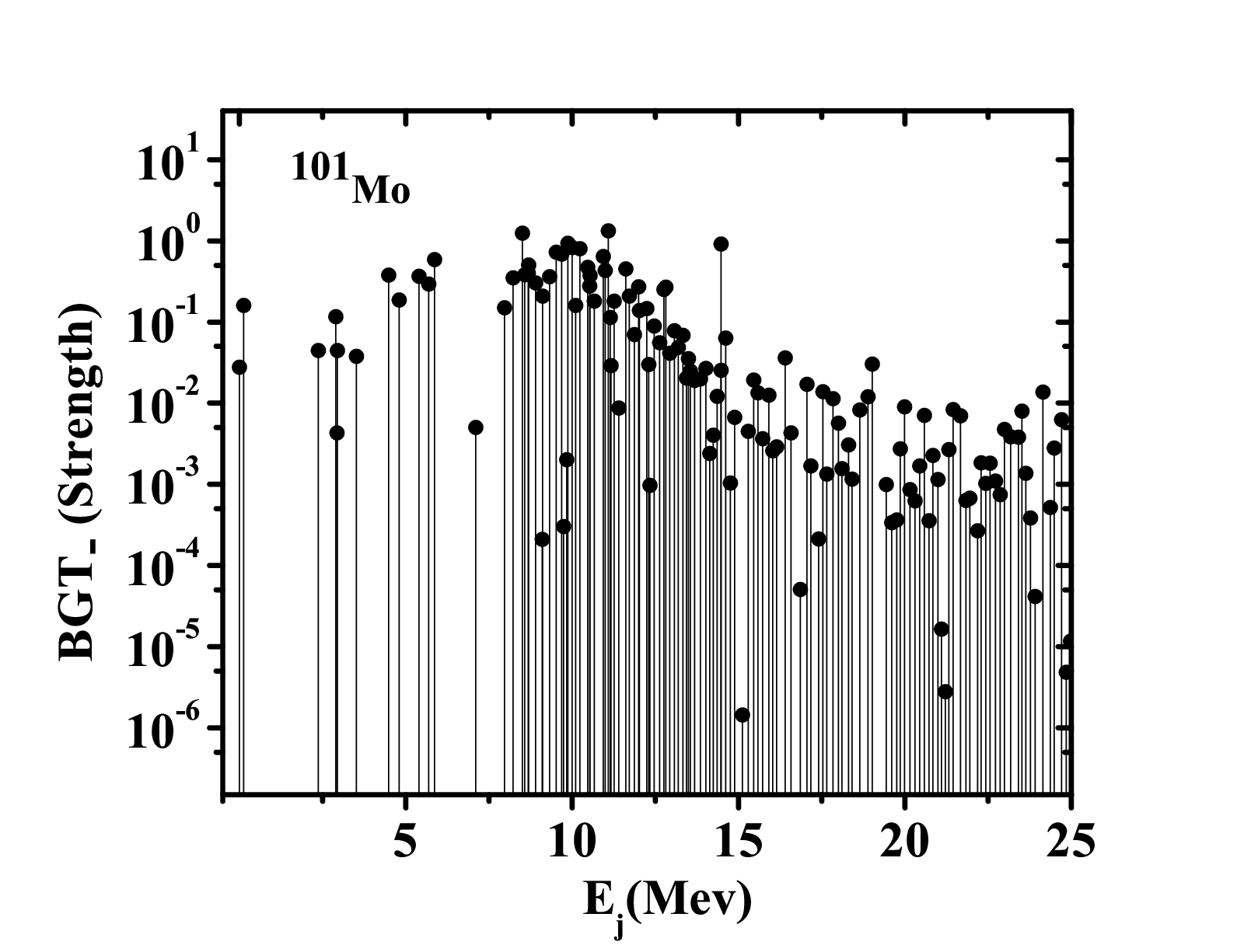}
\includegraphics[height=0.35\textwidth]{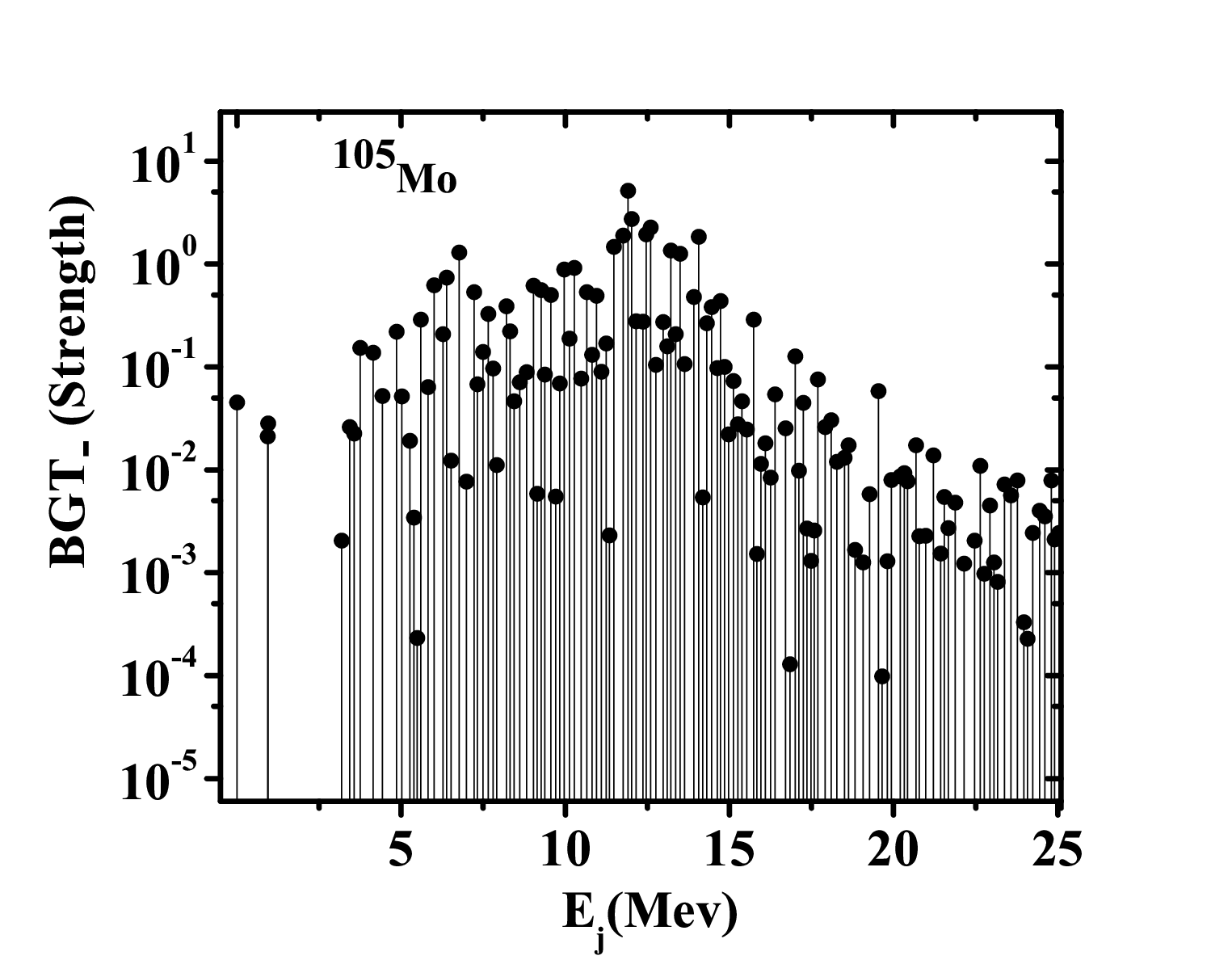}
\includegraphics[height=0.35\textwidth]{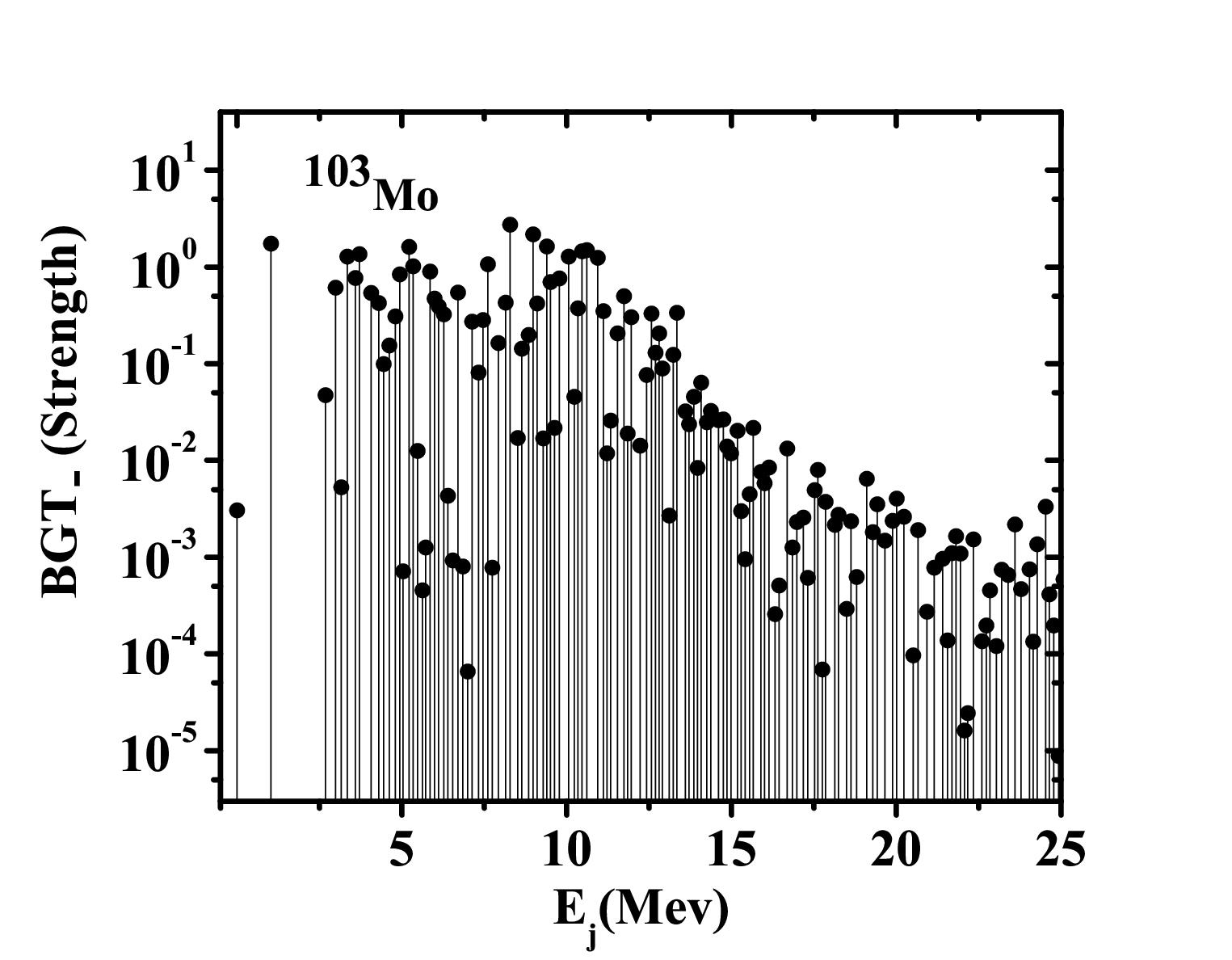}
\includegraphics[height=0.35\textwidth]{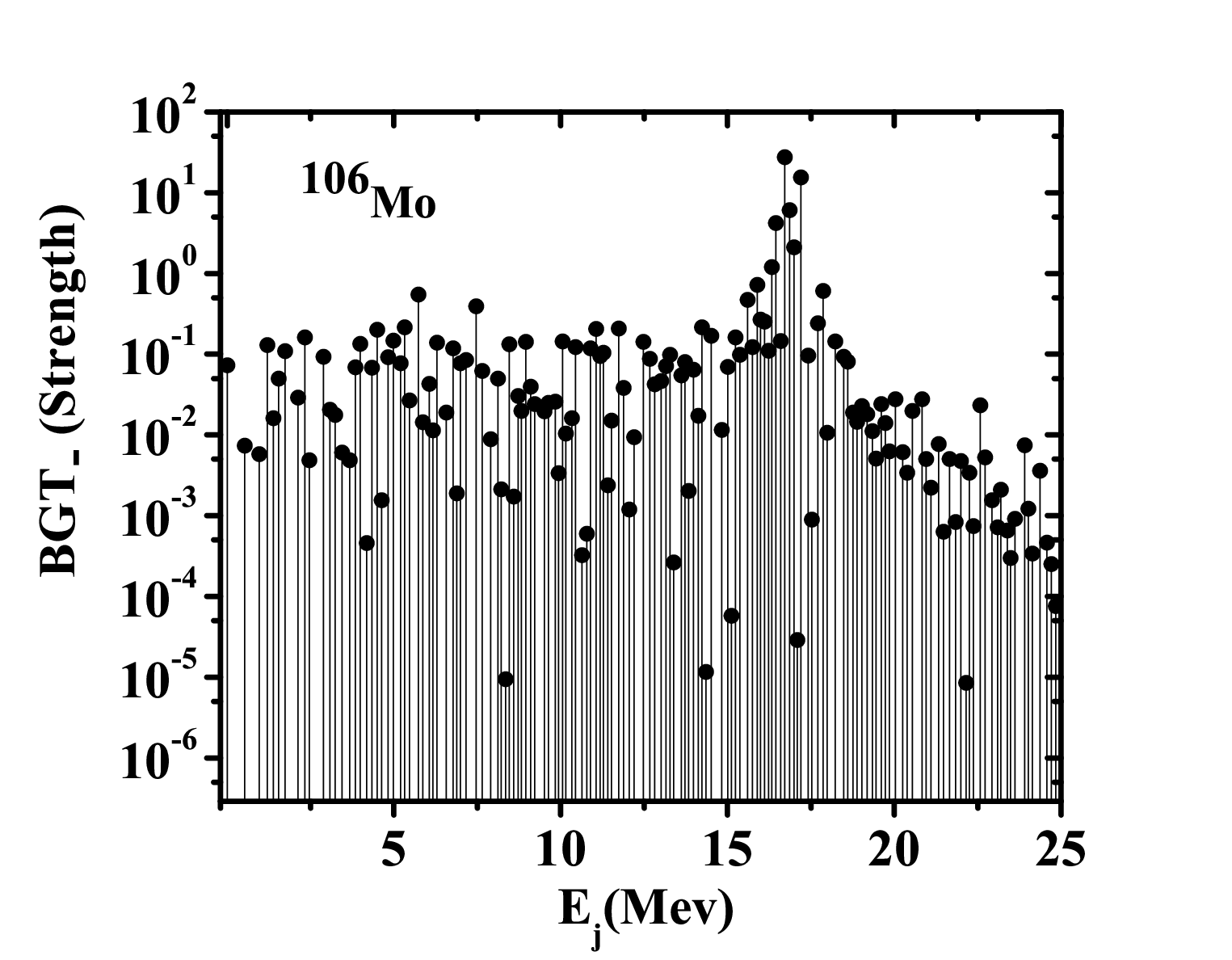}
\includegraphics[height=0.35\textwidth]{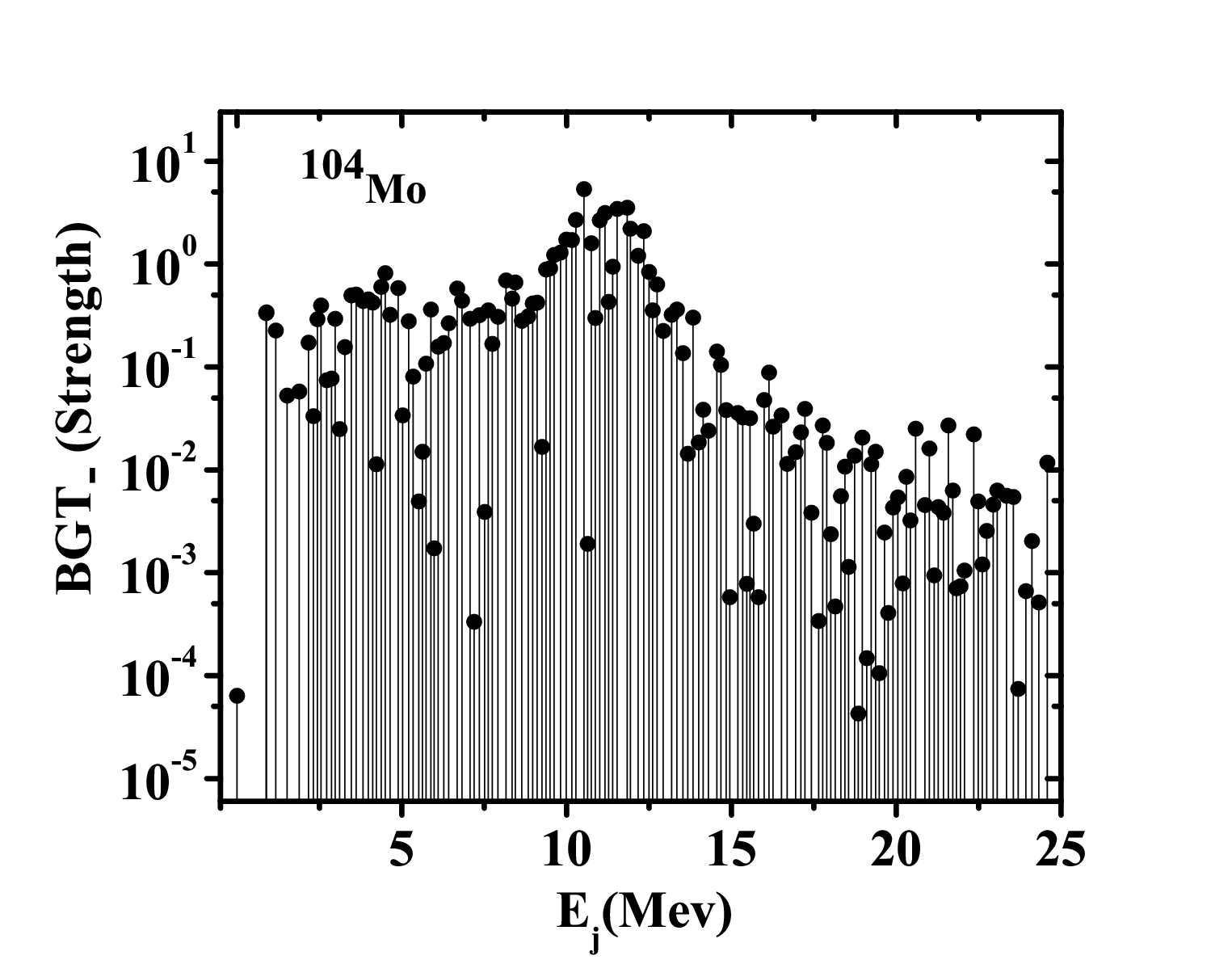}
\includegraphics[height=0.35\textwidth]{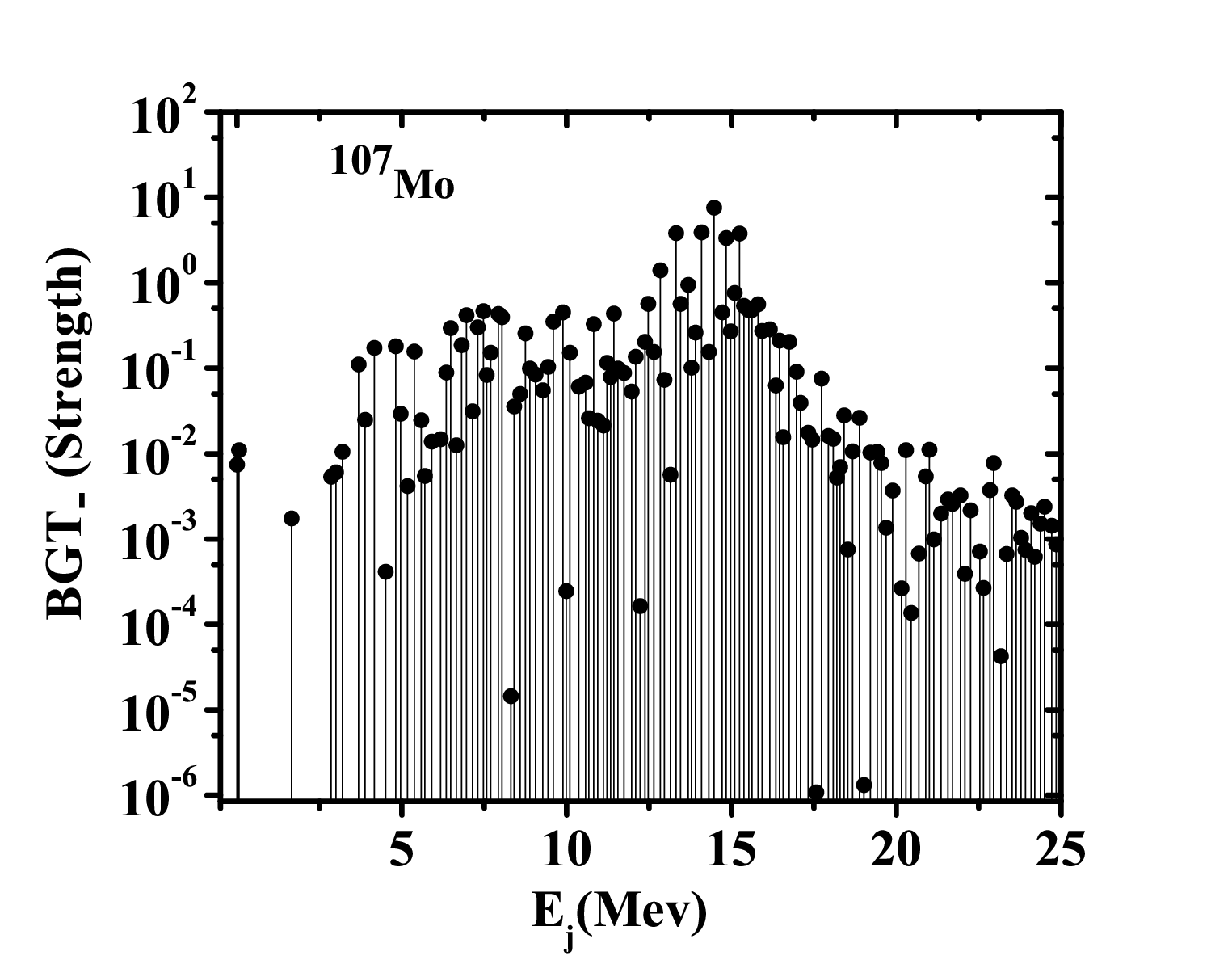}
\caption{The pn-QRPA calculated  GT strength distributions of $^{101}$Mo and $^{103-107}$Mo in daughter nuclei in the $\beta$-decay direction.} \label{GTBD}
\end{center}
\end{figure*}

The ground state charge-changing GT strength distributions for
 $^{83-88}$Mo along EC direction is shown in Fig.~\ref{GTEC}. The $BGT_{-}$ strength distributions for  $^{101,103,-107}$Mo isotopes along $\beta^{-}$ direction  are presented in  Fig.~\ref{GTBD}.   The
abscissa in Fig.~\ref{GTEC} shows energy of daughter $^{83-88}$Nb isotopes, respectively. Similarly 
abscissa in Fig.~\ref{GTBD} represents the energy of daughter
$^{101,103,-107}$Tc nuclei, respectively. The cutoff energy for the daughter nuclei
are 25 MeV in both directions. 
Fig.~\ref{GTEC} and Fig.~\ref{GTBD} clearly show the well fragmented $GT\pm$
transitions in nuclei of daughter states. It is to be noted that both ground and excited states GT strength distributions were calculated for the 55 Mo isotopes and electronic files of these strength distributions are available with the corresponding author. 

The calculated EC and $\beta^{-}$ decay rates are responsive to the GT centroid placement of the $BGT_{+}$ and $BGT_{-}$ distributions, respectively.  The centroid of calculated GT strength distributions, both along EC and $\beta^{-}$ decay directions, are given in Table~\ref{ebar}. All centroid values are stated in units of $MeV$. 

\begin{figure*}[htbp]
\begin{center}
\includegraphics[height=0.7\textwidth]{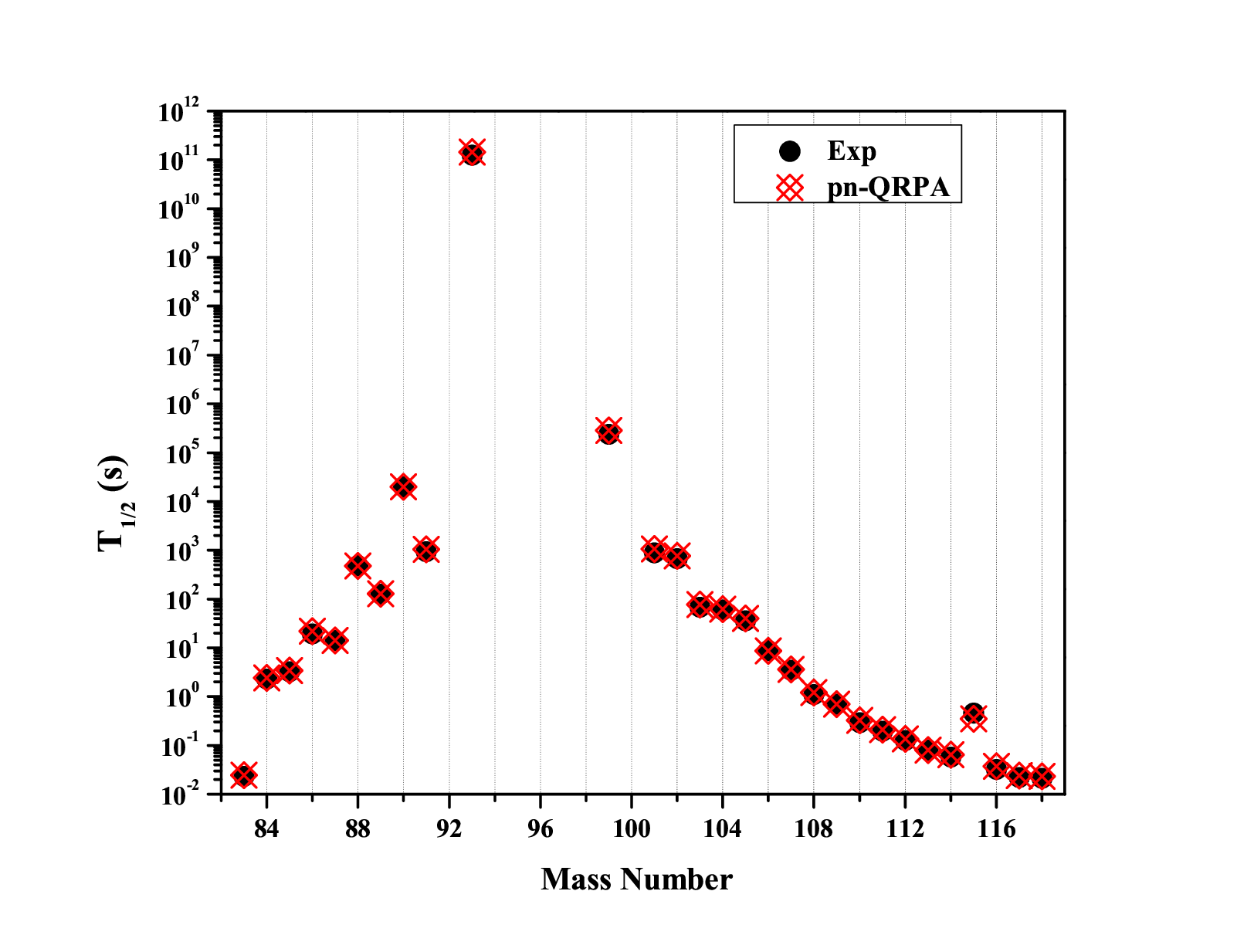}
\caption{ Comparison of pn-QRPA calculated $\beta$-decay half-life of Mo isotopes against measured data \citep{a47}.}
\label{HL}
\end{center}
\end{figure*}

 Fig.~\ref{HL} shows the
comparison of terrestrial half-lives of Mo isotopes calculated by the
pn-QRPA model with measured half-lives \citep{a47}. This excellent comparison was achieved due to a smart choice of model parameters as discussed earlier.

\begin{figure*}[htbp]
\begin{center}
\includegraphics[height=0.35\textwidth]{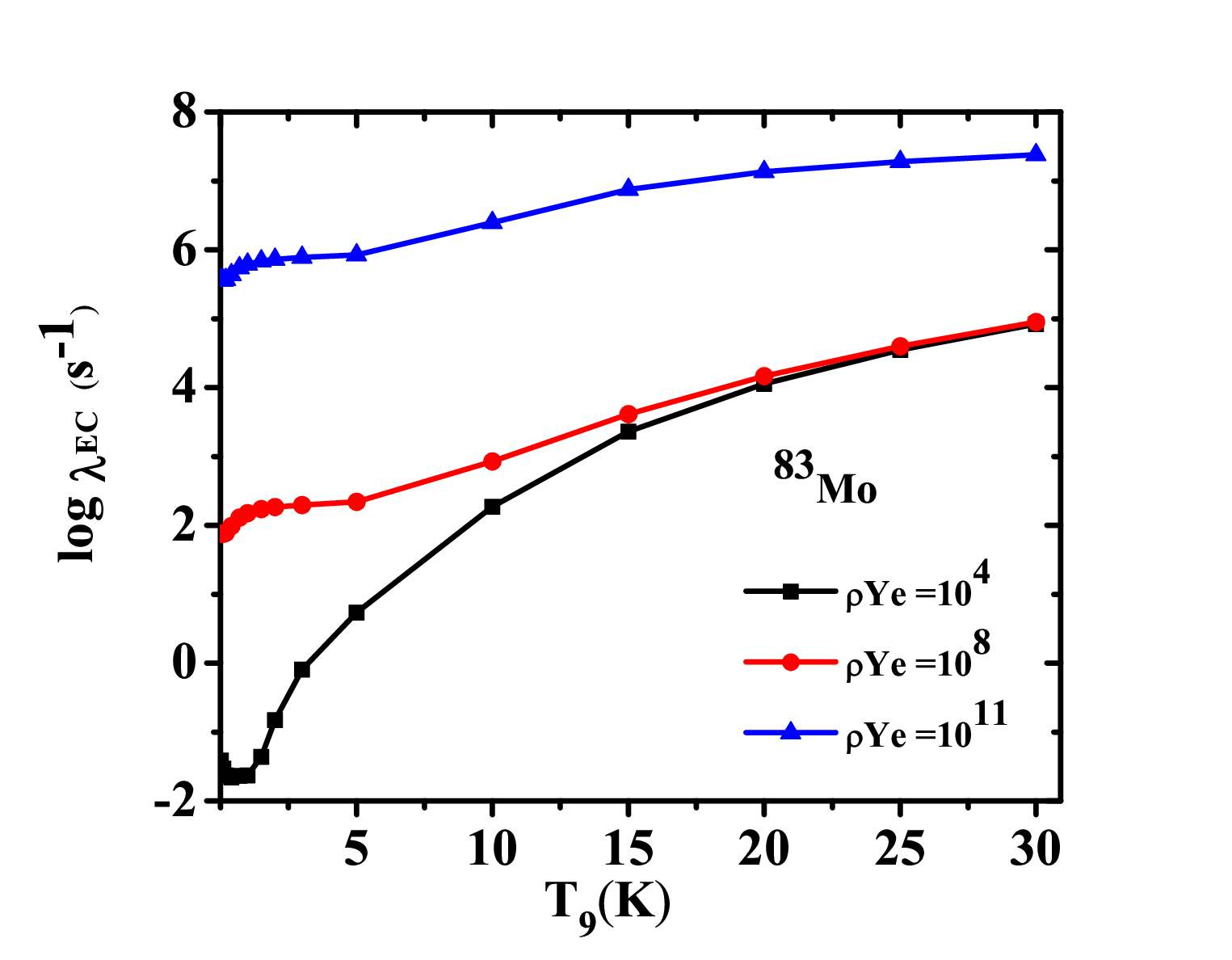}
\includegraphics[height=0.35\textwidth]{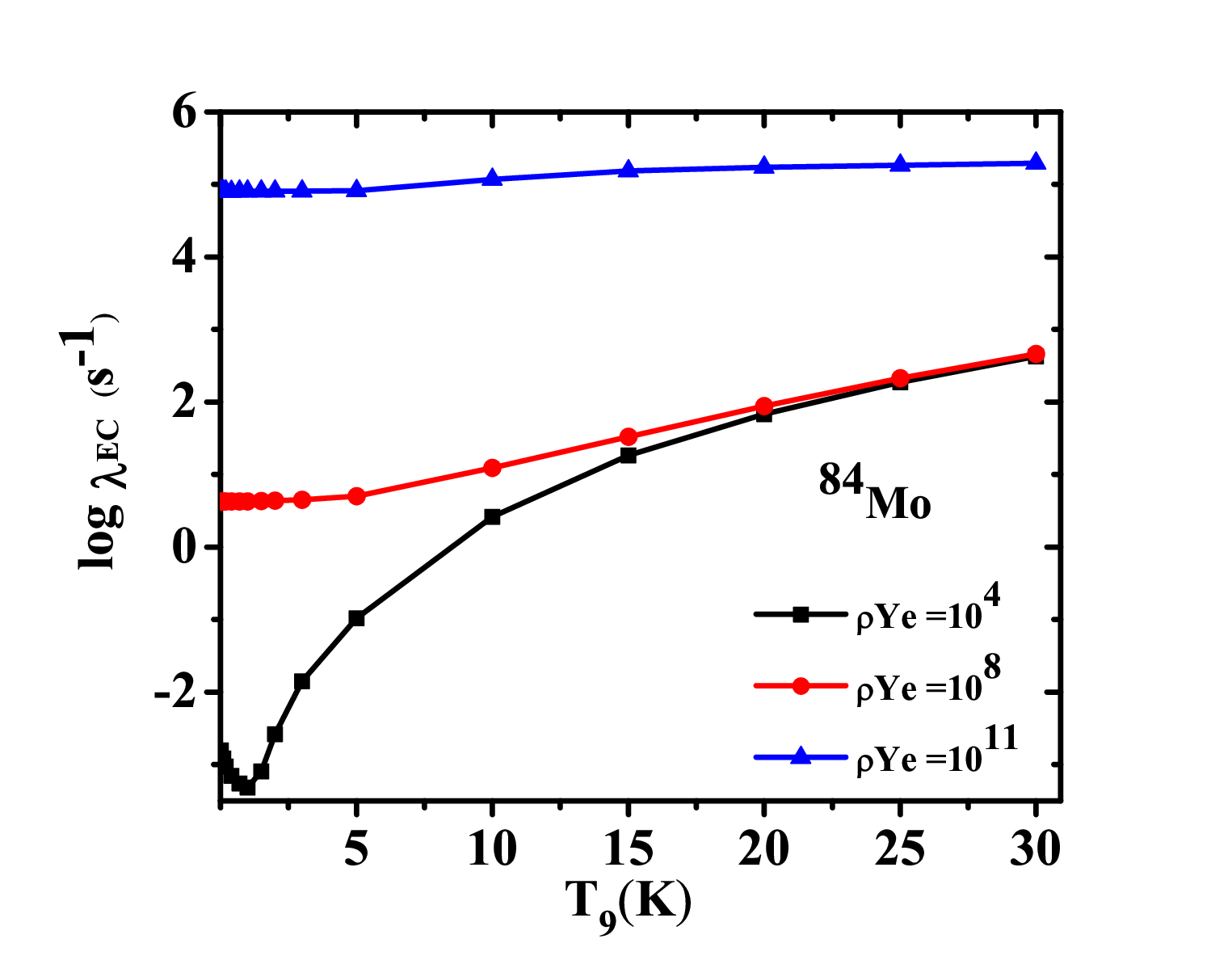}
\includegraphics[height=0.35\textwidth]{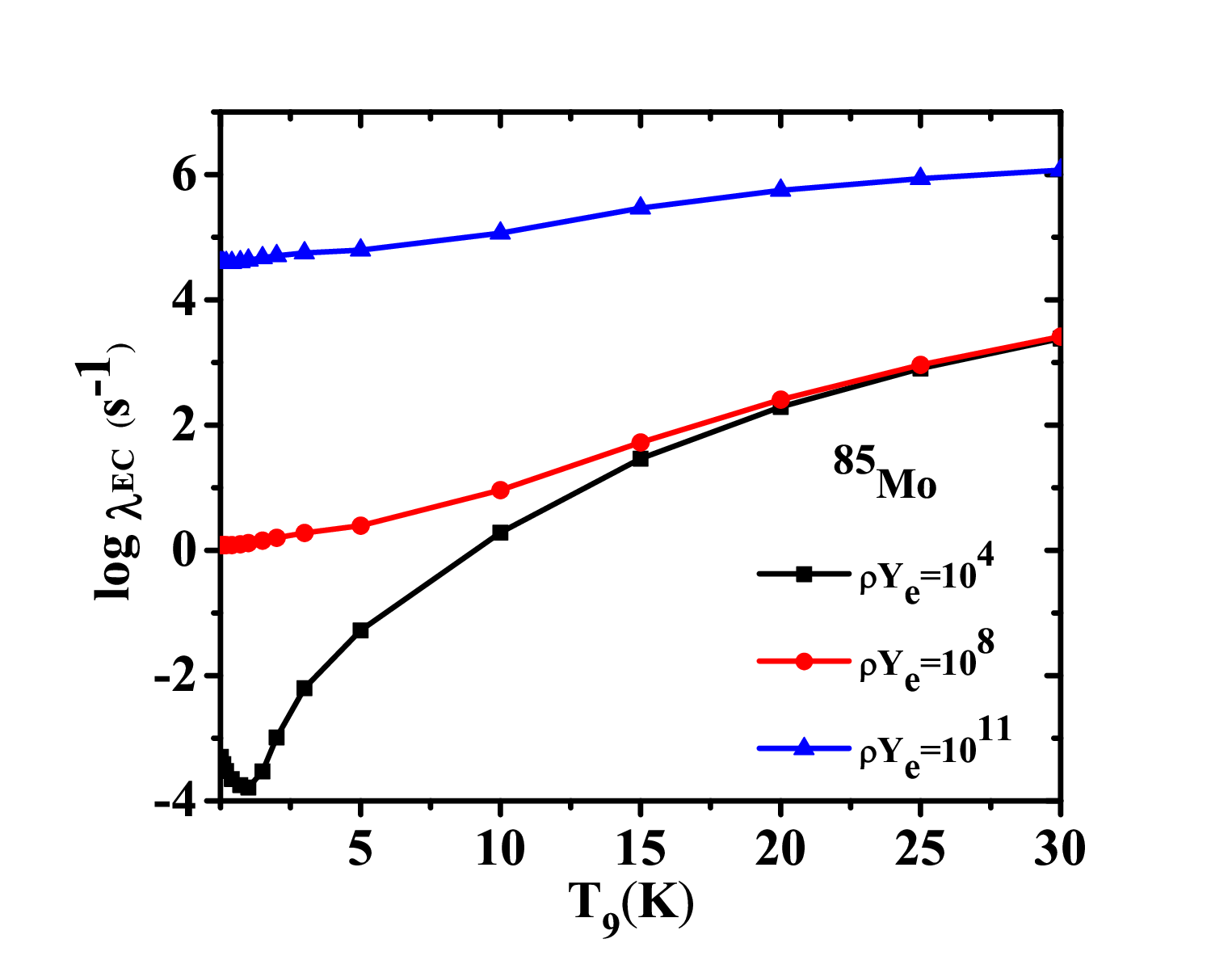}
\includegraphics[height=0.35\textwidth]{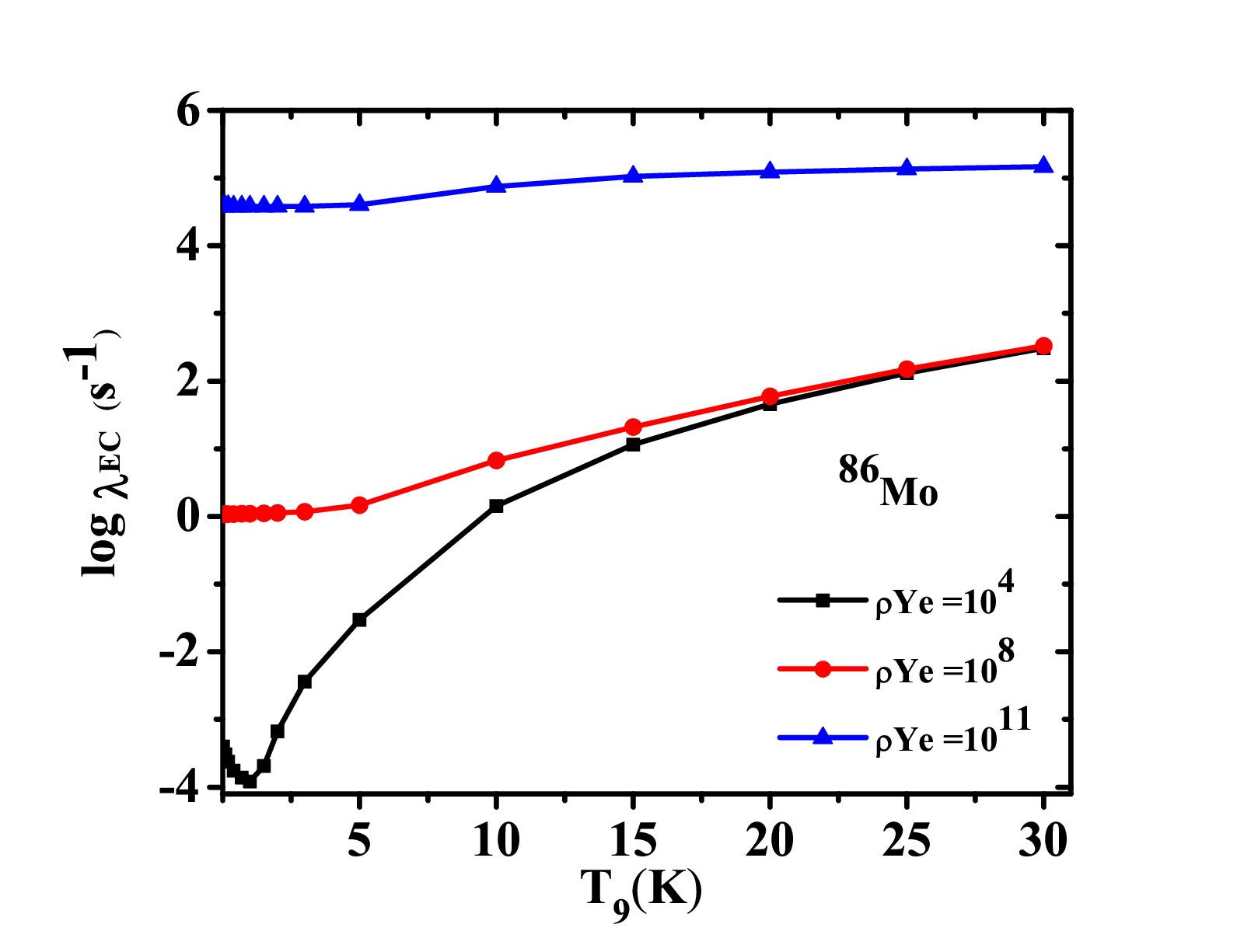}
\includegraphics[height=0.35\textwidth]{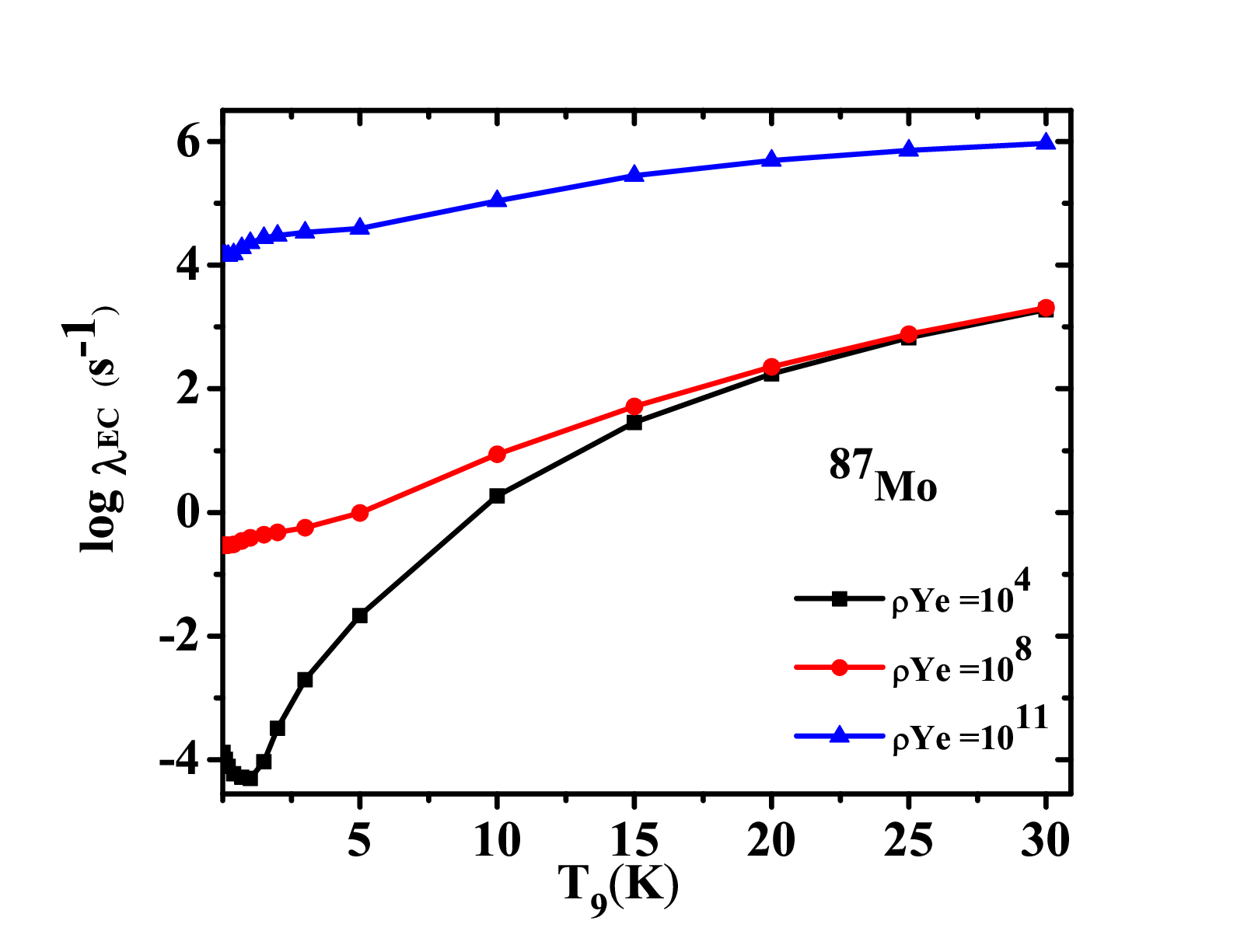}
\includegraphics[height=0.35\textwidth]{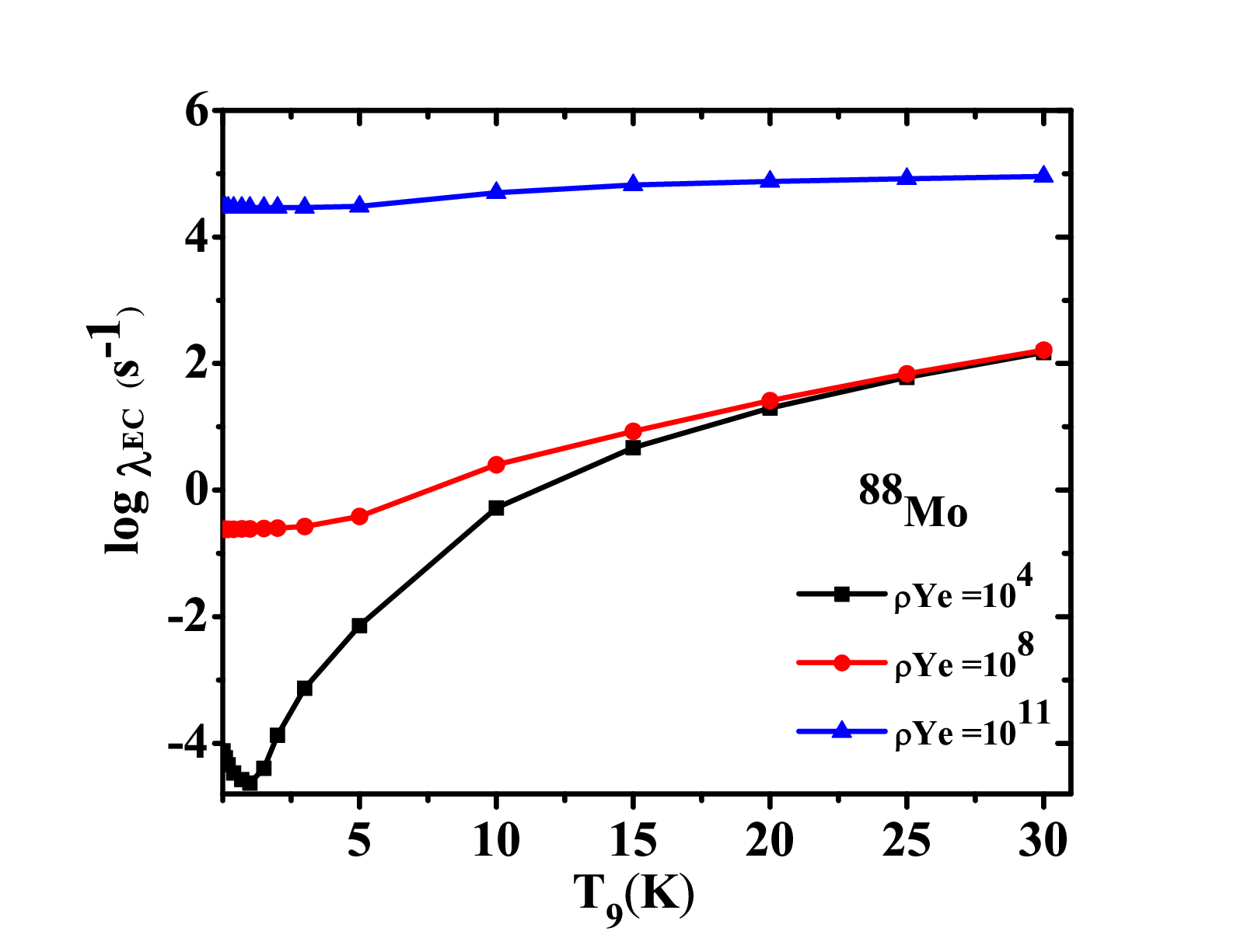}
\caption {The pn-QRPA calculated EC rates on $^{83-88}$Mo at selected density as a function of stellar temperature. The 
density in legend is represented by $\rho Y_{e}$ having units of $g/cm^{3 }$. T$_{9}$ is the core temperature in units of $10^{9}$K. The EC rates are given in $\log$
(to base 10) scale having units of $s^{-1}$.} \label{ECtemp}
\end{center}
\end{figure*}

\begin{figure*}[htbp]
\begin{center}
\includegraphics[height=0.35\textwidth]{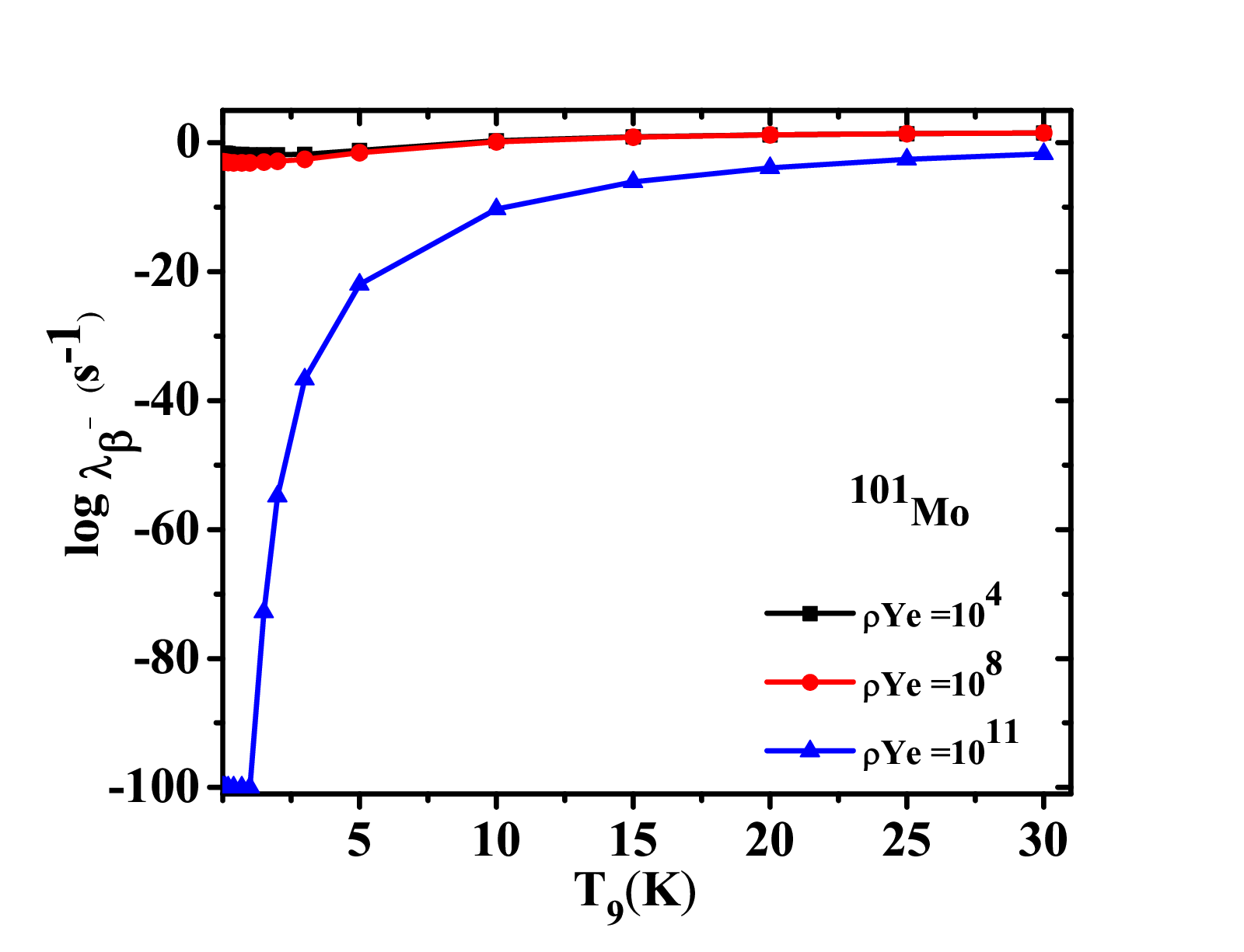}
\includegraphics[height=0.35\textwidth]{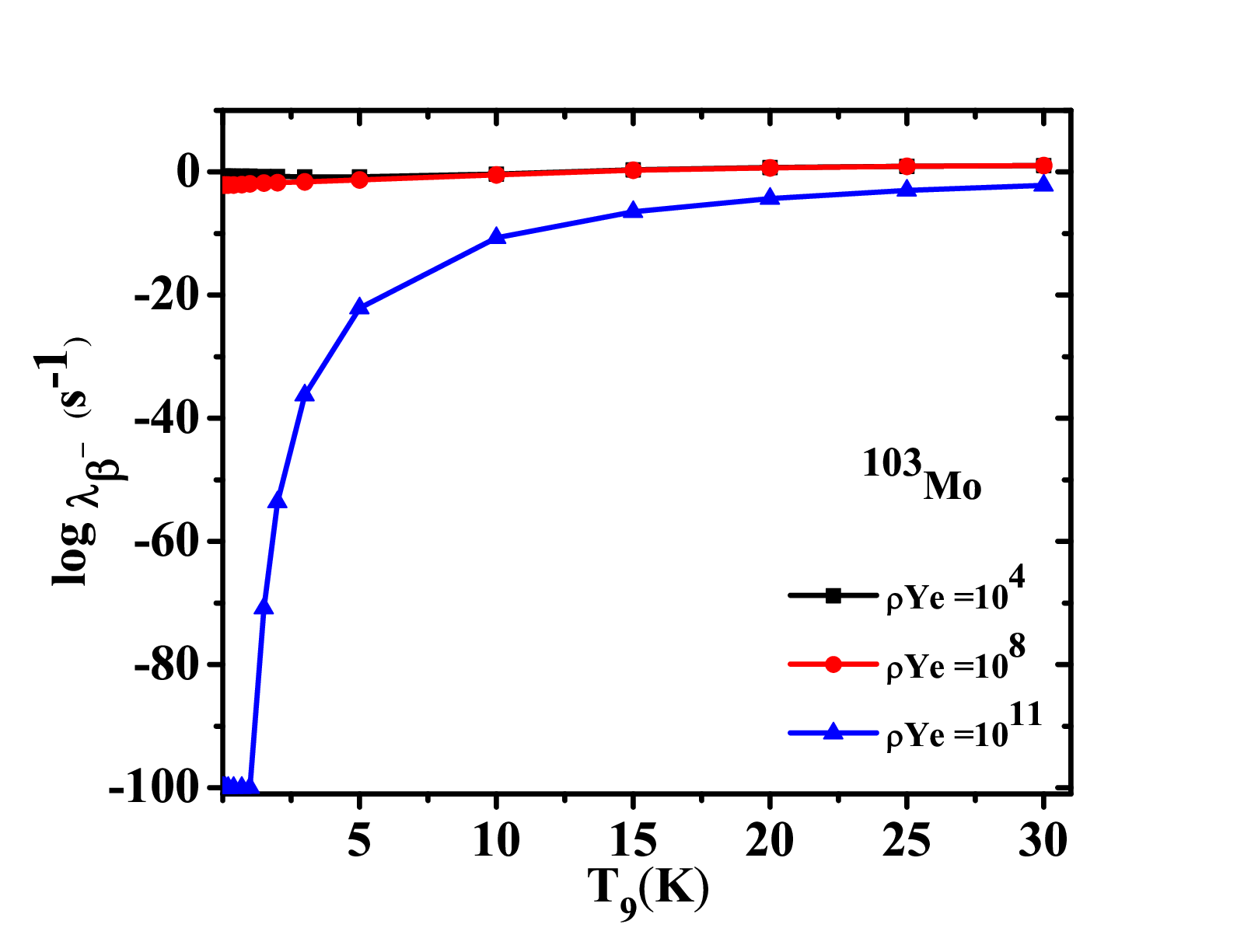}
\includegraphics[height=0.35\textwidth]{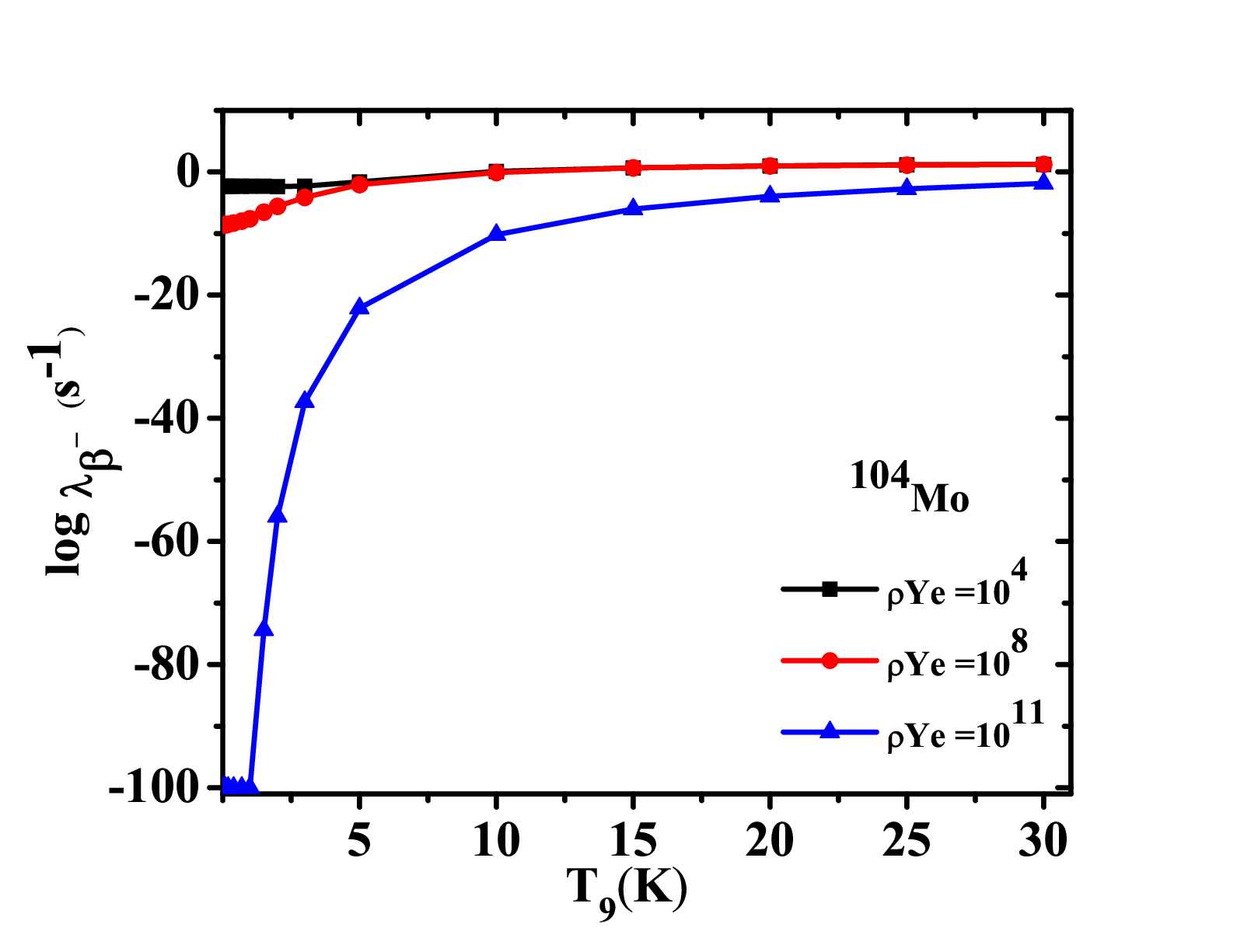}
\includegraphics[height=0.35\textwidth]{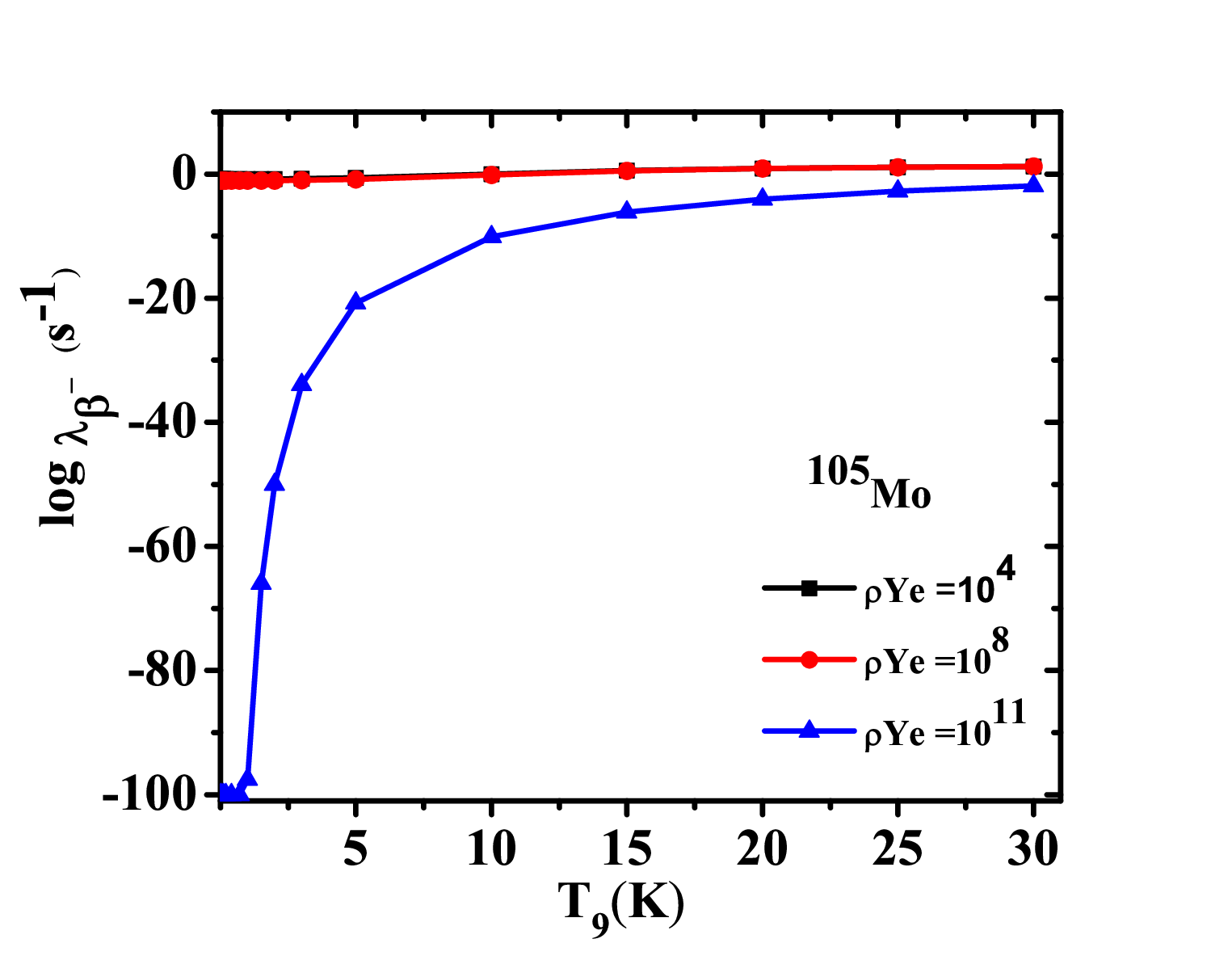}
\includegraphics[height=0.35\textwidth]{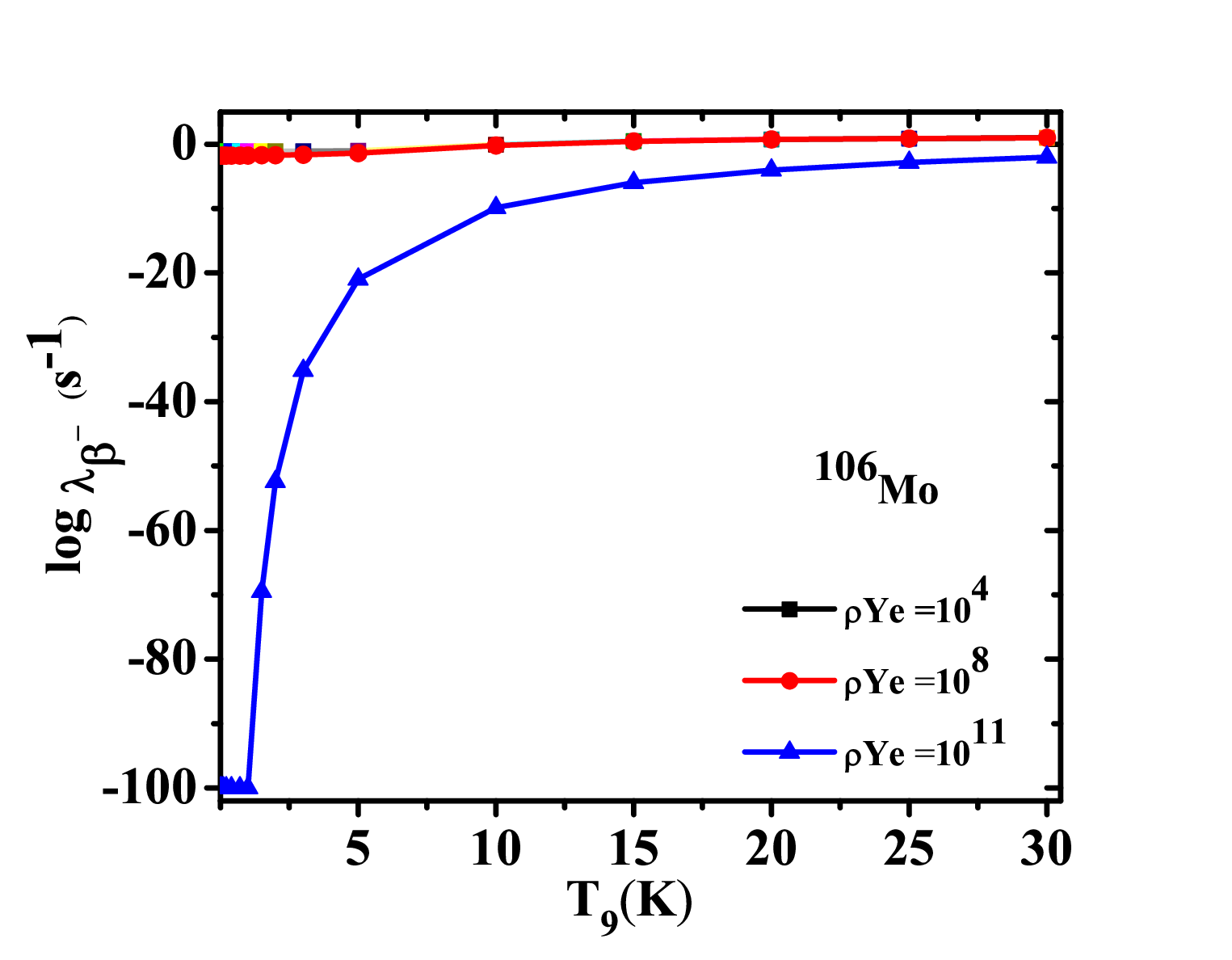}
\includegraphics[height=0.35\textwidth]{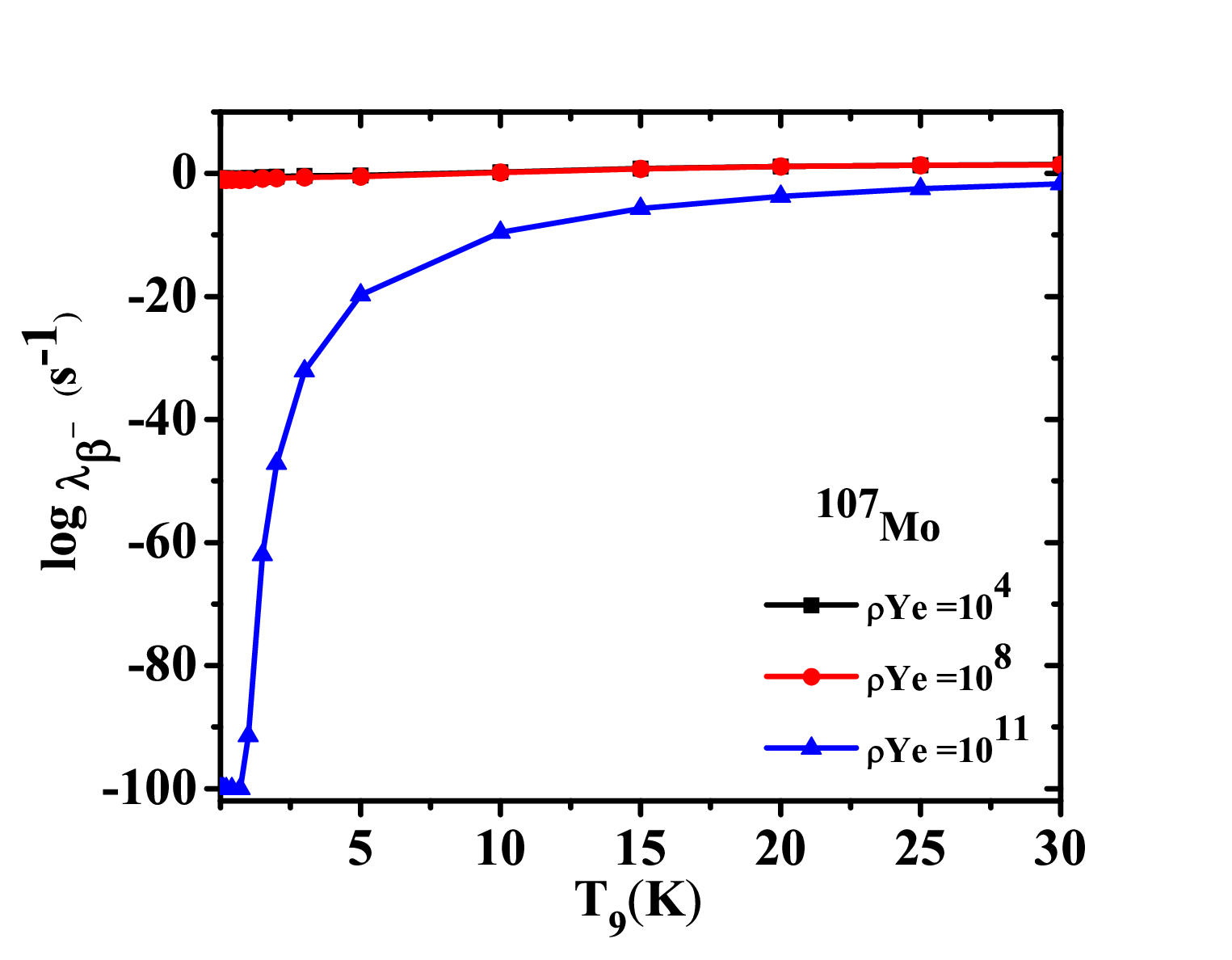}
\caption{The pn-QRPA calculated $\beta$-decay rates on $^{101}$Mo and $^{103-107}$Mo at selected density as a function of stellar temperature. The 
	density in legend is represented by $\rho Y_{e}$ having units of $g/cm^{3 }$. T$_{9}$ is the core temperature in units of $10^{9}$K. The $\beta$-decay rates are given in $\log$
	(to base 10) scale having units of $s^{-1}$.}
\label{BDtemp}
\end{center}
\end{figure*}

We next move from the terrestrial to stellar domain. Fig.~\ref{ECtemp} shows the electron capture rates on
$^{83-88}$Mo. Similarly the $\beta^{-}$ decay rates for $^{101}$Mo and
$^{103-107}$Mo isotopes are depicted in   Fig.~\ref{BDtemp}. The $T_9$ axis 
of Fig.~\ref{ECtemp} and Fig.~\ref{BDtemp} gives core temperature in units of 10$^{9}$K. We show calculated EC and $\beta^{-}$ decay rates at
low ($10^{4}gcm^{-3}$), intermediate ($10^{8}gcm^{-3}$) and high density regions of
$10^{11}gcm^{-3}$ in stellar matter. 
Fig.~\ref{ECtemp} shows that the EC rates get enhanced as the stellar density increases. This is because the Fermi energy shifts to higher energy as the core stiffens.   
It is noticed that $\beta^{-}$ decay  rates increase with a corresponding increase in core temperature.  The $\beta^{-}$ decay  rates increase with rise in core temperature due to contribution
of partial rates from low-lying parent levels. The rates however decrease  by orders of
magnitude as the  core material gets denser. Due to Pauli principle the available phase space for the reaction gets reduced  at higher densities.

\begin{figure*}[htbp]
\begin{center}
\includegraphics[height=0.35\textwidth]{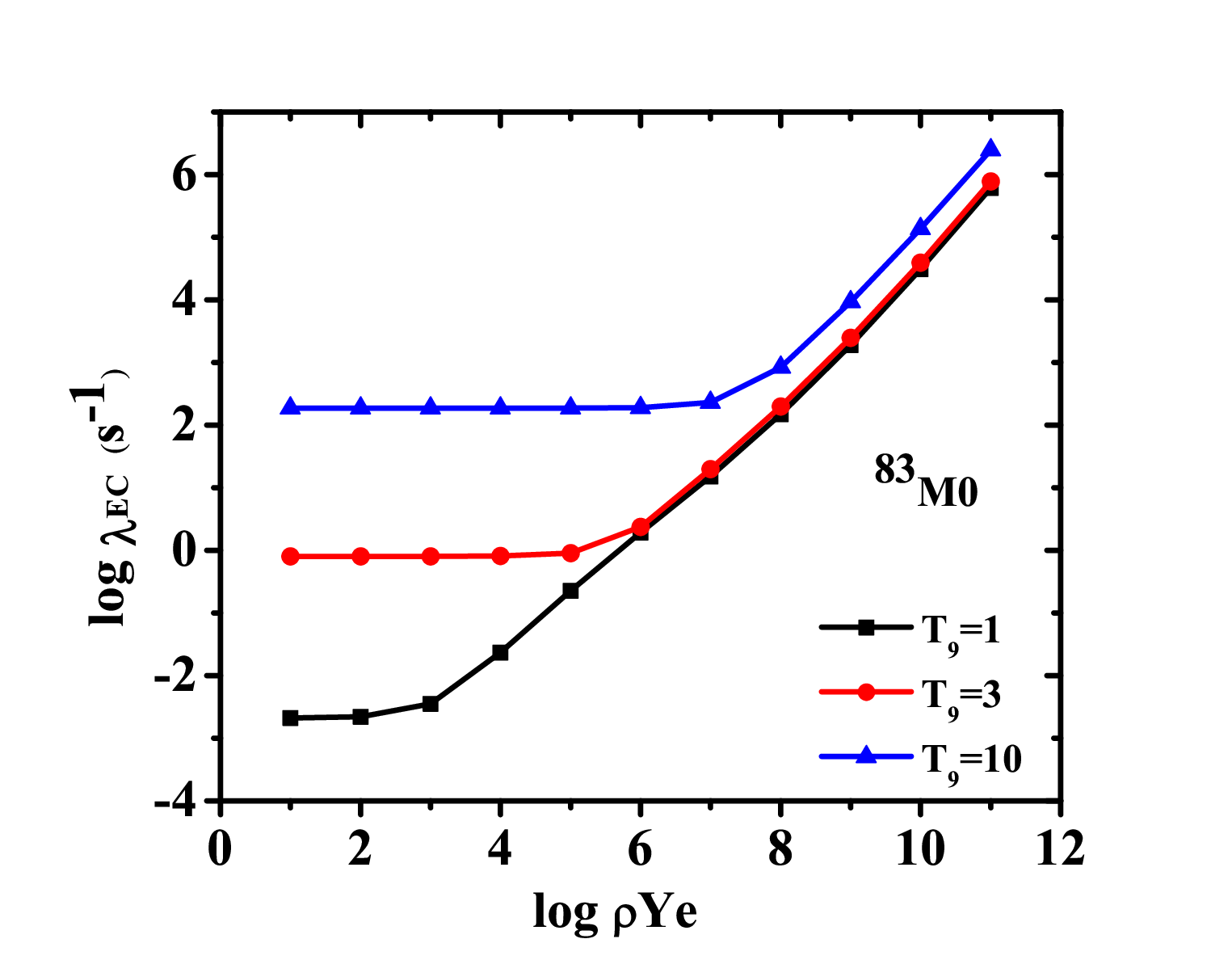}
\includegraphics[height=0.35\textwidth]{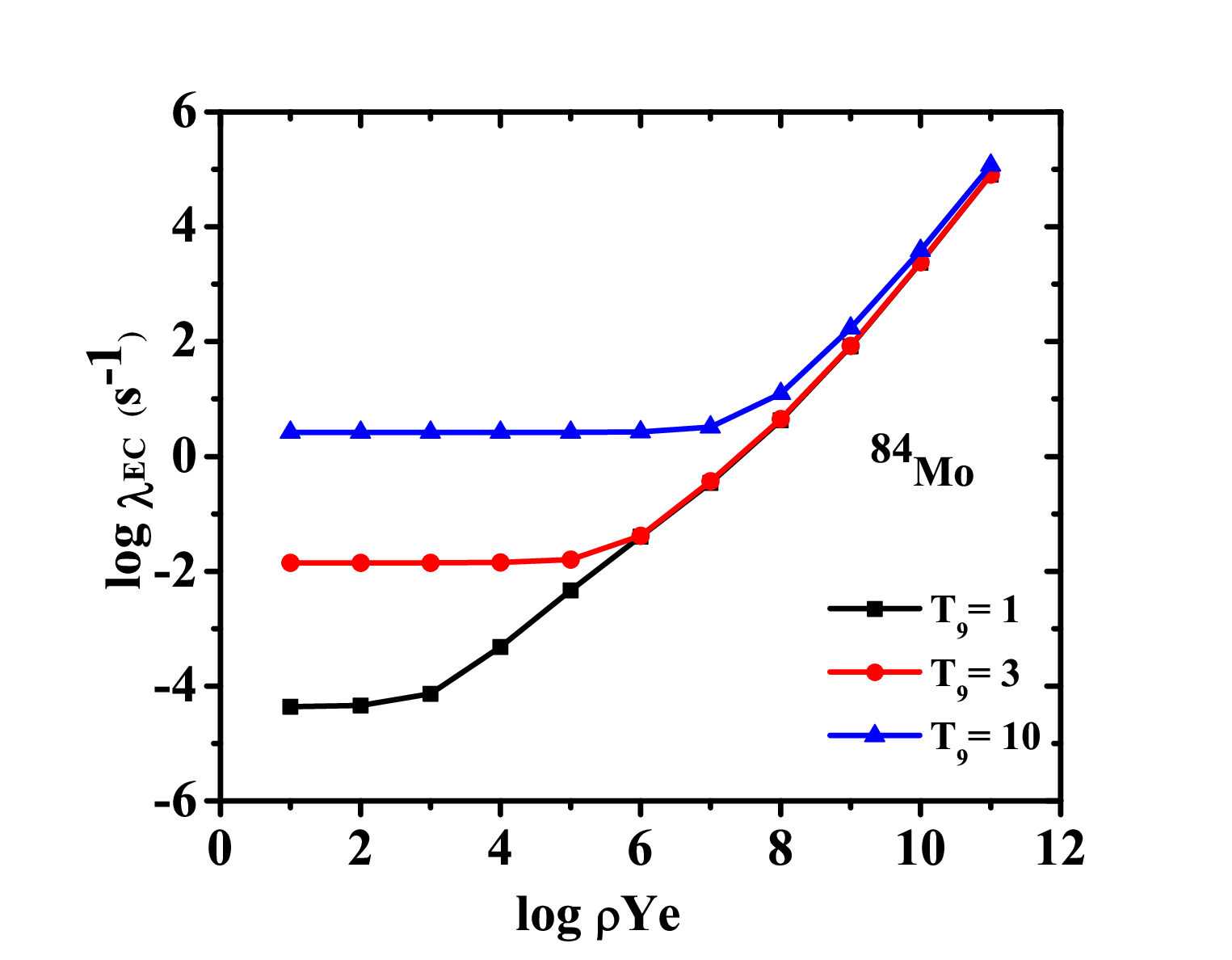}
\includegraphics[height=0.35\textwidth]{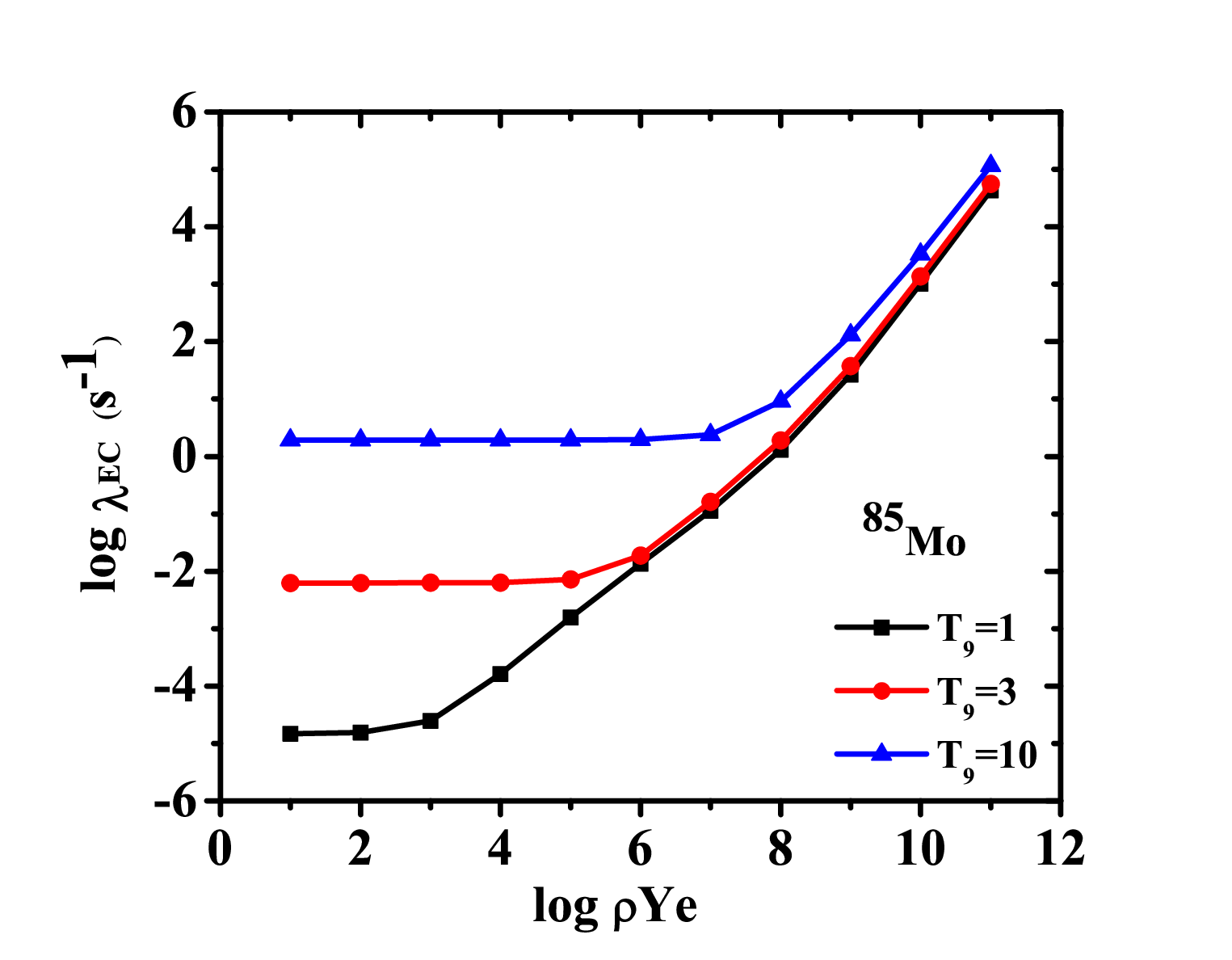}
\includegraphics[height=0.35\textwidth]{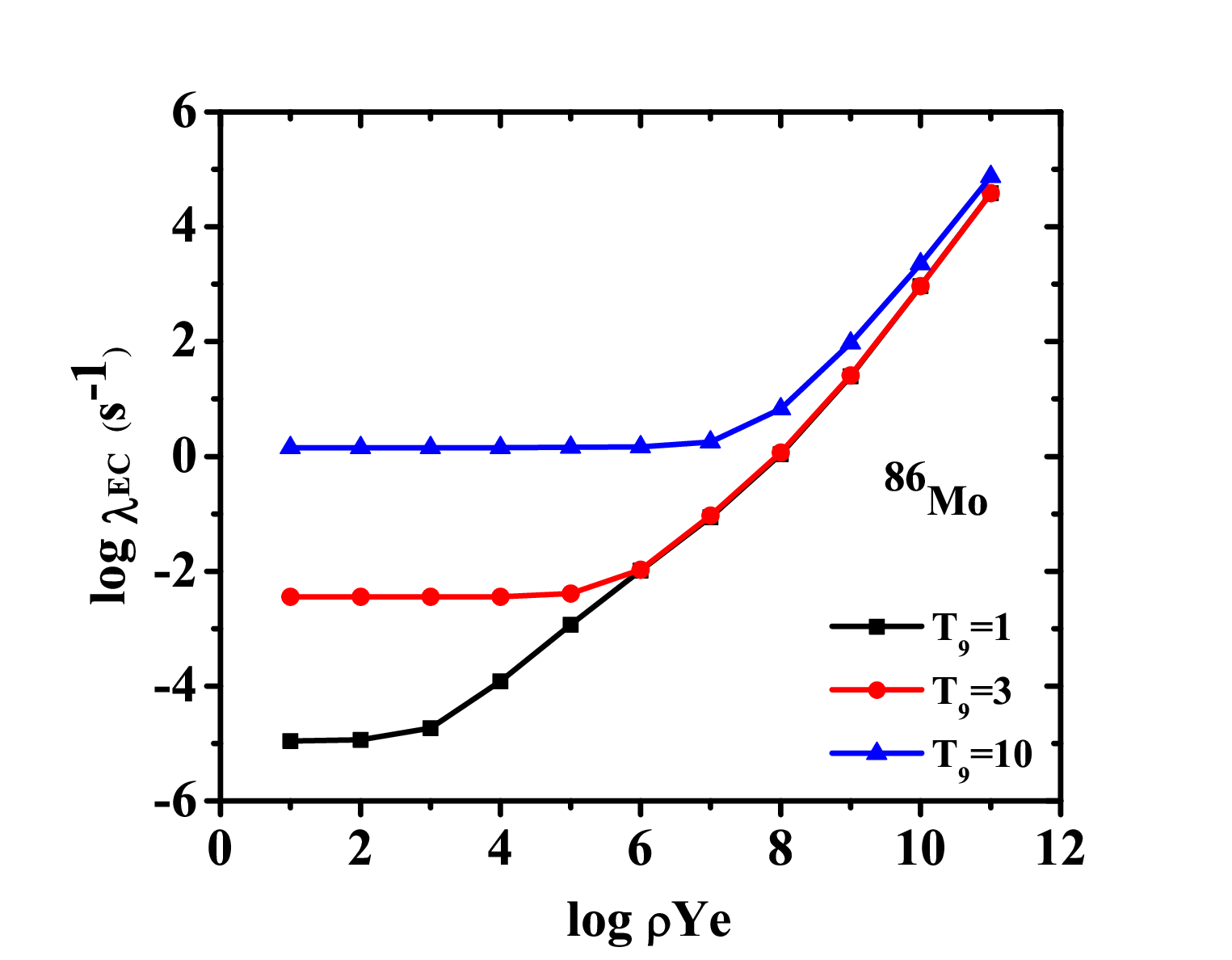}
\includegraphics[height=0.35\textwidth]{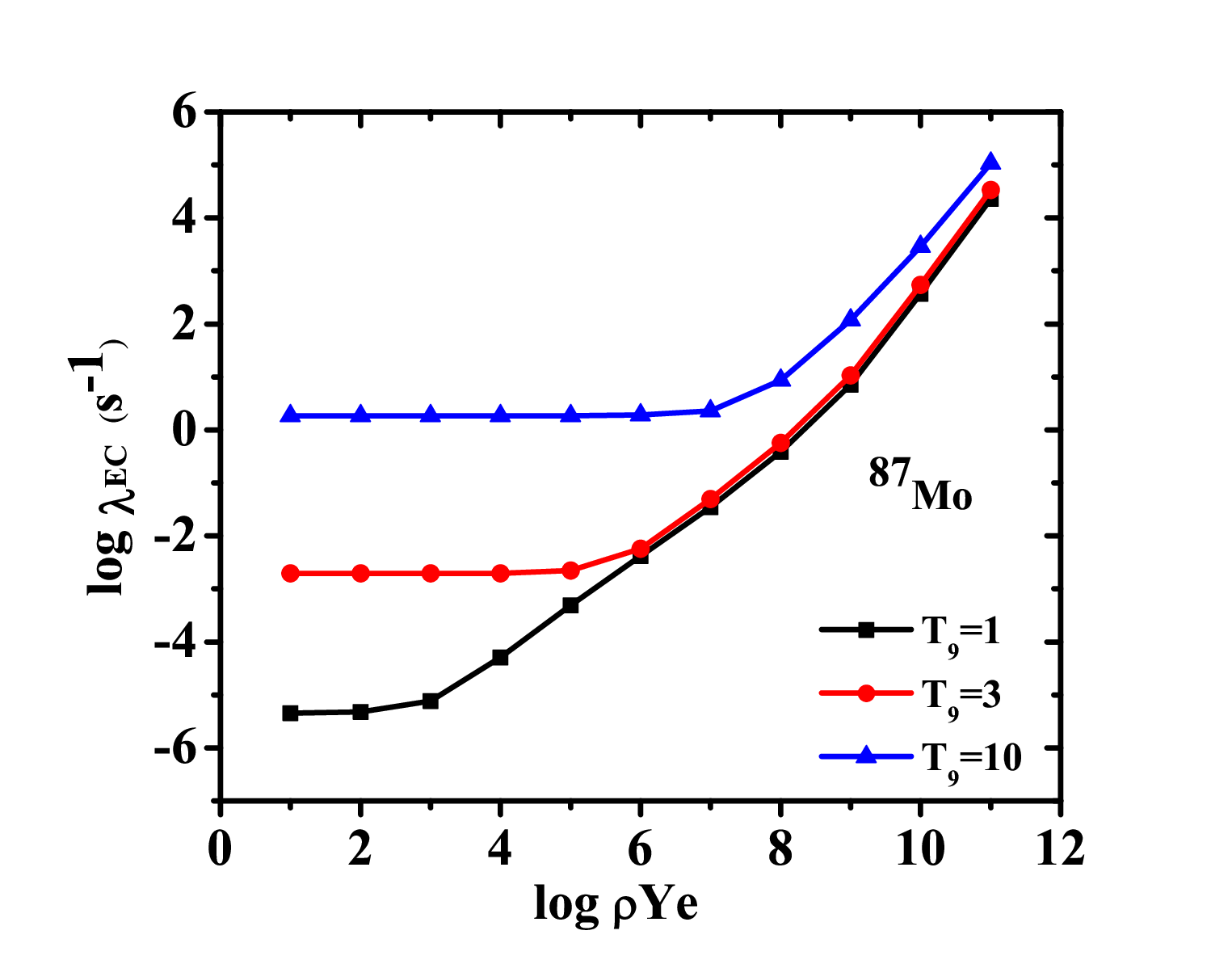}
\includegraphics[height=0.35\textwidth]{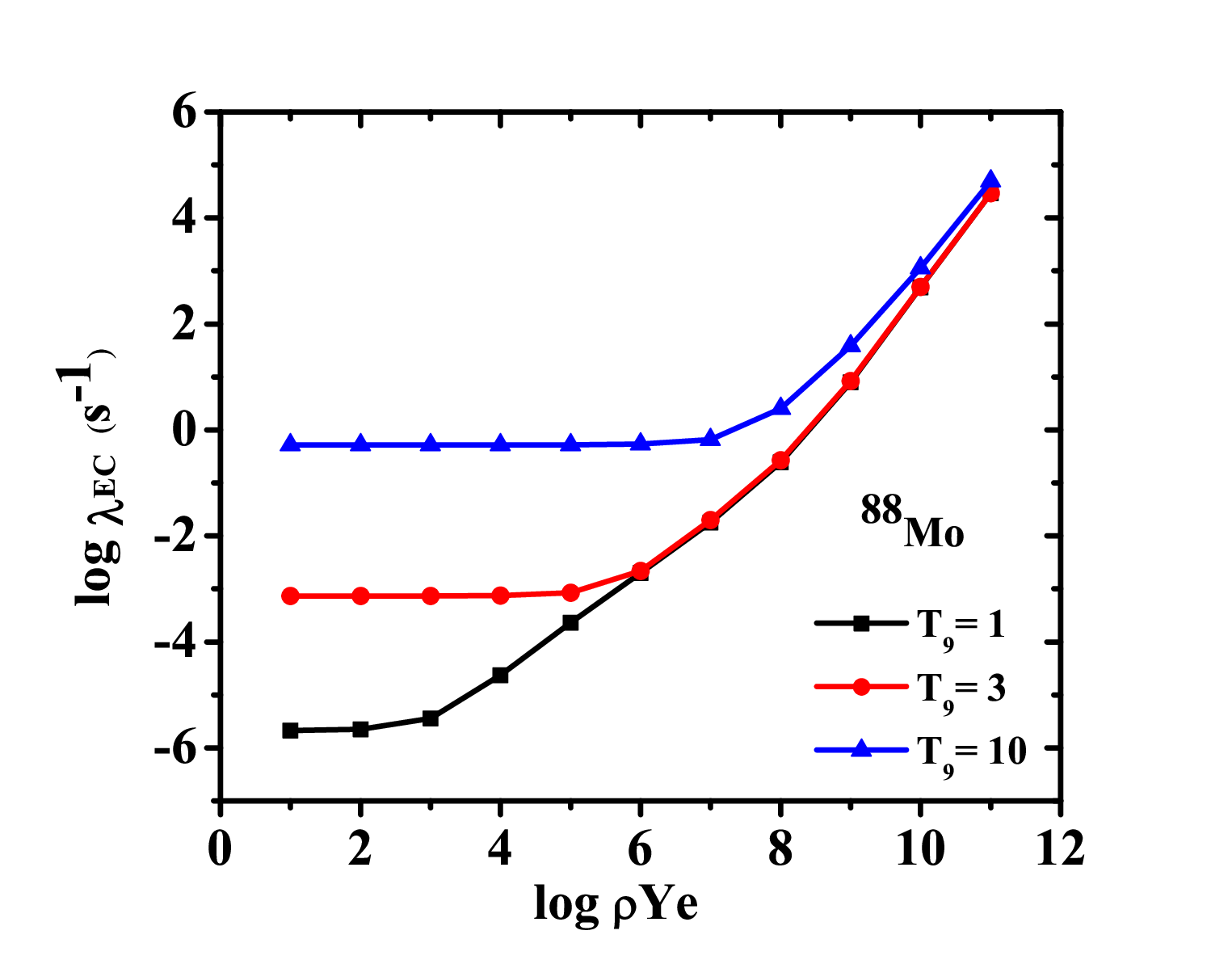}
\caption{The pn-QRPA calculated EC rates on $^{83-88}$Mo at selected temperature as a function of stellar density. Symbols have same meanings and units as discussed in Fig.~\ref{ECtemp}.} \label{ECden}
\end{center}
\end{figure*}

Fig.~\ref{ECden} shows the variation of calculated EC rates on $^{83-88}$Mo isotopes as a function of core density. The EC rates are computed at three different  values of core temperature shown in inset.
Initially the EC rates remain more or less constant as density increases to around $10^{4}gcm^{-3}$. Beyond this density there is a steep slope and the EC rates tend to merge at high stellar density of $10^{11}gcm^{-3}$.   In low density
regions the EC  rates compete well with $\beta^{-}$ decay rates. During the process of core collapse
the high EC rates make the composition of stellar environment
more neutron rich. During the final stages of core collapse, as
a result of Pauli blocking of phase space the
 $\beta$ decay rates become comparatively unimportant \citep{a43}.

\begin{figure*}[htbp]
\begin{center}
\includegraphics[height=0.35\textwidth]{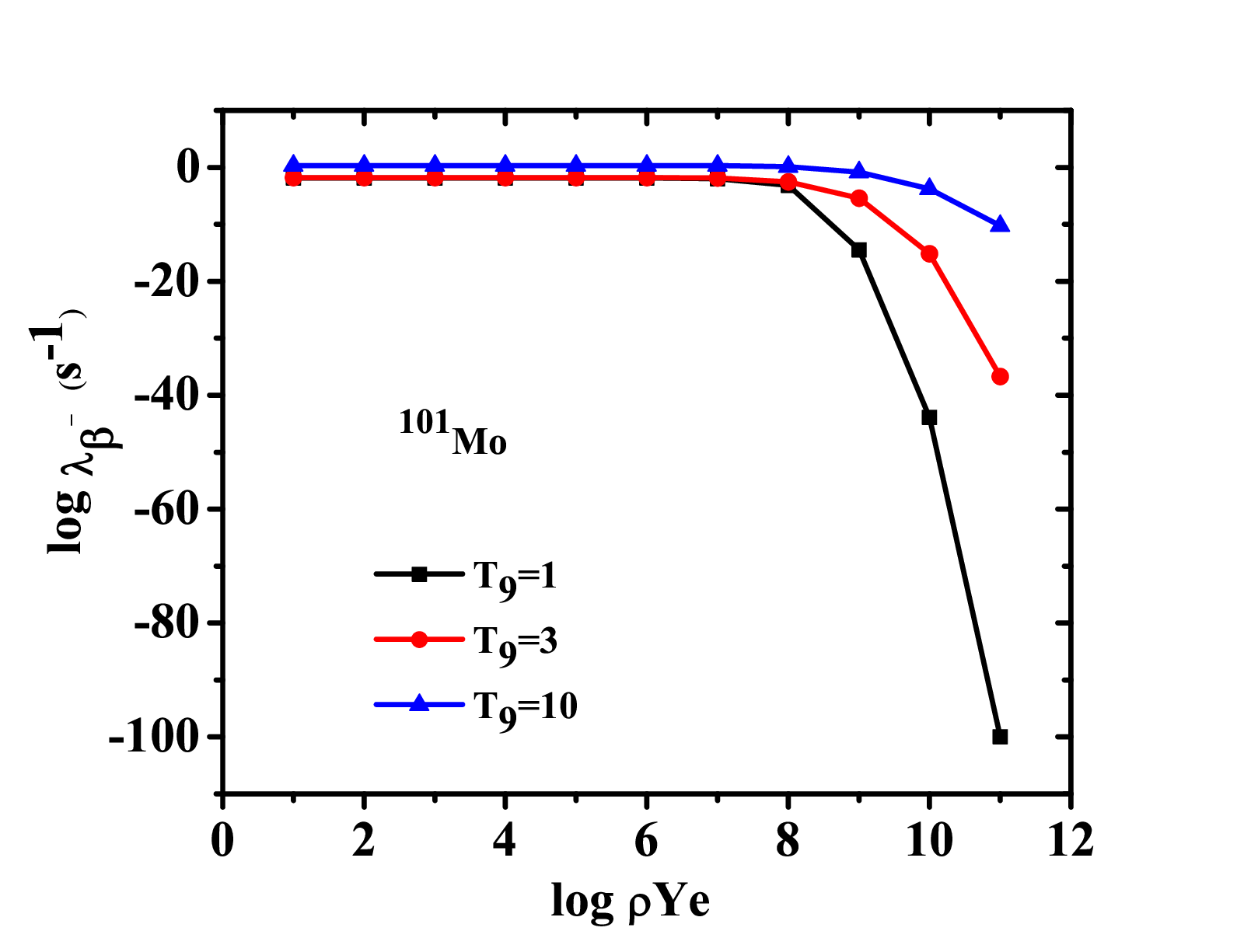}
\includegraphics[height=0.35\textwidth]{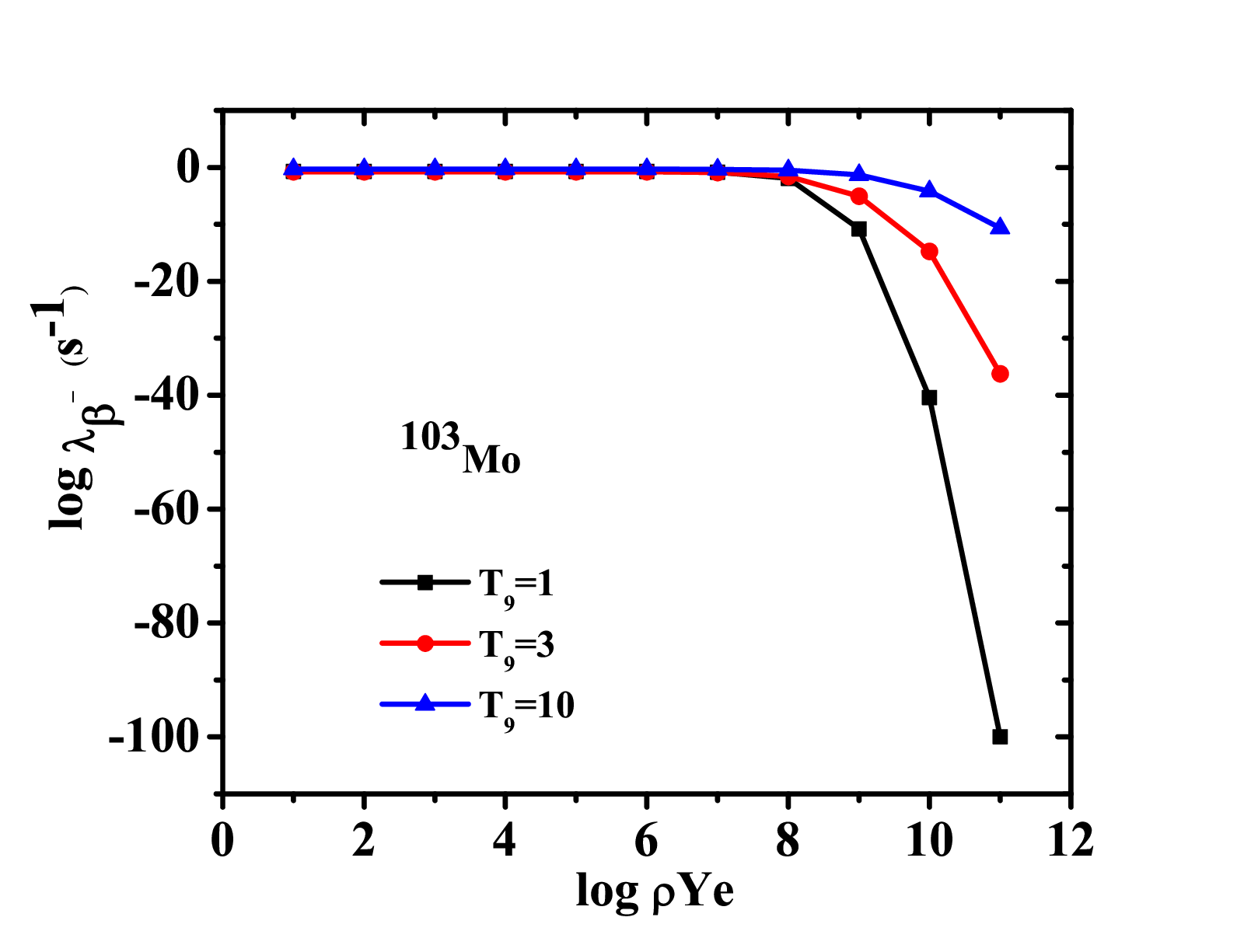}
\includegraphics[height=0.35\textwidth]{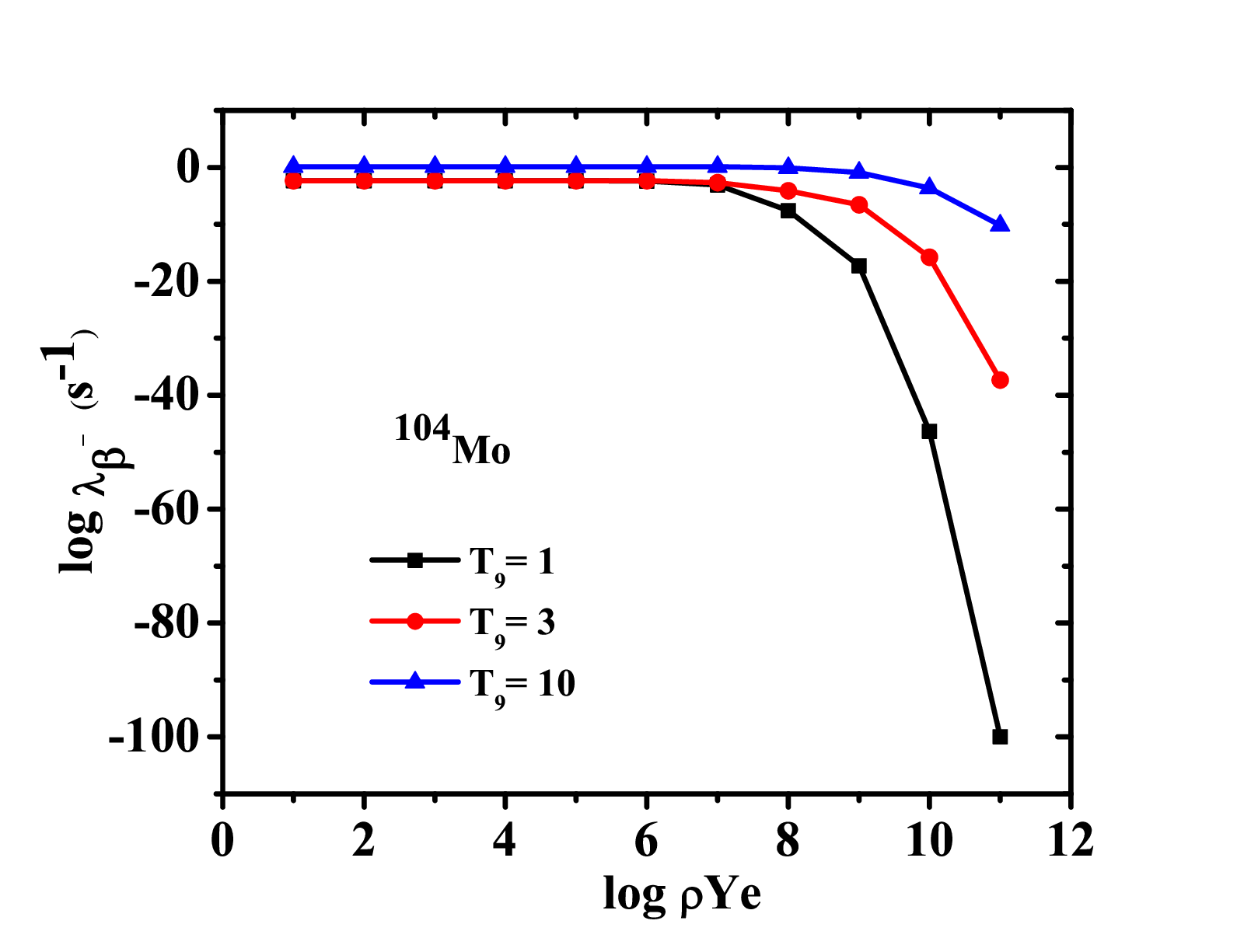}
\includegraphics[height=0.35\textwidth]{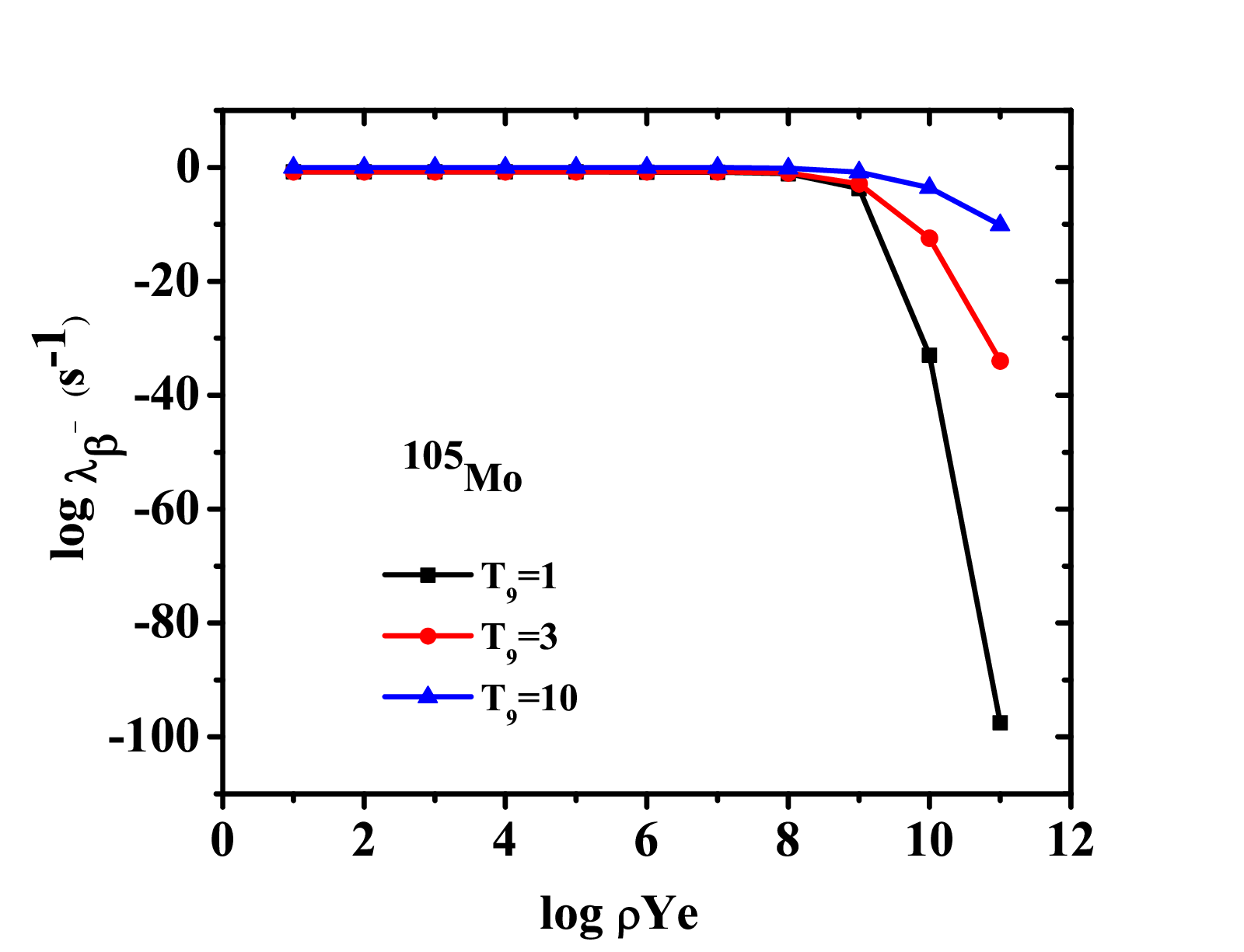}
\includegraphics[height=0.35\textwidth]{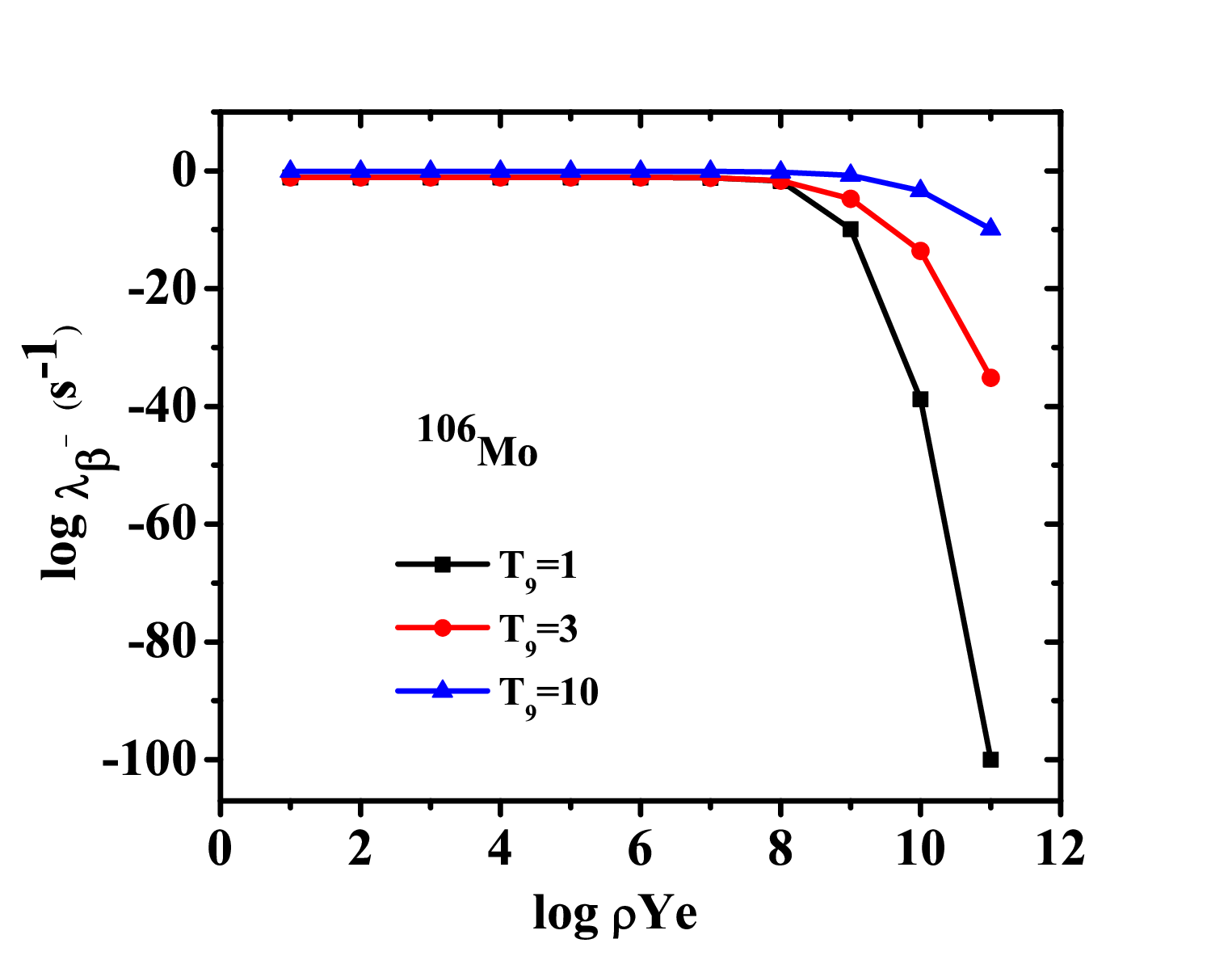}
\includegraphics[height=0.35\textwidth]{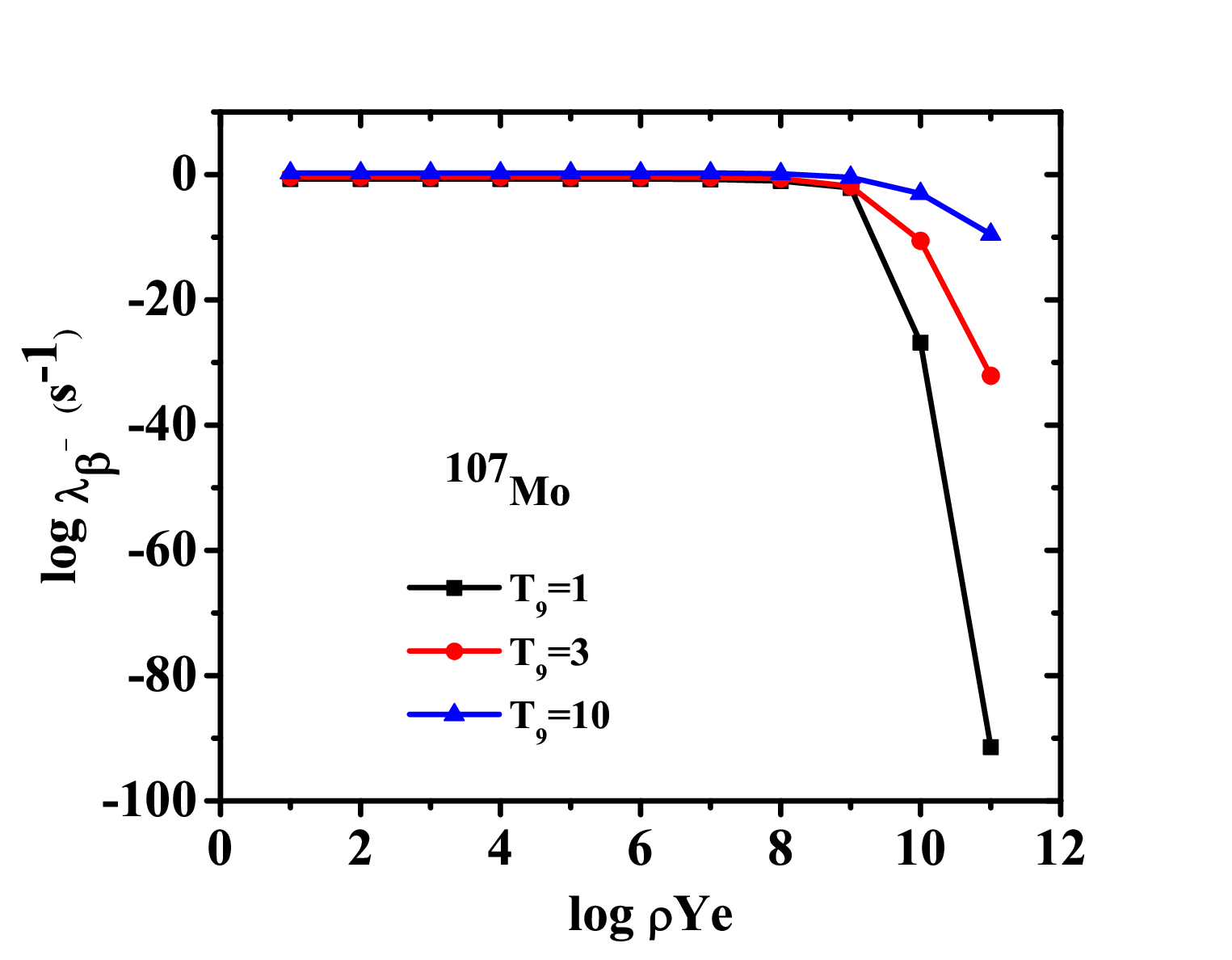}
\caption{The pn-QRPA calculated $\beta$-decay rates on $^{101}$Mo and $^{103-107}$Mo at selected temperature as a function of stellar density. Symbols have same meanings and units as discussed in Fig.~\ref{BDtemp}.}
\label{BDden}
\end{center}
\end{figure*}

Fig.~\ref{BDden} shows a similar snapshot for  $\beta^{-}$ decay rates on $^{101,103-107}$Mo at three selected values of temperature (1$\times10^{9}$ K, 3$\times10^{9}$ K
and 
10$\times10^{9}$ K). The decay rates are given in logarithmic
scale whereas $\rho Y_{e}$ along abscissa shows the logarithmic scale of density. The  $\beta^{-}$ decay rates are almost constant and superimposed on each other at low densities. They start to separate  from one another as the density increases. The  $\beta^{-}$ decay rates then decrease exponentially for reasons already mentioned above.

The calculated EC  and $\beta^{-}$ decay rates on
$^{82-94}$Mo, $^{96}$Mo and $^{98-138}$Mo are
shown on selected scale of temperature and density in Table~\ref{r1}.
The decay rates are recorded in $\log$
to base 10 scale. The electronic files of EC and  $\beta^{-}$ decay rates on a fine-grid density-temperature scale for all 55 isotopes of Mo may be requested from the authors. 

\section {Conclusions}
The mechanism of supernova explosion may be better understood if one can have a reliable estimate of EC and  $\beta^{-}$ decay rates.  These weak-interaction rates may also contribute to a deeper comprehension of the nucleosynthesis processes associated with the supernova explosions. 

Isotopes of Mo are relatively abundant in the core of massive stars and weak decay rates of these isotopes in stellar matter can assist the modeling and simulation of phases of stars prior to supernova explosion.   In the present study  
we first calculated some basic ground state properties of Mo isotopes by using RMF model with density dependent interactions starting from neutron drip-line to proton drip-line. Our results showed decent comparison with the available experimental data. The quadrupole deformation parameters in RMF model computed with DD-ME2 interaction were later used as input parameter in our pn-QRPA model calculation of terrestrial and stellar decay rates. The relative comparison of EC and  $\beta^{-}$ decay rates on 55 isotopes of Mo (shown in Table~\ref{r1}) may help core-collapse simulators to model the  evolution of stars in a more reliable fashion before they go supernova.

\section*{Acknowledgements}

The author wishes to acknowledge useful discussion with Mr. Kaleem Ullah.
J.-U. Nabi would like to
acknowledge the support of the Higher Education Commission Pakistan
through project numbers 5557/KPK/NRPU/R$\&$D/HEC/2016,
9-5(Ph-1-MG-7)\\/PAK-TURK/R$\&$D/HEC/2017 and Pakistan Science
Foundation through project number PSF-TUBITAK/\\KP-GIKI (02). T. Bayram would like to acknowledge the support of the Turkish Higher Education Council with Mevlana Exchange Program through project number MEV-2016-094.

\begin{table*}[ht]
\caption{The calculated ground-state properties of
Mo isotopes using RMF model with DD-ME2 interaction.} 
\centering 
\small
\begin{tabular}{|c|c|c|c|c|c|c|c|c|}
\hline Isotope & {\it BE/A} [MeV] & {\it S$_{n}$} [MeV] & {\it
S$_{p}$} [MeV] & {\it S$_{2n}$} [MeV] & {\it S$_{2p}$} [MeV] & {\it
r$_{c}$} [fm] & {\it $\beta_{2}$} & {\it Q$_{T}$} [barn]  \cr

\hline

$   ^   {   82  }   $Mo     &   $   666.734 $   &   $   -    $   & $-
$   &   $   0.262   $  &   $   2.300   $   &   $   4.305 $   &   $
0.000   $   &   $   -0.749  $   \cr $   ^   {   83  }$Mo     &   $
680.680 $   &   $   13.946  $   &   $   -    $   & $   0.882   $   &
$   3.772   $   &   $   4.318   $   &   $ -0.001  $   &   $   -1.022
$   \cr $   ^   {   84  }$Mo     & $   694.550 $   &   $   13.870
$   &   $   27.816  $   &   $ 1.725   $   &   $   5.202   $   &   $
4.324   $   &   $   -0.220 $   &   $   -386.224    $   \cr $   ^   {
85  }$Mo     &   $ 708.131 $   &   $   13.581  $   &   $   27.451
$   &   $   2.523 $   &   $   6.286   $   &   $   4.326   $   &   $
-0.219  $   & $   -391.789    $   \cr $   ^   {   86  }$Mo    &
$   721.162 $   &   $   13.031  $   &   $   26.612  $   &   $
2.984   $   & $   6.944   $   &   $   4.328   $   &   $   -0.216  $
&   $ -124.423    $   \cr $   ^   {   87  }$Mo   &   $
733.590 $ &   $   12.428  $   &   $   25.459  $   &   $   3.036   $
&   $ 8.544   $   &   $   4.327   $   &   $   -0.206  $   &   $
-382.741 $   \cr $   ^   {   88  }$Mo     &   $   746.794 $   &
$ 13.204  $   &   $   25.632  $   &   $   -2.630  $   &   $   4.507
$   &   $   4.313   $   &   $   0.089   $   &   $   168.503 $   \cr
$   ^   {   89  }$Mo    &   $   759.356 $   &   $   12.562  $ &
$   25.766  $   &   $   4.577   $   &   $   10.689  $   &   $ 4.312
$   &   $   0.072   $   &   $   139.889 $   \cr $   ^   { 90  }$Mo
   &   $   771.713 $   &   $   12.357  $   &   $ 24.919  $   &   $
5.060   $   &   $   11.794  $   &   $   4.310 $   &   $   0.002   $
&   $   4.520   $   \cr $   ^   {   91  }$Mo     &   $  783.996 $
&   $   12.283  $   &   $   24.640  $ &   $   5.616   $   &   $
12.903  $   &   $   4.311   $   &   $ 0.000   $   &   $   0.785   $
\cr $   ^   {   92  }$Mo     & $   796.152 $   &   $   12.156  $
&   $   24.439  $   &   $ 6.188   $   &   $   13.747  $   &   $
4.312   $   &   $   0.001 $   &   $   1.590   $   \cr $   ^   {   93
}$Mo     &   $ 803.533 $   &   $   7.381   $   &   $   19.537  $
&   $   6.604 $   &   $   14.764  $   &   $   4.323   $   &   $
0.001   $   & $   2.778   $   \cr $   ^   {   94  }$Mo     &   $
811.005 $ &   $   7.473   $   &   $   14.853  $   &   $   7.198   $
&   $ 16.265  $   &   $   4.351   $   &   $   0.141   $   &   $
298.515 $   \cr $   ^   {   95  }$Mo     &   $   818.910 $   &
$ 7.905   $   &   $   15.377  $   &   $   7.912   $   &   $   17.123
$   &   $   4.367   $   &   $   0.167   $   &   $   359.236 $   \cr
$   ^   {   96  }$Mo     &   $   826.751 $   &   $   7.842   $ &
$   15.746  $   &   $   8.346   $   &   $   18.149  $   &   $ 4.383
$   &   $   0.193   $   &   $   424.506 $   \cr $   ^   { 97  }$Mo
   &   $   834.582 $   &   $   7.831   $   &   $ 15.672  $   &   $
8.757   $   &   $   19.053  $   &   $   4.401 $   &   $   0.219   $
&   $   489.413 $   \cr $   ^   {   98  }$Mo     &   $   842.419 $
&   $   7.837   $   &   $   15.668  $ &   $   9.167   $   &   $
19.675  $   &   $   4.420   $   &   $ 0.246   $   &   $   557.900 $
\cr $   ^   {   99  }$Mo     & $   849.145 $   &   $   6.726   $
&   $   14.563  $   &   $ 9.558   $   &   $   20.173  $   &   $
4.430   $   &   $   0.247 $   &   $   570.316 $   \cr $   ^   {
100 }$Mo     &   $ 855.849 $   &   $   6.704   $   &   $   13.430
$   &   $   9.716 $   &   $   20.839  $   &   $   4.457   $   &   $
0.305   $   & $   715.077 $   \cr $   ^   {   101 }$Mo     &   $
862.776 $ &   $   6.927   $   &   $   13.630  $   &   $   10.041  $
&   $ 21.524  $   &   $   4.485   $   &   $   0.355   $   &   $
848.025 $   \cr $   ^   {   102 }$Mo     &   $   869.713 $   &
$ 6.937   $   &   $   13.864  $   &   $   10.407  $   &   $   22.269
$   &   $   4.518   $   &   $   0.408   $   &   $   989.265 $   \cr
$   ^   {   103 }$Mo     &   $   875.853 $   &   $   6.140   $ &
$   13.078  $   &   $   10.783  $   &   $   23.475  $   &   $ 4.521
$   &   $   0.392   $   &   $   967.161 $   \cr $   ^   { 104 }$Mo
   &   $   882.254 $   &   $   6.400   $   &   $ 12.541  $   &   $
11.465  $   &   $   24.635  $   &   $   4.477 $   &   $   -0.221  $
&   $   -553.322    $   \cr $   ^   {   105 }$Mo     &   $
888.479 $   &   $   6.226   $   &   $   12.626 $   &   $   11.920  $
&   $   24.492  $   &   $   4.488   $   & $   -0.223  $   &   $
-569.234    $   \cr $   ^   {   106 }$Mo    &   $   893.462 $   &
$   4.982   $   &   $   11.208  $   & $   17.400  $   &   $   33.934
$   &   $   4.541   $   &   $ 0.367   $   &   $   950.012 $   \cr $
^   {   107 }$Mo     & $   900.406 $   &   $   6.944   $   &   $
11.926  $   &   $ 12.827  $   &   $   27.375  $   &   $   4.509   $
&   $   -0.230 $   &   $   -605.484    $   \cr $   ^   {   108 }$Mo  
   &   $ 906.022 $   &   $   5.617   $   &   $   12.561  $   &
$   13.274 $   &   $   28.334  $   &   $   4.519   $   &   $
-0.233  $   & $   -621.177    $   \cr $   ^   {   109 }$Mo     &
$   911.311 $   &   $   5.288   $   &   $   10.905  $   &   $
16.005  $   & $   29.088  $   &   $   4.529   $   &   $   -0.236  $
&   $ -640.544    $   \cr $   ^   {   110 }$Mo     &   $
916.258 $ &   $   4.947   $   &   $   10.235  $   &   $   14.240  $
&   $ 28.472  $   &   $   4.539   $   &   $   -0.239  $   &   $
-657.079 $   \cr $   ^   {   111 }$Mo     &   $   919.168 $   &
$ 2.910   $   &   $   7.857   $   &   $   13.055  $   &   $   30.790
$   &   $   4.587   $   &   $   0.360   $   &   $   1006.303    $
\cr $   ^   {   112 }$Mo     &   $   925.055 $   &   $   5.887 $
&   $   8.797   $   &   $   14.998  $   &   $   31.422  $   & $
4.550   $   &   $   -0.217  $   &   $   -615.003    $   \cr $ ^   {
113 }$Mo     &   $   929.247 $   &   $   4.192   $   & $   10.079
$   &   $   15.268  $   &   $   32.098  $   &   $ 4.555   $   &   $
-0.203  $   &   $   -584.753    $   \cr $   ^ {   114 }$Mo     &
$   933.348 $   &   $   4.102   $   &   $ 8.293   $   &   $   15.660
$   &   $   32.505  $   &   $   4.560 $   &   $   -0.191  $   &   $
-557.073    $   \cr $   ^   {   115 }$Mo     &   $   937.330 $
&   $   3.981   $   &   $   8.083 $   &   $   16.040  $   &   $
33.401  $   &   $   4.567   $   & $   -0.181  $   &   $   -536.993
$   \cr $   ^   {   116 }$Mo    &   $   941.105 $   &   $   3.776
$   &   $   7.757   $   & $   16.457  $   &   $   34.043  $   &   $
4.573   $   &   $ -0.175  $   &   $   -526.136    $   \cr $   ^   {
117 }$Mo   &   $   944.543 $   &   $   3.438   $   &   $   7.214
$   &   $ 17.022  $   &   $   34.524  $   &   $   4.580   $   &   $
-0.170 $   &   $   -518.856    $   \cr $   ^   {   118 }$Mo     &
$ 947.748 $   &   $   3.205   $   &   $   6.643   $   &   $   17.071
$   &   $   35.475  $   &   $   4.564   $   &   $   0.098   $   & $
302.628 $   \cr $   ^   {   119 }$Mo     &   $   951.358 $ &   $
3.610   $   &   $   6.815   $   &   $   17.373  $   &   $ 36.428  $
&   $   4.572   $   &   $   0.103   $   &   $   323.918 $   \cr $
^   {   120 }$Mo    &   $   954.908 $   &   $ 3.550   $   &   $
7.160   $   &   $   17.903  $   &   $   37.075 $   &   $   4.578   $
&   $   0.100   $   &   $   100.229 $   \cr $   ^   {   121 }$Mo
   &   $   958.093 $   &   $   3.185   $ &   $   6.735   $   &   $
18.363  $   &   $   37.653  $   &   $ 4.583   $   &   $   0.081   $
&   $   261.880 $   \cr $   ^   { 122 }$Mo    &   $   961.156 $
&   $   3.062   $   &   $ 6.247   $   &   $   18.527  $   &   $
38.476  $   &   $   4.584 $   &   $   0.005   $   &   $   15.149  $
\cr $   ^   {   123 }$Mo     &   $   964.419 $   &   $   3.264   $
&   $   6.326   $ &   $   14.001  $   &   $   39.302  $   &   $
4.591   $   &   $ 0.001   $   &   $   3.485   $   \cr $   ^   {
124 }$Mo     & $   967.627 $   &   $   3.208   $   &   $   6.472
$   &   $ 19.361  $   &   $   39.687  $   &   $   4.597   $   &   $
-0.001 $   &   $   -3.717  $   \cr $   ^   {   125 }$Mo    &   $
968.130 $   &   $   0.503   $   &   $   3.711   $   &   $   19.543 $
&   $   40.062  $   &   $   4.604   $   &   $   0.005   $   & $
16.734  $   \cr $   ^   {   126 }$Mo     &   $   968.613 $ &   $
0.483   $   &   $   0.985   $   &   $   19.727  $   &   $ 40.447  $
&   $   4.610   $   &   $   0.016   $   &   $   56.053 $   \cr $   ^
{   127 }$Mo     &   $   969.088 $   &   $ 0.476   $   &   $
0.958   $   &   $   19.931  $   &   $   40.857 $   &   $   4.617   $
&   $   0.039   $   &   $   136.351 $   \cr $   ^   {   128 }$Mo
   &   $   969.562 $   &   $   0.474   $ &   $   0.950   $   &   $
20.164  $   &   $   41.247  $   &   $ 4.626   $   &   $   0.066   $
&   $   233.689 $   \cr $   ^   { 129 }$Mo     &   $   969.988 $
&   $   0.425   $   &   $ 0.899   $   &   $   20.372  $   &   $
41.586  $   &   $   4.634 $   &   $   0.081   $   &   $   289.411 $
\cr $   ^   {   130 }$Mo     &   $   970.346 $   &   $   0.359   $
&   $   0.784   $ &   $   20.560  $   &   $   41.869  $   &   $
4.641   $   &   $ 0.088   $   &   $   321.733 $   \cr $   ^   {
131 }$Mo     & $   970.728 $   &   $   0.382   $   &   $   0.741
$   &   $ 20.814  $   &   $   41.943  $   &   $   4.655   $   &   $
-0.116 $   &   $   -426.223    $   \cr $   ^   {   132 }$Mo     &
$ 970.869 $   &   $   0.141   $   &   $   0.523   $   &   $   20.534
$   &   $   42.880  $   &   $   4.655   $   &   $   0.091   $   & $
339.110 $   \cr $   ^   {   133 }$Mo     &   $   971.822 $ &   $
0.953   $   &   $   1.094   $   &   $   21.198  $   &   $ 43.404  $
&   $   4.701   $   &   $   -0.194  $   &   $   -734.129 $   \cr $
^   {   134 }$Mo    &   $   972.293 $   &   $ 0.471   $   &   $
1.424   $   &   $   15.062  $   &   $   44.730 $   &   $   4.714   $
&   $   -0.204  $   &   $   -781.920    $ \cr $   ^   {   135 }$Mo
   &   $   972.635 $   &   $   0.342 $   &   $   0.813   $   &   $
21.701  $   &   $   44.355  $   & $   4.724   $   &   $   -0.209  $
&   $   -810.890    $   \cr
$ ^   {   136 }$Mo     &   $
972.856 $   &   $   0.220   $   & $   0.562   $   &   $   23.038  $
&   $   44.819  $   &   $ 4.734   $   &   $   -0.212  $   &   $
-829.707    $   \cr $   ^ {   137 }$Mo     &   $   972.977 $   &
$   0.121   $   &   $ 0.341   $   &   $   22.168  $   &   $   45.289
$   &   $   4.743 $   &   $   -0.213  $   &   $   -843.955    $
\cr $   ^   {   138 }$Mo    &   $   973.014 $   &   $   0.037
$   &   $   0.158 $   &   $   22.405  $   &   $   45.768  $   &   $
4.752   $   & $   -0.213  $   &   $   -855.740    $   \cr
\hline
\end{tabular}
\label{me2}
\end{table*}

\begin{table*}[ht]
\caption{Same as Table \ref{me2} but for RMF model with DD-PC1 interaction.} 
\centering 
\small
\begin{tabular}{|c|c|c|c|c|c|c|c|c|}
\hline Isotope & {\it BE/A} [MeV] & {\it S$_{n}$} [MeV] & {\it
S$_{p}$} [MeV] & {\it S$_{2n}$} [MeV] & {\it S$_{2p}$} [MeV] & {\it
r$_{c}$} [fm] & {\it $\beta_{2}$} & {\it Q$_{T}$} [barn]  \cr
\hline $   ^   {   82  }$Mo & $    668.279 $   &   $     -  $
&   $  -     $   &   $   0.478   $   &   $    -   $   &   $   4.283 $
&   $   -0.001  $   &   $   -0.020  $   \cr $   ^   {   83  }$Mo  
&   $   682.364 $   &   $   14.084  $   &   $    -   $   & $   1.064
$   &   $   2.646   $   &   $   4.285   $   &   $ -0.002  $   &   $
-0.031  $   \cr $   ^   {   84  }$Mo & $    696.096 $   &   $
13.732  $   &   $   27.816  $   &   $ 1.638   $   &   $   3.797   $
&   $   4.288   $   &   $   -0.003 $   &   $   -0.047  $   \cr $   ^
{   85  }$Mo & $ 709.559 $   &   $   13.464  $   &   $
27.196  $   &   $   -1.987 $   &   $   4.931   $   &   $   4.290   $
&   $   -0.004  $   & $   -0.071  $   \cr $   ^   {   86  }$Mo  
&   $   722.787 $ &   $   13.228  $   &   $   26.691  $   &   $
-1.166  $   &   $ 6.049   $   &   $   4.293   $   &   $   -0.006  $
&   $   -0.106 $   \cr $   ^   {   87  }$Mo & $   735.797 $
&   $ 13.010  $   &   $   26.238  $   &   $   -6.949  $   &   $
7.152 $   &   $   4.297   $   &   $   -0.008  $   &   $   -0.144  $
\cr $   ^   {   88  }$Mo & $   748.601 $   &   $   12.804
$ &   $   25.814  $   &   $   -2.784  $   &   $   8.239   $   &   $
4.300   $   &   $   -0.008  $   &   $   -0.158  $   \cr $   ^   { 89
}$Mo & $   761.205 $   &   $   12.604  $   &   $ 25.408  $
&   $   4.420   $   &   $   4.192   $   &   $   4.303 $   &   $
-0.007  $   &   $   -0.139  $   \cr $   ^   {   90  }$Mo     &   $
773.616 $   &   $   12.410  $   &   $   25.015  $ &   $   4.959   $
&   $   10.372  $   &   $   4.307   $   &   $ -0.005  $   &   $
-0.105  $   \cr $   ^   {   91  }$Mo & $    785.835 $   &   $
12.219  $   &   $   24.630  $   &   $ 5.492   $   &   $   11.419  $
&   $   4.310   $   &   $   -0.004 $   &   $   -0.075  $   \cr $   ^
{   92  }$Mo & $ 797.862 $   &   $   12.027  $   &   $
24.246  $   &   $   6.017 $   &   $   12.455  $   &   $   4.314   $
&   $   -0.003  $   & $   -0.053  $   \cr $   ^   {   93  }$Mo  
&   $   805.335 $ &   $   7.473   $   &   $   19.500  $   &   $
8.168   $   &   $ 13.303  $   &   $   4.322   $   &   $   -0.005  $
&   $   -0.108 $   \cr $   ^   {   94  }$Mo & $   812.696 $
&   $ 7.361   $   &   $   14.834  $   &   $   6.863   $   &   $
14.157 $   &   $   4.331   $   &   $   -0.015  $   &   $   -0.316  $
\cr $   ^   {   95  }$Mo & $   820.845 $   &   $   8.149
$ &   $   15.510  $   &   $   7.943   $   &   $   15.908  $   &   $
4.365   $   &   $   0.162   $   &   $   3.492   $   \cr $   ^   { 96
}$Mo & $   827.802 $   &   $   6.957   $   &   $ 15.106  $
&   $   7.436   $   &   $   16.010  $   &   $   4.382 $   &   $
-0.169  $   &   $   -3.704  $   \cr $   ^   {   97  }$Mo     &   $
836.346 $   &   $   8.543   $   &   $   15.501  $ &   $   8.528   $
&   $   17.596  $   &   $   4.401   $   &   $ 0.213   $   &   $
4.755   $   \cr $   ^   {   98  }$Mo & $    844.095 $   &   $
7.749   $   &   $   16.293  $   &   $ 9.261   $   &   $   18.588  $
&   $   4.420   $   &   $   0.238 $   &   $   5.402   $   \cr $   ^
{   99  }$Mo & $ 851.069 $   &   $   6.973   $   &   $
14.723  $   &   $   9.246 $   &   $   19.001  $   &   $   4.430   $
&   $   0.240   $   & $   5.545   $   \cr $   ^   {   100 }$Mo  
&   $   857.896 $ &   $   6.827   $   &   $   13.801  $   &   $
9.299   $   &   $ 19.481  $   &   $   4.437   $   &   $   -0.224  $
&   $   -5.250 $   \cr $   ^   {   101 }$Mo & $   864.909 $
&   $ 7.013   $   &   $   13.840  $   &   $   9.743   $   &   $
19.550 $   &   $   4.448   $   &   $   -0.228  $   &   $   -5.433  $
\cr $   ^   {   102 }$Mo & $   871.813 $   &   $   6.904
$ &   $   13.917  $   &   $   10.067  $   &   $   20.676  $   &   $
4.529   $   &   $   0.413   $   &   $   10.028  $   \cr $   ^   {
103 }$Mo & $   878.462 $   &   $   6.649   $   &   $ 13.553
$   &   $   10.636  $   &   $   21.637  $   &   $   4.469 $   &   $
-0.230  $   &   $   -5.677  $   \cr $   ^   {   104 }$Mo     &   $
885.030 $   &   $   6.568   $   &   $   13.217  $ &   $   11.078  $
&   $   22.742  $   &   $   4.479   $   &   $ -0.231  $   &   $
-5.792  $   \cr $   ^   {   105 }$Mo & $    891.452 $   &   $
6.423   $   &   $   12.990  $   &   $ 11.517  $   &   $   23.870  $
&   $   4.488   $   &   $   -0.232 $   &   $   -5.915  $   \cr $   ^
{   106 }$Mo & $ 897.712 $   &   $   6.260   $   &   $
12.683  $   &   $   17.668 $   &   $   24.837  $   &   $   4.498   $
&   $   -0.233  $   & $   -6.040  $   \cr $   ^   {   107 }$Mo  
&   $   903.780 $ &   $   6.068   $   &   $   12.328  $   &   $
12.175  $   &   $ 32.639  $   &   $   4.508   $   &   $   -0.234  $
&   $   -6.159 $   \cr $   ^   {   108 }$Mo & $   908.079 $
&   $ 4.299   $   &   $   10.367  $   &   $   11.273  $   &   $
25.019 $   &   $   4.572   $   &   $   0.374   $   &   $   9.997   $
\cr $   ^   {   109 }$Mo & $   915.171 $   &   $   7.091
$ &   $   11.390  $   &   $   13.253  $   &   $   27.436  $   &   $
4.529   $   &   $   -0.238  $   &   $   -6.455  $   \cr $   ^   {
110 }$Mo & $   920.388 $   &   $   5.217   $   &   $ 12.308
$   &   $   13.699  $   &   $   28.298  $   &   $   4.539 $   &   $
-0.240  $   &   $   -6.615  $   \cr $   ^   {   111 }$Mo     &   $
925.163 $   &   $   4.776   $   &   $   9.993   $ &   $   14.112  $
&   $   29.141  $   &   $   4.546   $   &   $ -0.235  $   &   $
-6.563  $   \cr $   ^   {   112 }$Mo & $    929.693 $   &   $
4.530   $   &   $   9.306   $   &   $ 14.481  $   &   $   29.882  $
&   $   4.551   $   &   $   -0.224 $   &   $   -6.340  $   \cr $   ^
{   113 }$Mo & $ 934.109 $   &   $   4.416   $   &   $
8.946   $   &   $   14.814 $   &   $   30.532  $   &   $   4.556   $
&   $   -0.210  $   & $   -6.048  $   \cr $   ^   {   114 }$Mo  
&   $   938.449 $ &   $   4.339   $   &   $   8.756   $   &   $
15.140  $   &   $ 31.161  $   &   $   4.561   $   &   $   -0.197  $
&   $   -5.751 $   \cr $   ^   {   115 }$Mo & $   942.700 $
&   $ 4.251   $   &   $   8.591   $   &   $   15.479  $   &   $
31.826 $   &   $   4.566   $   &   $   -0.185  $   &   $   -5.494  $
\cr $   ^   {   116 }$Mo & $   946.805 $   &   $   4.105
$ &   $   8.357   $   &   $   15.844  $   &   $   32.547  $   &   $
4.573   $   &   $   -0.177  $   &   $   -5.320  $   \cr $   ^   {
117 }$Mo & $   950.668 $   &   $   3.863   $   &   $ 7.968
$   &   $   16.223  $   &   $   33.199  $   &   $   4.580 $   &   $
-0.171  $   &   $   -5.231  $   \cr $   ^   {   118 }$Mo     &   $
954.243 $   &   $   3.575   $   &   $   7.438   $ &   $   16.480  $
&   $   33.430  $   &   $   4.586   $   &   $ -0.166  $   &   $
-5.151  $   \cr $   ^   {   119 }$Mo & $    958.225 $   &   $
3.982   $   &   $   7.557   $   &   $ 16.800  $   &   $   34.148  $
&   $   4.551   $   &   $   -0.012 $   &   $   -0.364  $   \cr $   ^
{   120 }$Mo & $ 962.191 $   &   $   3.966   $   &   $
7.948   $   &   $   17.182 $   &   $   34.922  $   &   $   4.560   $
&   $   -0.009  $   & $   -0.271  $   \cr $   ^   {   121 }$Mo  
&   $   966.079 $ &   $   3.888   $   &   $   7.854   $   &   $
17.561  $   &   $ 35.691  $   &   $   4.568   $   &   $   -0.005  $
&   $   -0.169 $   \cr $   ^   {   122 }$Mo & $   969.872 $
&   $ 3.793   $   &   $   7.681   $   &   $   17.918  $   &   $
36.434 $   &   $   4.577   $   &   $   0.000   $   &   $   0.002   $
\cr $   ^   {   123 }$Mo & $   973.622 $   &   $   3.750
$ &   $   7.543   $   &   $   18.306  $   &   $   37.205  $   &   $
4.586   $   &   $   0.000   $   &   $   -0.017  $   \cr $   ^   {
124 }$Mo & $   977.291 $   &   $   3.669   $   &   $ 7.419
$   &   $   18.687  $   &   $   37.955  $   &   $   4.593 $   &   $
0.000   $   &   $   -0.014  $   \cr $   ^   {   125 }$Mo     &   $
977.505 $   &   $   0.214   $   &   $   3.883   $ &   $   18.894  $
&   $   38.402  $   &   $   4.600   $   &   $ 0.004   $   &   $
0.133   $   \cr $   ^   {   126 }$Mo & $    977.686 $   &   $
0.181   $   &   $   0.395   $   &   $ 19.104  $   &   $   38.835  $
&   $   4.606   $   &   $   0.012 $   &   $   0.411   $   \cr $   ^
{   127 }$Mo & $ 977.846 $   &   $   0.160   $   &   $
0.341   $   &   $   19.327 $   &   $   39.273  $   &   $   4.613   $
&   $   0.032   $   & $   1.121   $   \cr $   ^   {   128 }$Mo  
&   $   978.189 $ &   $   0.343   $   &   $   0.503   $   &   $
19.771  $   &   $ 39.913  $   &   $   4.635   $   &   $   0.114   $
&   $   4.050 $   \cr $   ^   {   129 }$Mo & $   978.717 $
&   $ 0.528   $   &   $   0.871   $   &   $   20.242  $   &   $
40.747 $   &   $   4.656   $   &   $   0.153   $   &   $   5.483   $
\cr $   ^   {   130 }$Mo & $   979.261 $   &   $   0.544
$ &   $   1.072   $   &   $   20.571  $   &   $   41.410  $   &   $
4.674   $   &   $   0.178   $   &   $   6.456   $   \cr $   ^   {
131 }$Mo & $   979.673 $   &   $   0.412   $   &   $ 0.956
$   &   $   20.845  $   &   $   41.946  $   &   $   4.689 $   &   $
0.192   $   &   $   7.058   $   \cr $   ^   {   132 }$Mo     &   $
979.962 $   &   $   0.289   $   &   $   0.701   $ &   $   21.096  $
&   $   42.403  $   &   $   4.701   $   &   $ 0.201   $   &   $
0.002   $   \cr $   ^   {   133 }$Mo & $    980.165 $   &   $
0.203   $   &   $   0.491   $   &   $ 21.337  $   &   $   42.596  $
&   $   4.713   $   &   $   0.208 $   &   $   7.847   $   \cr $   ^
{   134 }$Mo & $ 980.298 $   &   $   0.133   $   &   $
0.336   $   &   $   21.333 $   &   $   42.784  $   &   $   4.725   $
&   $   0.214   $   & $   8.170   $   \cr $   ^   {   135 }$Mo  
&   $   980.367 $ &   $   0.069   $   &   $   0.202   $   &   $
21.297  $   &   $ 42.990  $   &   $   4.731   $   &   $   -0.209  $
&   $   -8.103 $   \cr $   ^   {   136 }$Mo & $   980.675 $
&   $ 0.309   $   &   $   0.378   $   &   $   21.562  $   &   $
43.508 $   &   $   4.743   $   &   $   -0.216  $   &   $   -8.456  $
\cr $   ^   {   137 }$Mo & $   980.936 $   &   $   0.261
$ &   $   0.569   $   &   $   21.832  $   &   $   44.038  $   &   $
4.755   $   &   $   -0.221  $   &   $   -8.775  $   \cr $   ^   {
138 }$Mo & $   981.152 $   &   $   0.216   $   &   $ 0.477
$   &   $   23.573  $   &   $   46.258  $   &   $   4.767 $   &   $
-0.226  $   &   $   -9.082  $   \cr
\hline
\end{tabular}
\label{pc1}
\end{table*}

\begin{table*}[ht]
\caption{The pn-QRPA calculated centroid of GT strength distributions of Mo isotopes along EC ($\bar{E}_{+}$) and $\beta$-decay ($\bar{E}_{-}$) directions.} 
\centering 
\begin{tabular}{|c| c| c| c| c| c| c| c} 

Nuclei & $\bar{E}_{+}$ & $\bar{E}_{-}$& Nuclei & $\bar{E}_{+}$ & $\bar{E}_{-}$ \\ [0.5ex] 
\hline 
$\rm^{82}$Mo&  1.58&4.48& $\rm^{112}$Mo& 3.24&13.6   \\
$\rm^{83}$Mo&  4.64&4.23& $\rm^{113}$Mo& 5.85&14.5   \\
$\rm^{84}$Mo&  3.50&3.39& $\rm^{114}$Mo& 9.90&21.2   \\
$\rm^{85}$Mo&  2.87&5.31& $\rm^{115}$Mo& 12.0&24.9   \\
$\rm^{86}$Mo&  3.13&3.59& $\rm^{116}$Mo& 4.12&20.3   \\
$\rm^{87}$Mo&  3.98&6.87& $\rm^{117}$Mo& 5.23&14.6   \\
$\rm^{88}$Mo&  2.32&3.38& $\rm^{118}$Mo& 10.8&7.14   \\
$\rm^{89}$Mo&  3.51&6.26& $\rm^{119}$Mo& 12.8&9.06   \\
$\rm^{90}$Mo&  2.01&3.44& $\rm^{120}$Mo& 11.4&7.59   \\
$\rm^{91}$Mo&  4.56&3.06& $\rm^{121}$Mo& 13.2&9.18   \\
$\rm^{92}$Mo&  9.21&15.1& $\rm^{122}$Mo& 14.2&10.9   \\
$\rm^{93}$Mo&  15.1&16.4& $\rm^{123}$Mo& 15.3&11.0   \\
$\rm^{94}$Mo&  8.38&18.0& $\rm^{124}$Mo& 14.3&13.2   \\
$\rm^{96}$Mo&  7.07&20.6& $\rm^{125}$Mo& 17.5&12.1   \\
$\rm^{98}$Mo&  5.84&16.0& $\rm^{126}$Mo& 12.6&8.73   \\
$\rm^{99}$Mo&  22.5&21.9& $\rm^{127}$Mo& 15.5&12.1   \\
$\rm^{100}$Mo& 4.93&19.3& $\rm^{128}$Mo& 13.2&9.31   \\
$\rm^{101}$Mo& 8.67&9.93& $\rm^{129}$Mo& 15.7&13.2   \\
$\rm^{102}$Mo& 4.39&5.64& $\rm^{130}$Mo& 14.9&4.95   \\
$\rm^{103}$Mo& 5.21&7.60& $\rm^{131}$Mo& 17.1&11.7   \\
$\rm^{104}$Mo& 2.76&9.77& $\rm^{132}$Mo& 19.0&10.3   \\
$\rm^{105}$Mo& 4.32&11.4& $\rm^{133}$Mo& 15.5&11.1   \\
$\rm^{106}$Mo& 1.24&16.1& $\rm^{134}$Mo& 13.9&9.36   \\
$\rm^{107}$Mo& 2.59&13.4& $\rm^{135}$Mo& 15.9&11.4   \\
$\rm^{108}$Mo& 2.58&17.3& $\rm^{136}$Mo& 13.8&9.42   \\
$\rm^{109}$Mo& 1.41&18.9& $\rm^{137}$Mo& 8.56&8.77   \\
$\rm^{110}$Mo& 2.24&10.7& $\rm^{138}$Mo& 14.3&9.66   \\
$\rm^{111}$Mo& 4.37&10.2&  -  &- &- \\

\end{tabular}
\label{ebar} 
\end{table*}

\begin{table*}[ht]\label{a}
\caption{The pn-QRPA calculated EC and $\beta$-decay rates
of Mo isotopes at fixed density of 10$^7$
$g/cm^{3}$ as a function of core temperature. The decay rates are tabulated in log to base 10 scale in units of $s^{-1}$.  } 
\centering 
\begin{tabular}{|c|c|c|c|c|c|c|c|c|c| } 
\hline 
Nuclei & ${\rho}Y_{e}$  &  $T_{9}$&  EC  & $\beta$-decay &  Nuclei  &${\rho}Y_{e}$  &  $T_{9}$&  EC  & $\beta$-decay  \\ [0.5ex] 
\hline 
$\rm^{82}$Mo&  $10^{7}$ &  1.00 $\times10^{9}$ & -4.86$
\times10^{-1}$& $<$  - 1.00$\times10^{2}$ & $\rm^{112}$Mo& $10^{7}$ &
1.00 $\times10^{9}$ & -6.39 $\times10^{1}$ & 6.62$\times 10^{-1}$
\\$$&$10^{7}$ & 10.0 $\times10^{9}$ & 6.57$\times 10^{-1}$
&$<$  - 1.00$\times10^{2}$&$$&$10^{7}$ &
10.0$\times10^{9}$&-3.89$\times10^{0}$&
1.19$\times10^{0}$\\ &$10^{7}$ & 30.0
$\times10^{9}$ &3.23$\times10^{0}$& $<$  - 1.00$\times10^{2}$& &$10^{7}$
& 30.0 $\times10^{9}$&2.38$\times10^{0}$&
2.01$\times10^{0}$\\$\rm^{83}$Mo& $10^{7}$ & 1.00 $\times10^{9}$ &
1.18$\times10^{0}$& $<$  - 1.00$\times10^{2}$&$\rm^{113}$Mo& $10^{7}$ &
1.00 $\times10^{9}$ & -5.87$\times 10^{1}$&
9.58$\times10^{-1}$\\$$&$10^{7}$ & 10.0 $\times10^{9}$&2.36$\times
10^{0}$& $<$  -  1.00$\times10^{2}$&$$&$10^{7}$ & 10.0
$\times10^{9}$&-3.89$\times 10^{0}$& 1.19$\times10^{0}$\\$$&$10^{7}$
& 30.0 $\times10^{9}$&4.92$\times 10^{0}$& $<$  -  1.00$\times10^{2}$&$$&$10^{7}$ & 30 $\times10^{9}$&2.86$\times
10^{0}$& 2.5$\times 10^{0}$\\$\rm^{84}$Mo& $10^{7}$ &1.00
$\times10^{9}$ & -4.60$\times10^{-1}$& $<$  -  1.00$\times10^{2}$&$\rm^{114}$Mo& $10^{7}$ & 1.00 $\times10^{9}$ &
-7.19$\times10^{1}$& 1.04$\times 10^{0}$\\$$&$10^{7}$ & 10.0
$\times10^{9}$&5.11$\times 10^{-1}$ &
$<$  -  1.00$\times10^{2}$&$$&$10^{7}$ & 10.0 $\times10^{9}$
&-4.23$\times 10^{0}$& 1.46$\times 10^{0}$\\$$&$10^{7}$ & 30.0
$\times10^{9}$&2.63$\times10^{0}$& $<$  -  1.00$\times10^{2}$&$$&$10^{7}$
& 30.0 $\times10^{9}$&2.48$\times 10^{0}$& 2.18$\times
10^{0}$\\$\rm^{85}$Mo& $10^{7}$ & 1.00 $\times10^{9}$ &
-9.46$\times10^{-1}$& $<$  - 1.00$\times10^{2}$&$\rm^{115}$Mo& $10^{7}$
& 1.00 $\times10^{9}$ & -6.50$\times10^{1}$& 8.58$\times
10^{-1}$\\$$&$10^{7}$ & 10.0 $\times10^{9}$&3.76$\times 10^{-1}$&
$<$  - 1.00$\times10^{2}$&$$&$10^{7}$ & 10.0 $\times10^{9}$&-3.59$\times
10^{0}$& 1.36$\times 10^{0}$\\$$&$10^{7}$ & 30.0
$\times10^{9}$&3.38$\times 10^{0}$&
$<$  - 1.00$\times10^{2}$&$$&$10^{7}$ & 30.0$\times10^{9}$&2.84$\times
10^{0}$& 2.23$\times10^{0}$\\$\rm^{86}$Mo&  $10^{7}$ & 1.00
$\times10^{9}$ & -1.06$\times10^{0}$&
$<$  - 1.00$\times10^{2}$&$\rm^{116}$Mo& $10^{7}$ & 1.00 $\times10^{9}$
& -7.66$\times10^{1}$& 1.50$\times10^{0}$\\$$&$10^{7}$ & 10.0
$\times10^{9}$&2.47$\times10^{-1}$&
$<$  - 1.00$\times10^{2}$&$$&$10^{7}$ & 10.0
$\times10^{9}$&-4.80$\times10^{0}$ & 2.00$\times10^{0}$\\$$&$10^{7}$
& 30.0 $\times10^{9}$&2.49$\times10^{0}$ &
$<$  - 1.00$\times10^{2}$&$$&$10^{7}$ &
30.0$\times10^{9}$&2.20$\times10^{0}$&
2.68$\times10^{0}$\\$\rm^{87}$Mo& $10^{7}$ & 1.00 $\times10^{9}$ &
-1.46$\times10^{0}$& $<$  - 1.00$\times10^{2}$&$\rm^{117}$Mo& $10^{7}$ &
1.00 $\times10^{9}$ & -6.95$\times10^{1}$&
2.53$\times10^{-1}$\\$$&$10^{7}$ &
10.0$\times10^{9}$&3.61$\times10^{-1}$&
$<$  - 1.00$\times10^{2}$&$$&$10^{7}$ &
10.0$\times10^{9}$&-3.90$\times10^{0}$&
8.65$\times10^{-1}$\\$$&$10^{7}$ &
30.0$\times10^{9}$&3.28$\times10^{0}$&
$<$  - 1.00$\times10^{2}$&$$&$10^{7}$ &
30.0$\times10^{9}$&2.35$\times10^{0}$&
1.81$\times10^{0}$\\$\rm^{88}$Mo& $10^{7}$ & 1.00$\times10^{9}$ &
-1.74$\times10^{1}$& $<$  - 1.00$\times10^{2}$&$\rm^{118}$Mo& $10^{7}$ &
1.00$\times10^{9}$ & -8.16$\times10^{1}$&
8.48$\times10^{-1}$\\$$&$10^{7}$ &
10.0$\times10^{9}$&-1.87$\times10^{-1}$&
$<$  - 1.00$\times10^{2}$&$$&$10^{7}$ &
10.0$\times10^{9}$&-5.00$\times10^{0}$&
1.30$\times10^{0}$\\$$&$10^{7}$ & 30.0$\times10^{9}$&
2.18$\times10^{0}$ & $<$  - 1.00$\times10^{2}$&$$&$10^{7}$ &
30.0$\times10^{9}$&2.35$\times 10^{0}$&
2.01$\times10^{0}$\\$\rm^{89}$Mo& $10^{7}$ & 1.00$\times10^{9}$ &
-1.87$\times10^{0}$& $<$  - 1.00$\times10^{2}$&$\rm^{119}$Mo& $10^{7}$ &
1.00$\times10^{9}$ & $<$  - 1.00$\times10^{2}$&
-2.96$\times10^{1}$\\$$&$10^{7}$ &
10.0$\times10^{9}$&-1.00$\times10^{-2}$&
$<$  - 1.00$\times10^{2}$&$$&$10^{7}$ &
10.0$\times10^{9}$&-1.63$\times10^{1}$&
-6.20$\times10^{0}$\\$$&$10^{7}$ &
30.0$\times10^{9}$&3.04$\times10^{0}$&
$<$  - 1.00$\times10^{2}$&$$&$10^{7}$ &
30.0$\times10^{9}$&-3.39$\times10^{0}$&
-3.88$\times10^{0}$\\$\rm^{90}$Mo& $10^{7}$ & 1.00$\times10^{9}$ &
-2.58$\times10^{0}$& $<$  - 1.00$\times10^{2}$&$\rm^{120}$Mo& $10^{7}$ &
1.00$\times10^{9}$& -8.57$\times10^{1}$&
1.02$\times10^{0}$\\$$&$10^{7}$ &
10.0$\times10^{9}$&-3.07$\times10^{-1}$& $<$  - 1.00$\times10^{2}$
&$$&$10^{7}$ & 10.0$\times10^{9}$&-5.19$\times10^{0}$&
1.52$\times10^{0}$\\$$&$10^{7}$ &
30.0$\times10^{9}$&2.26$\times10^{0}$&
$<$  - 1.00$\times10^{2}$&$$&$10^{7}$ &
30.0$\times10^{9}$&2.32$\times10^{0}$&
2.26$\times10^{0}$\\$\rm^{91}$Mo& $10^{7}$ & 1.00$\times10^{9}$ &
-2.43$\times10^{0}$& $<$  - 1.00$\times10^{2}$&$\rm^{121}$Mo& $10^{7}$ &
1.00$\times10^{9}$ & -7.77$\times10^{1}$&
1.10$\times10^{0}$\\$$&$10^{7}$ &
10.0$\times10^{9}$&-5.35$\times10^{-1}$&
$<$  - 1.00$\times10^{2}$&$$&$10^{7}$
&10.0$\times10^{9}$&-4.42$\times10^{0}$&
1.48$\times10^{0}$\\$$&$10^{7}$ &
30.0$\times10^{9}$&2.90$\times10^{0}$&
$<$  - 1.00$\times10^{2}$&$$&$10^{7}$ &
30.0$\times10^{9}$&2.68$\times10^{0}$&
2.34$\times10^{0}$\\$\rm^{92}$Mo& $10^{7}$ & 1.00$\times10^{9}$ &
-1.08$\times10^{1}$& -4.97$\times10^{1}$&$\rm^{122}$Mo&  $10^{7}$ &
1.00$\times10^{9}$ & -8.73$\times10^{1}$& 1.23$\times10^{0}$
\\$$&$10^{7}$ & 10.0$\times10^{9}$&-1.45$\times10^{0}$ & -6.18$\times10^{0}$&$$&$10^{7}$ & 10.0$\times10^{9}$&-5.28$\times10^{0}$& 1.69$\times10^{0}$\\$$&$10^{7}$ & 30.0$\times10^{9}$&2.32$\times10^{0}$& -2.76$\times10^{0}$&$$&$10^{7}$ & 30.0$\times10^{9}$&2.38$\times10^{0}$&
2.56$\times10^{0}$\\$\rm^{93}$Mo&  $10^{7}$ & 1.00$\times10^{9}$ &
-5.00$\times10^{0}$& -2.52$\times10^{1}$&$\rm^{123}$Mo& $10^{7}$ &
1.00$\times10^{9}$ & -8.22$\times10^{1}$&
1.07$\times10^{0}$\\$$&$10^{7}$ &
10.0$\times10^{9}$&-1.84$\times10^{0}$&
-5.00$\times10^{0}$&$$&$10^{7}$ &
10.0$\times10^{9}$&-5.28$\times10^{0}$&
1.69$\times10^{0}$\\$$&$10^{7}$ &
30.0$\times10^{9}$&2.46$\times10^{0}$&
-2.79$\times10^{0}$&$$&$10^{7}$ &
30.0$\times10^{9}$&2.58$\times10^{0}$&
2.69$\times10^{0}$\\$\rm^{94}$Mo& $10^{7}$ & 1.00$\times10^{9}$ &
-1.90$\times10^{1}$& -2.83$\times10^{1}$&$\rm^{124}$Mo&  $10^{7}$ &
1.00$\times10^{9}$ & $<$  - 1.00$\times10^{2}$&
-1.17$\times10^{1}$\\$$&$10^{7}$ &
10.0$\times10^{9}$&-1.32$\times10^{0}$&
-3.58$\times10^{0}$&$$&$10^{7}$ &
10.0$\times10^{9}$&-7.15$\times10^{0}$&
-4.62$\times10^{-1}$\\$$&$10^{7}$ &
30.0$\times10^{9}$&2.62$\times10^{0}$&
-1.17$\times10^{0}$&$$&$10^{7}$ &
30.0$\times10^{9}$&1.94$\times10^{0}$&2.45$\times10^{0}$
\\
\hline
\end{tabular}
\label{r1} 
\end{table*}

\begin{table*}
\centering 
\begin{tabular}{|c| c| c| c| c| c| c| c| c| c |} 
\hline 
Nuclei & ${\rho}Y_{e}$  &  $T_{9}$&  EC & $\beta$-decay &  Nuclei  &${\rho}Y_{e}$  &  $T_{9}$&  EC  & $\beta$-decay  \\ [0.5ex] 
\hline
$\rm^{96}$Mo&  $10^{7}$ & 1.00 $\times10^{9}$ & -2.09$\times10^{1}$&
-2.36$\times10^{1}$&$\rm^{125}$Mo&  $10^{7}$ & 1.00 $\times10^{9}$ &
$<$  - 1.00$\times10^{2}$& 1.50$\times10^{0}$\\$$&$10^{7}$ & 10.0
$\times10^{9}$&-2.18$\times10^{0}$& -3.07$\times10^{0}$&$$&$10^{7}$
& 10.0 $\times10^{9}$&-1.10$\times10^{1}$&
2.04$\times10^{0}$\\$$&$10^{7}$ & 30.0
$\times10^{9}$&2.24$\times10^{0}$& -9.85$\times10^{-1}$&$$&$10^{7}$
& 30.0 $\times10^{9}$&2.64$\times10^{0}$&
2.74$\times10^{0}$\\$\rm^{98}$Mo& $10^{7}$ & 1.00 $\times10^{9}$ &
-2.74$\times10^{1}$& -1.71$\times10^{0}$&$\rm^{126}$Mo& $10^{7}$ &
1.00 $\times10^{9}$ &$<$  - 1.00$\times10^{2}$&
1.53$\times10^{0}$\\$$&$10^{7}$ & 10.0
$\times10^{9}$&-2.70$\times10^{0}$& -3.26$\times10^{0}$&$$&$10^{7}$
& 10.0 $\times10^{9}$&-7.50$\times10^{0}$&
2.07$\times10^{0}$\\$$&$10^{7}$ & 30.0
$\times10^{9}$&2.60$\times10^{0}$& -4.32$\times10^{-1}$&$$&$10^{7}$
& 30.0 $\times10^{9}$&1.75$\times10^{0}$&
2.84$\times10^{0}$\\$\rm^{99}$Mo& $10^{7}$ & 1.00 $\times10^{9}$ &
-1.78$\times10^{1}$& -5.37$\times10^{0}$&$\rm^{127}$Mo& $10^{7}$ &
1.00 $\times10^{9}$ &$<$  - 1.00$\times10^{2}$&
1.47$\times10^{0}$\\$$&$10^{7}$ & 10.0
$\times10^{9}$&-2.74$\times10^{0}$& -2.20$\times10^{0}$&$$&$10^{7}$
& 10.0 $\times10^{9}$&-1.18$\times10^{1}$&
1.90$\times10^{0}$\\$$&$10^{7}$ & 30.0
$\times10^{9}$&2.60$\times10^{0}$& -4.32$\times10^{-1}$&$$&$10^{7}$
& 30.0 $\times10^{9}$&-2.30$\times10^{0}$&
2.80$\times10^{0}$\\$\rm^{100}$Mo& $10^{7}$ & 1.00 $\times10^{9}$ &
-3.17$\times10^{1}$& -1.07$\times10^{1}$&$\rm^{128}$Mo& $10^{7}$ &
1.00 $\times10^{9}$ & $<$  - 1.00$\times10^{2}$&
1.60$\times10^{0}$\\$$&$10^{7}$ & 10.0
$\times10^{9}$&-1.90$\times10^{0}$& -1.07$\times10^{0}$&$$&$10^{7}$
& 10.0 $\times10^{9}$&-1.18$\times10^{1}$ &
1.90$\times10^{0}$\\$$&$10^{7}$ & 30.0
$\times10^{9}$&2.56$\times10^{0}$& 0.00$\times10^{0}$&$$&$10^{7}$ &
30.0$\times10^{9}$&1.61$\times10^{0}$&
2.90$\times10^{0}$\\$\rm^{101}$Mo& $10^{7}$ & 1.00 $\times10^{9}$ &
-2.20$\times10^{1}$& -2.01$\times10^{0}$&$\rm^{129}$Mo& $10^{7}$ &
1.00 $\times10^{9}$ & $<$  - 1.00$\times10^{2}$&
1.25$\times10^{0}$\\$$&$10^{7}$ & 10.0
$\times10^{9}$&-1.13$\times10^{0}$& 2.86$\times10^{-1}$&$$&$10^{7}$
& 10.0 $\times10^{9}$&-1.11$\times10^{1}$&
2.23$\times10^{0}$\\$$&$10^{7}$ & 30.0
$\times10^{9}$&3.40$\times10^{0}$& 1.51$\times10^{0}$&$$&$10^{7}$ &
30.0 $\times10^{9}$&-1.59$\times10^{0}$&
3.20$\times10^{0}$\\$\rm^{102}$Mo& $10^{7}$ & 1.00 $\times10^{9}$ &
-3.48$\times10^{1}$& -4.00$\times10^{0}$&$\rm^{130}$Mo& $10^{7}$ &
1.00 $\times10^{9}$ & $<$  - 1.00$\times10^{2}$&
1.31$\times10^{0}$\\$$&$10^{7}$ & 10.0
$\times10^{9}$&-2.64$\times10^{0}$& -6.02$\times10^{-1}$&$$&$10^{7}$
& 10.0 $\times10^{9}$&-1.80$\times10^{1}$&
2.30$\times10^{0}$\\$$&$10^{7}$ & 30.0
$\times10^{9}$&2.43$\times10^{0}$& 8.53$\times10^{-1}$&$$&$10^{7}$ &
30.0$\times10^{9}$&1.50$\times10^{0}$&
3.10$\times10^{0}$\\$\rm^{103}$Mo& $10^{7}$ & 1.00 $\times10^{9}$ &
-2.87$\times10^{1}$& -8.60$\times10^{0}$&$\rm^{131}$Mo& $10^{7}$ &
1.00 $\times10^{9}$ & $<$  - 1.00$\times10^{2}$&
-1.37$\times10^{0}$\\$$&$10^{7}$ & 10.0
$\times10^{9}$&-2.50$\times10^{0}$& -3.41$\times10^{-1}$&$$&$10^{7}$
& 10.0 $\times10^{9}$&-1.12$\times10^{1}$&
2.48$\times10^{0}$\\$$&$10^{7}$ & 30.0 $\times10^{9}$&
2.93$\times10^{0}$ & 1.05$\times10^{0}$&$$&$10^{7}$ &
30.0$\times10^{9}$ &-1.55$\times10^{0}$&
3.29$\times10^{0}$\\$\rm^{104}$Mo& $10^{7}$ & 1.00 $\times10^{9}$ &
-4.15$\times10^{1}$& -3.09$\times10^{0}$&$\rm^{132}$Mo& $10^{7}$ &
1.00 $\times10^{9}$ & $<$  - 1.00$\times10^{2}$&
1.41$\times10^{0}$\\$$&$10^{7}$ & 10.0
$\times10^{9}$&-3.15$\times10^{0}$& 7.50$\times10^{-2}$&$$&$10^{7}$
& 10.0 $\times10^{9}$&-8.60$\times10^{0}$&
2.55$\times10^{0}$\\$$&$10^{7}$ & 30.0
$\times10^{9}$&2.30$\times10^{0}$& 1.24$\times10^{0}$&$$&$10^{7}$ &
30.0 $\times10^{9}$&1.46$\times10^{0}$&
3.33$\times10^{0}$\\$\rm^{105}$Mo& $10^{7}$ & 1.00 $\times10^{9}$ &
-3.53$\times10^{1}$& -8.53$\times10^{-1}$&$\rm^{133}$Mo& $10^{7}$ &
1.00 $\times10^{9}$ & $<$  - 1.00$\times10^{2}$&
1.50$\times10^{0}$\\$$&$10^{7}$ & 10.0 $\times10^{9}$&
-3.17$\times10^{0}$& 5.80$\times106{-2}$&$$&$10^{7}$ & 10.0
$\times10^{9}$&-8.60$\times10^{0}$&2.55$\times10^{0}$\\$$&$10^{7}$ &
30.0 $\times10^{9}$&2.79$\times10^{0}$&
1.22$\times10^{0}$&$$&$10^{7}$ & 30.0
$\times10^{9}$&-1.70$\times10^{0}$&
3.44$\times10^{0}$\\$\rm^{106}$Mo& $10^{7}$ & 1.00 $\times10^{9}$ &
-4.860$\times10^{1}$& -1.21$\times10^{0}$&$\rm^{134}$Mo& $10^{7}$ &
1.00 $\times10^{9}$ & $<$  - 1.00$\times10^{2}$&
1.52$\times10^{0}$\\$$&$10^{7}$ & 10.0
$\times10^{9}$&-3.80$\times10^{0}$& -1.11$\times10^{-1}$&$$&$10^{7}$
& 10.0 $\times10^{9}$&-8.52$\times10^{0}$&
2.51$\times10^{0}$\\$$&$10^{7}$ & 30.0
$\times10^{9}$&2.24$\times10^{0}$& 9.97$\times10^{-1}$&$$&$10^{7}$ &
30.0 $\times10^{9}$&1.41$\times10^{0}$&
3.22$\times10^{0}$\\$\rm^{107}$Mo& $10^{7}$ & 1.00 $\times10^{9}$ &
-4.24$\times10^{1}$& -7.71$\times10^{-1}$&$\rm^{135}$Mo& $10^{7}$
&1.00 $\times10^{9}$& $<$  - 1.00$\times10^{2}$&
1.70$\times10^{0}$\\$$&$10^{7}$ & 10.0 $\times10^{9}$&
-3.47$\times10^{0}$ & 2.27$\times10^{-1}$&$$&$10^{7}$ & 10.0
$\times10^{9}$&-1.19$\times10^{1}$ & 2.29$\times10^{0}$\\$$&$10^{7}$
& 30.0 $\times10^{9}$&2.77$\times10^{0}$&
1.43$\times10^{0}$&$$&$10^{7}$ & 30.0
$\times10^{9}$&-1.83$\times10^{0}$&
3.40$\times10^{0}$\\$\rm^{108}$Mo& $10^{7}$ & 1.00 $\times10^{9}$ &
-5.44$\times10^{1}$& -3.32$\times10^{-1}$&$\rm^{136}$Mo& $10^{7}$ &
1.00 $\times10^{9}$ & $<$  -1.00$\times10^{2}$&
1.61$\times10^{1}$\\$$&$10^{7}$ & 10.0
$\times10^{9}$&-4.00$\times10^{0}$& 3.47$\times10^{-1}$&$$&$10^{7}$
& 10.0  $\times10^{9}$&-8.72$\times10^{0}$&
2.60$\times10^{0}$\\$$&$10^{7}$ & 30.0
$\times10^{9}$&2.29$\times10^{0}$& 1.42$\times10^{0}$&$$&$10^{7}$ &
30.0 $\times10^{9}$&1.34$\times10^{0}$&
3.34$\times10^{0}$\\$\rm^{109}$Mo& $10^{7}$ & 1.00 $\times10^{9}$ &
-4.82$\times10^{1}$& -8.60$\times10^{-2}$&$\rm^{137}$Mo& $10^{7}$ &
1.00 $\times10^{9}$ & $<$  - 1.00$\times10^{2}$&
2.49$\times10^{0}$\\$$&$10^{7}$ &
10.0$\times10^{9}$&-3.47$\times10^{0}$&
4.18$\times10^{-1}$&$$&$10^{7}$ & 10.0
$\times10^{9}$&-1.21$\times10^{1}$& 2.80$\times10^{0}$\\$$&$10^{7}$
& 30.0 $\times10^{9}$&2.79$\times10^{0}$&
1.54$\times10^{0}$&$$&$10^{7}$ &
30.0$\times10^{9}$&-1.90$\times10^{0}$&
3.72$\times10^{0}$\\$\rm^{110}$Mo& $10^{7}$ & 1.00 $\times10^{9}$ &
-5.86$\times10^{1}$& 2.53$\times10^{-1}$&$\rm^{138}$Mo& $10^{7}$ &
1.00 $\times10^{9}$ & $<$  - 1.00$\times10^{2}$&
1.66$\times10^{0}$\\$$&$10^{7}$ & 10.0
$\times10^{9}$&-3.90$\times10^{0}$& 8.65$\times10^{-1}$&$$&$10^{7}$
& 10.0 $\times10^{9}$&-9.10$\times10^{0}$&
2.77$\times10^{0}$\\$$&$10^{7}$ & 30.0
$\times10^{9}$&2.35$\times10^{0}$& 1.80$\times10^{0}$&$$&$10^{7}$ &
30.0 $\times10^{9}$&1.20$\times10^{0}$& 3.70$\times10^{0}$
\\
\hline
\end{tabular}

\label{r2} 
\end{table*}

\end{document}